\documentclass[twocolumn]{aastex63}
\usepackage{graphicx}
\usepackage{verbatim}
\usepackage{slashed}
\usepackage{hyperref}
\usepackage{longtable}
\usepackage{multirow,booktabs}
\usepackage[margin=2.54cm,includeheadfoot=True]{geometry}
\usepackage{epstopdf}
\usepackage{amsmath}
\usepackage{color}
\usepackage{txfonts}
\usepackage{textcomp}
\usepackage[T1]{fontenc}

\shorttitle{LGRB progenitors and magnetar formation}
\shortauthors{Song et al.}
\graphicspath{{./}{figures/}}

\begin{document}

\title{Long-duration Gamma-ray Burst Progenitors and Magnetar Formation}

\author[0000-0001-8390-9962]{Cui-Ying Song}
\affiliation{Tsung-Dao Lee Institute, Shanghai Jiao Tong University, Shanghai 200240, China; songcuiying@sjtu.edu.cn}

\author[0000-0001-8678-6291]{Tong Liu}
\affiliation{Department of Astronomy, Xiamen University, Xiamen 361005, China; tongliu@xmu.edu.cn}

\begin{abstract}
Millisecond magnetars produced in the center of dying massive stars are one prominent model to power gamma-ray bursts (GRBs). However, their detailed nature remains a mystery. To explore the effects of the initial mass, rotation rate, wind mass loss, and metallicity on the GRB progenitors and the newborn magnetar properties, we evolve 227 of $10-30~M_\odot$ single star models from the pre-main-sequence to core collapse by using the stellar evolution code MESA. The pre-supernova properties, the compactness parameter, and magnetar characteristics of models with different initial parameters are presented. The compactness parameter remains a non-monotonic function of the initial mass and initial rotation rate when the effects of vary metallicity and ``Dutch'' wind scale factor are taken into account. We find that the initial rotation rate and mass play the dominant roles in whether a star can evolve into a GRB progenitor. The minimum rotation rate necessary to generate a magnetar gradually reduces as the initial mass increases. The greater the initial metallicity and ``Dutch'' wind scale factor, the larger the minimum rotation rate required to produce a magnetar. In other words, massive stars with low metallicity are more likely to harbor magnetars. Furthermore, we present the estimated period, magnetic field strength, and masses of magnetars in all cases. The typical rotational energy of these millisecond magnetars is sufficient to power long-duration GRBs.
\end{abstract}


\section{Introduction}

Observations of long-duration gamma-ray bursts (LGRBs) associated with core-collapse supernovae \citep[CCSNe, e.g.,][]{Galama1998,Bloom1999,Hjorth2003,Stanek2003} suggest that some fraction of LGRBs originate from the death of massive stars \citep[see reviews by][]{Piran2004,Woosley2006,Kumar2015,Cano2017}. After the massive star collapses, a black hole (BH) hyperaccretion system  \citep[e.g.,][]{Woosley1993,MacFadyen1999,Liu2017} or a rapidly rotating neutron star (NS) with a strong magnetic field \citep[magnetar, e.g.,][]{Usov1992,Duncan1992,Wheeler2000,Zhang2001} might be formed. In the BH hyperaccretion scenario, the annihilation of neutrinos and anti-neutrinos escaped from neutrino-dominated accretion disks \citep[e.g.,][see a review by \citealt{Liu2017}]{Popham1999,Narayan2001,Chen2007,Gu2006,Liu2007,Liu2015,Zalamea2011,Kawanaka2013} or the large-scale magnetic fields extract the rotational energy of BHs, i.e., Blandford-Znajek mechanism, \citep[ e.g.,][]{Blandford1977,Lee2000,Wang2002,McKinney2005,McKinney2012,Komissarov2009,Tchekhovskoy2011,Lei2013,Liu2015} to launch ultrarelativistic jets. The release energy from the spin-down of newborn magnetars could also account for gamma-ray bursts \citep[GRBs, e.g.,][]{Usov1992,Wheeler2000,Metzger2011,Metzger2017,Rowlinson2013,Kaspi2017,Hou2021}. The collimated jet production mechanism of magnetars has been studied by some literatures \citep{komissarov2007,Bucciantini2008,Bucciantini2009,Bucciantini2012,Kiuchi2012,Siegel2014,Shankar2021}. If the ultrarelativistic jet produced by the central engine could break out from the progenitor envelope and circumstellar medium and be along the line of sight, the observable LGRB could be triggered.

The characteristics of LGRB progenitors are still unclear since no direct observational evidence has been presented to reveal the association of any progenitor and the observed LGRBs. The event rate of LGRBs is much lower than that of CCSNe \citep[e.g.,][]{Podsiadlowski2004,Izzard2004,Guetta2005,Soderberg2006,Wanderman2010}, implying that only a small fraction of massive stars could produce LGRBs and CCSNe at the end of their lives. \cite{Soderberg2010} estimated that the ratio of LGRBs to Ibc supernovae (SN) is approximately 1$\%$ based on a radio survey. \cite{Georgy2012} obtained this fraction from rotating stellar models and found that it could exceed 25$\%$. The absence of H/He lines in the spectra of SNe associated with LGRBs suggests that their progenitors have undergone violent mass loss or chemical mixing of elements. Wolf-Rayet (WR) stars with striping H/He envelopes before explosion have been proposed as LGRB progenitors \citep[e.g.,][]{Woosley1993,Woosley2006a,Crowther2007}. WR stars can be subdivided into three classes based on the exhibition of broad emission lines in spectra. These groups include N-rich WR stars (WN) and C-rich and O-rich WR stars (WC/WO). The origin of WR is an ongoing study. It is proposed that rapidly rotating single massive stars could undergo quasi-chemically homogeneous evolution \citep[e.g.,][]{Yoon2005,Yoon2006,Ekstrom2012,Maeder2012}, allowing the systems to lose their envelopes while still retaining enough angular momentum to produce LGRBs. Furthermore, WR stars can also be born in massive close binaries through tidal interactions with their companion stars \citep[e.g.,][]{Podsiadlowski1992,Cantiello2007,Detmers2008,Yoon2010,Eldridge2011,Langer2012,deMink2013}.

Several studies have investigated LGRB progenitor evolution models and corresponding remnants \citep[e.g.,][]{Heger2003,Hirschi2005,Obergaulinger2017,Aguilera2018,Aloy2021,Li2023}. \cite{Yoon2005} found that quasi-chemically homogeneous evolving massive stars could fulfill the requirements of collapsar. \cite{Woosley2006} explored the evolution of stripped-down He cores and single star (12, 16, and 35 $M_{\odot}$) with rapid rotating velocity, and they concluded that such stars could retain enough angular momentum produce GRBs at core-collapse. Recently, \cite{Roy2020} showed the effects of mass, rotation rate, and metallicity of massive stars on their evolution into WN and subsequently into WC, which are possible Type Ic SNe/GRB progenitors. By studying the surface helium and nitrogen mass fraction evolution, they found that an O star with rotation rate $\Omega/ \Omega_{\rm cirt} \geq 0.4$ could evolve into the WN phase. This might denote that rapidly rotating massive stars could satisfy the requirement of Type Ic SNe/LGRB production. \cite{Aguilera2020} assumed that these successful neutrino-driven explosion models would produce magnetar and power superluminous SNe, and that failed explosion models would lead to the BH form and power LGRBs. The subnormal distribution of the initial SN explosion energy could naturally build the ``lower mass gap'' in the mass distribution of compact objects \citep{Liu2021}. However, in addition to the collapsar, the magnetar might also be the central engine of LGRBs. It is difficult to determine whether the central engine of GRB is a BH or magnetar. The shallow decay phase \citep{Dai2004,Zhang2006,Dall2011,Li2016} and ``internal plateau'' \citep{Troja2007,Lyons2010,Stratta2018} in some GRB afterglow light curves have been proposed as the magnetar signature, but the BH hyperaccretion model cannot be ruled out, especially for long-lasting plateaus \citep[e.g.,][]{Yi2022}. Regardless of the type of central engine, the shallow decay phase might be related to a precessing jet \citep{Huang2021}. In this paper, we focus on the progenitors of LGRBs and the formation of magnetars.

\begin{figure*}
\centering
\includegraphics[width=0.4\textwidth,height=0.4\textwidth]{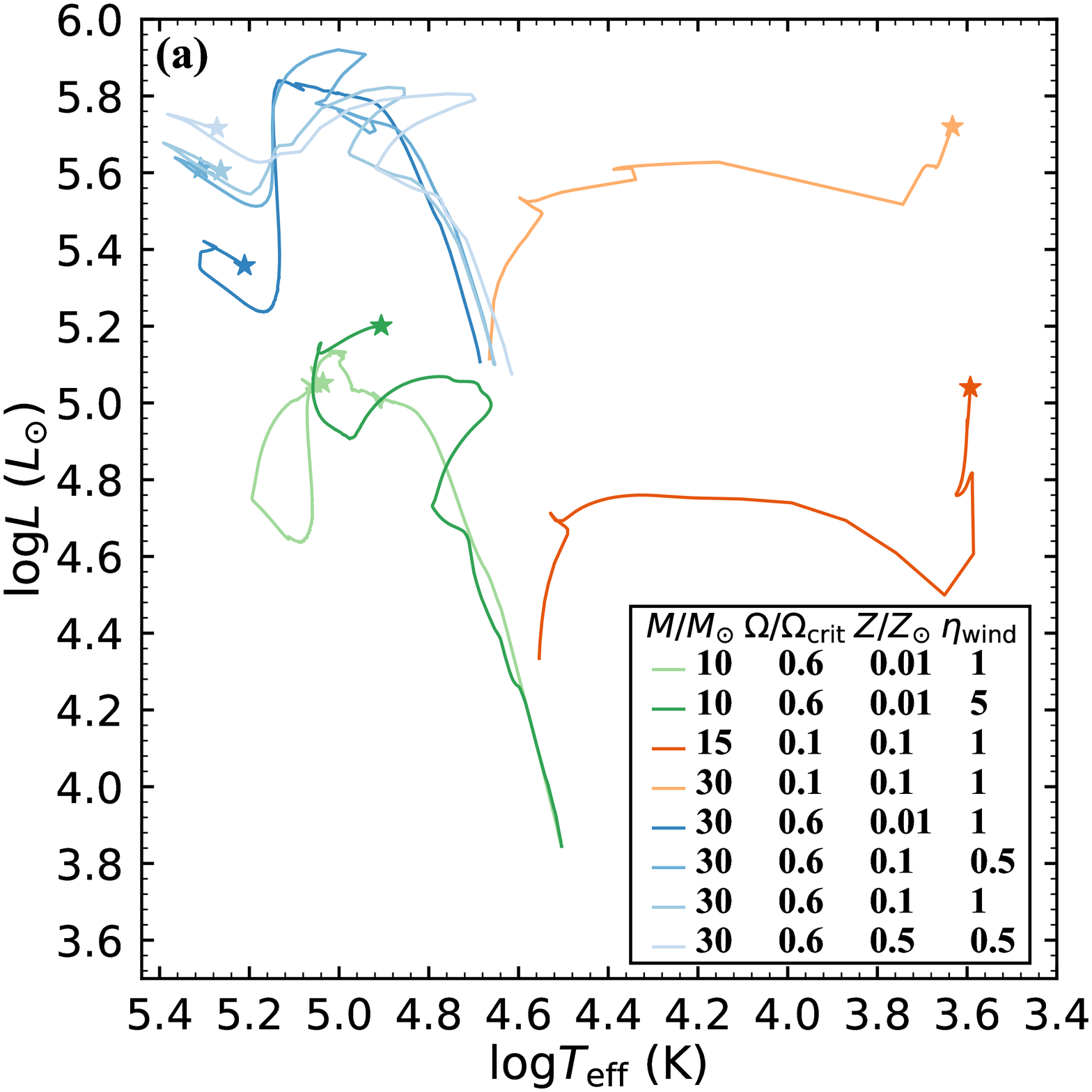}
\includegraphics[width=0.4\textwidth,height=0.4\textwidth]{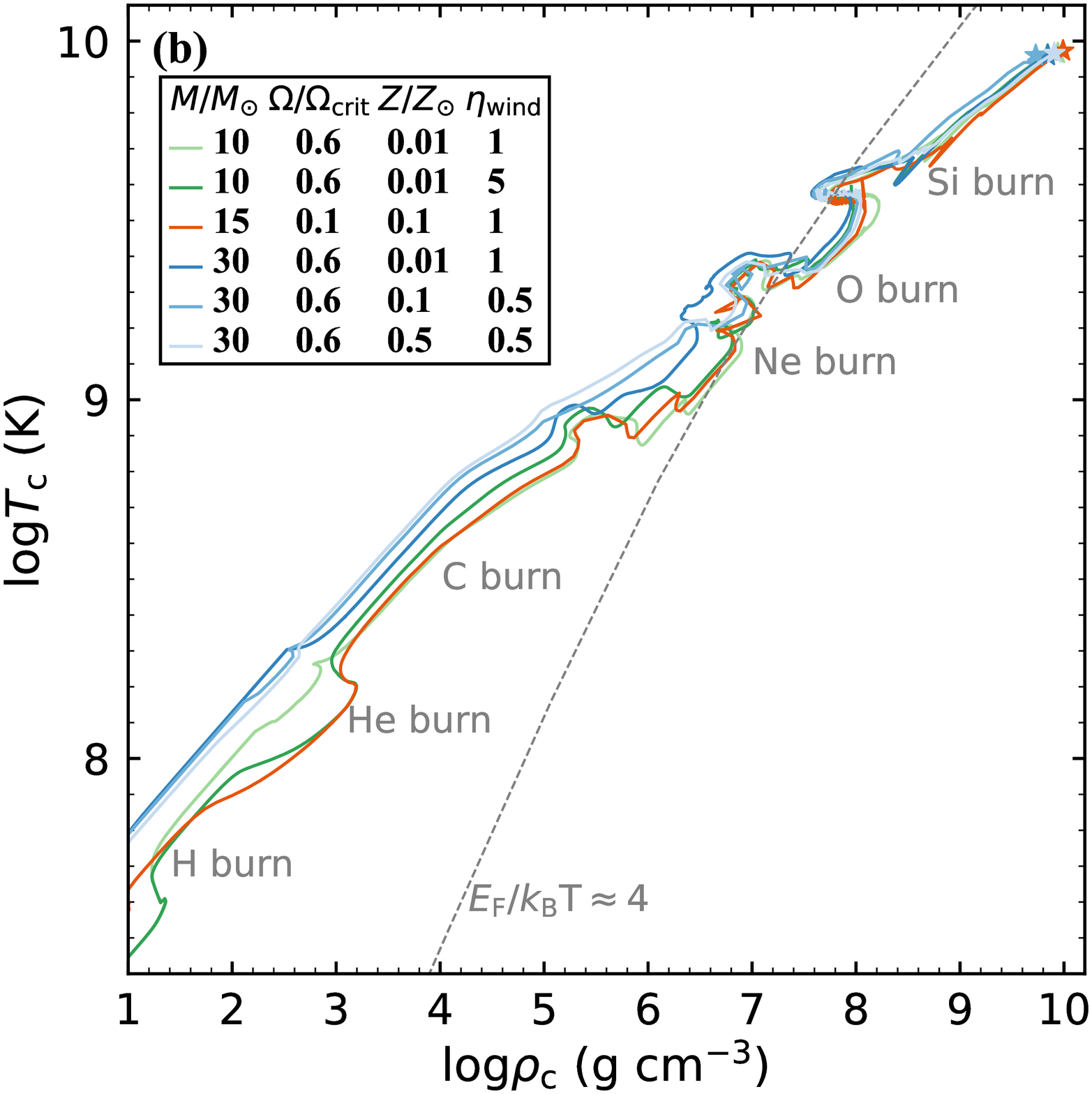}
\caption{(a) H-R diagram of the massive stars with different initial masses $M/M_{\odot}=10, 15$, and $30$, initial rotation rates $\Omega/\Omega_{\rm crit}$ = 0.1 and 0.6, metallicity $Z/Z_{\odot}=0.01, 0.1$, and $0.5$, and scaling factors of the ``Dutch'' wind loss rate $\eta_{\rm wind}$ = 0.5, 1, and 5. (b) Evolution of the central density and temperature in the models. Locations of hydrogen, helium, carbon, neon, oxygen, and silicon burning are labeled. The grey dashed line indicates the $E_{F}/k_{B}T \approx 4$ electron degeneracy curve. The endpoint of the line is marked by the pentagram, corresponding to the time of the iron core collapse.}
\label{HR}
\end{figure*}

The physical conditions for dying massive stars to produce GRBs remain an open question. Developments to enhance theoretical models are required to study how various physical parameters could replicate the formation conditions and characteristics of the GRB central engines. Conversely, to predict the final fate of massive stars and remnants, researchers should use observational GRB and SN data to constrain stellar evolution models.

In the framework of the 1D single star evolution scenario, we explore the effect of initial rotation, mass, metallicity, and mass loss on the evolution path and pre-SN properties of the massive stars. More importantly, the initial signatures of the magnetars are discussed. This paper is structured as follows. In Section 2, we describe the physical ingredients of our stellar evolution models. We present our analysis of pre-SNe and the characteristics of the protomagnetar in Section 3. Conclusions and discussion are presented in Section 4.

\section{Progenitor Modeling in MESA} \label{method}

\begin{figure*}
\centering
\includegraphics[width=0.4\textwidth,height=0.4\textwidth]{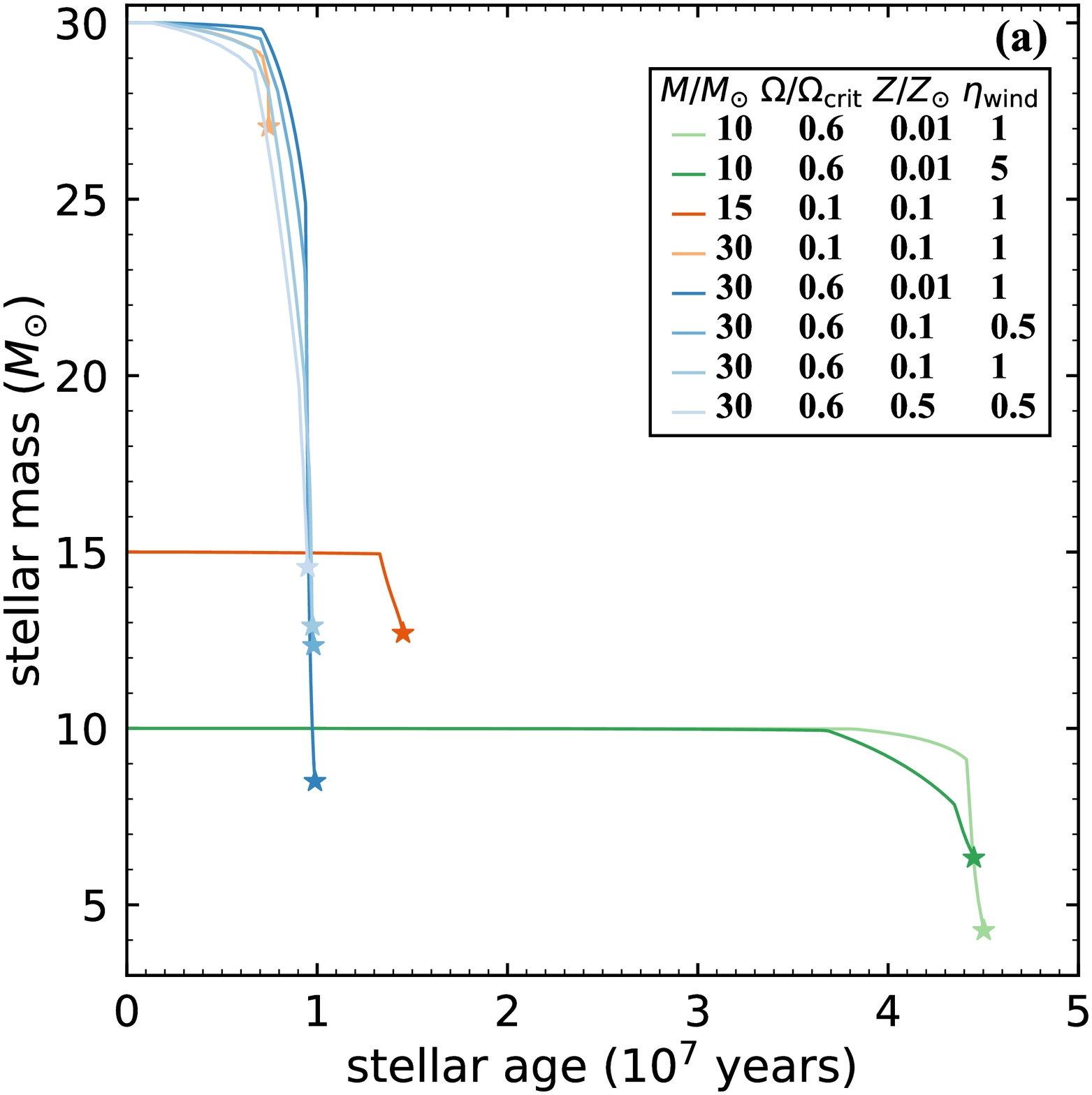}
\includegraphics[width=0.4\textwidth,height=0.4\textwidth]{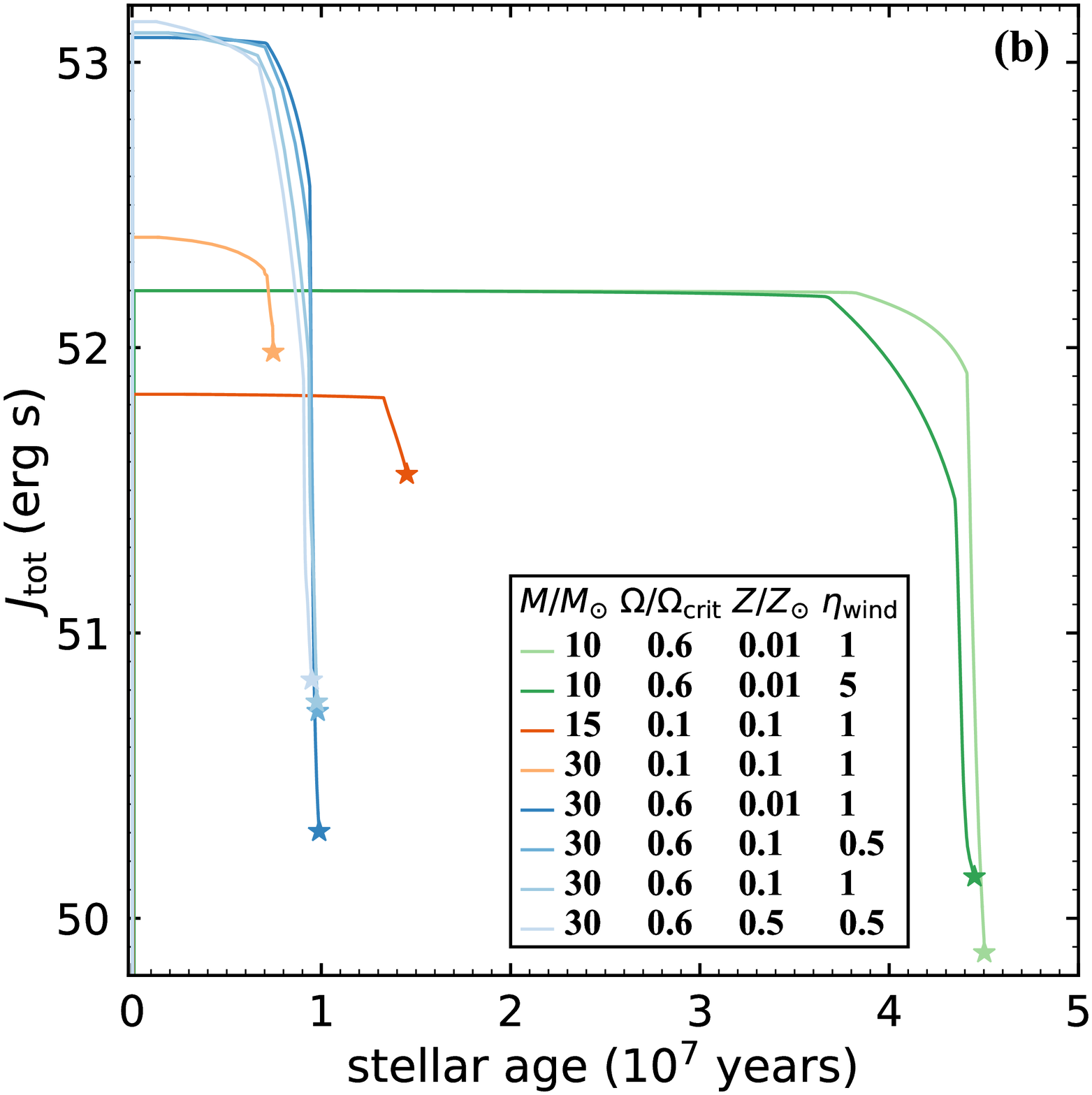}
\caption{The evolution of stellar mass and total angular momentum for models with different initial parameters.}
\label{ages}
\end{figure*}

We use the Modules for Experiments in Stellar Astrophysics software package \citep[MESA,][version 21.12.1]{Paxton2011,Paxton2013,Paxton2015,Paxton2018,Paxton2019,Jermyn2023} and MESA SDK 21.4.1 \citep{Townsend2021} to simulate a large set of  227 massive star models from the pre-main-sequence stage until the onset of iron core collapse. The software packages are available on the website \dataset[https://docs.mesastar.org]{https://docs.mesastar.org/en/release-r22.11.1/} and on Zenodo under an open-source Creative Commons Attribution license: \dataset[doi:10.5281/zenodo.7457681]{https://doi.org/10.5281/zenodo.7457681}. Here, we define collapse as the moment when the infall velocity of any point inside the stellar model exceeds 1,000 km s$^{-1}$. We take the test suite \texttt{20M$\_$pre$\_$ms$\_$to$\_$core$\_$collapse} as our baseline model. All inlists of the model are available on Zenodo\dataset[(doi:10.5281/zenodo.7839968)]{https://doi.org/10.5281/zenodo.7839968}. Our stellar models cover an initial mass range between 10 and 30 $M_{\odot}$. Here we simply expect stars in that mass range will all form NSs upon collapse. It should be noted that there is a non-mononic pattern of NS and BH formation as a function of initial mass \citep[e.g.,][]{Woosley2007,Schneider2021}.

The initial metallicity is $Z=0.01, 0.1,$ and $0.5~Z_{\odot}$, where $Z$ is the mass fraction of elements heavier than helium and $Z_{\odot}$ is the solar metallicity \citep[e.g.,][]{Grevesse1998,Von2016}. Following \cite{Tout1996} and \cite{Pols1998}, we use the initial helium fraction $Y=0.24 + 2Z$ and initial hydrogen mass fraction $X=1-Y-Z$. The initial rotational rate $\Omega/\Omega_{\rm crit}=0.1-0.8$, which corresponds to the initial rotational velocity at the equator of the star $\sim 80-770$ km s$^{-1}$. Here, $\Omega_{\rm crit}=(GM/R_{\rm e}^3)^{1/2}$ is the critical angular velocity at the equator, where $M$ and $R_{\rm e}$ denote the mass of the star and the equator radius, respectively. For stellar winds, all models are evolved with the ``Dutch'' wind-loss scheme \citep[e.g.,][]{deJager1988,Nieuwenhuijzen1990,Nugis2000,Vink2001,Glebbeek2009}, and the scaling factor $\eta_{\rm wind}=$ 0.5, 1, and 5. We treat convection using the \texttt{MLT++} scheme of MESA. According to the Ledoux criterion and standard mixing length theory (MLT) approximation \citep{Cox1968}, we choose a mixing length parameter $\alpha_{\rm MLT}$ = 1.5. The effect of semiconvection \citep{Langer1991,Yoon2006} is included, and we adopt a baseline choice of $\alpha_{\rm sc}=0.01$ except after core helium depletion, where $\alpha_{\rm sc}=0$. Considering thermohaline mixing \citep{Kippenhahn1980}, we set $\alpha_{\rm th}=2$ and 0 before and after core helium depletion, respectively. The hydrodynamic mixing instabilities at convective boundaries, called overshoot mixing, are treated as an exponential decay process \citep{Herwig2000} beyond the Schwarzschild boundary. We adopt overshoot mixing parameters $f_{\rm ov}=0.005$ and $f_{0}=0.001$ \citep{Paxton2011}. The uncertainty of the convective mixing and overshooting also affect stellar evolution \citep{Arnett2015,Farmer2016}, but are beyond the scope of this paper. The detailed definition and range of these parameters are described and discussed by \cite{Paxton2011,Paxton2013}, \cite{Farmer2016} and on the MESA website\footnote{https://docs.mesastar.org/en/release-r22.05.1/}.

We use the \texttt{approx21$\_$cr60$\_$plus$\_$co56.net} network to compute nuclear reactions in the stellar interior \citep{Timmes1999,Paxton2011}. This approximate nuclear reaction network consists of 21 base isotopes plus $^{60}$Cr and $^{56}$Co, and it is a traditional workhorse in massive star models. Nuclear reaction rates are from \texttt{JINA REACLIB} \citep{Cyburt2010}, where \texttt{type 2 opacity tables} \citep{Iglesias1996} and the \texttt{Skye} \citep{Jermyn2021} equation of state are adopted.

We start by running models at the pre-main-sequence stage and activating rotation when near the zero-age-main-sequence. The implementation of rotation in MESA closely follows \cite{Heger2000} and \cite{Heger2005}. Six different rotationally induced mixing processes are included: dynamical shear instability $D_{\rm DSI}$, Solberg-H{\o}iland instability $D_{\rm SHI}$, secular shear instability $D_{\rm SSI}$, Eddington-Sweet circulation $D_{\rm ES}$, Goldreich-Schubert-Fricke $D_{\rm GSF}$, and Spruit-Tayler dynamo $D_{\rm ST}$, which could lead to angular momentum redistribution and chemical mixing. It is assumed that dynamical shear instability is weak and thus is set to 0. It should be noted that the Spruit-Tayler dynamo only operates in radiative regions of the star and produces poloidal magnetic fields. The toroidal magnetic fields generated by differential winding are orders of magnitude larger than the poloidal magnetic fields \citep{Heger2005}. The turbulent viscosity $\nu=D_{\rm conv}+D_{\rm sem}+D_{\rm DSI}+D_{\rm SHI}+D_{\rm SSI}+D_{\rm ES}+D_{\rm GSF}+D_{\rm ST}$, which is determined as the sum of the diffusion coefficients for convection $D_{\rm conv}$, semiconvection $D_{\rm sem}$, and rotationally induced instabilities. The factor $f_{\rm c}$ denotes the contribution of the rotationally induced instabilities to the diffusion coefficient $D=D_{\rm conv}+D_{\rm sem}+f_{\rm c}( D_{\rm DSI}+D_{\rm SHI}+D_{\rm SSI}+D_{\rm ES}+D_{\rm GSF}+D_{\rm ST})$. We assume that $f_{\rm c}=1/30$ in this paper \citep{Heger2000}.

\begin{figure*}
\centering
\includegraphics[width=0.4\textwidth,height=0.4\textwidth]{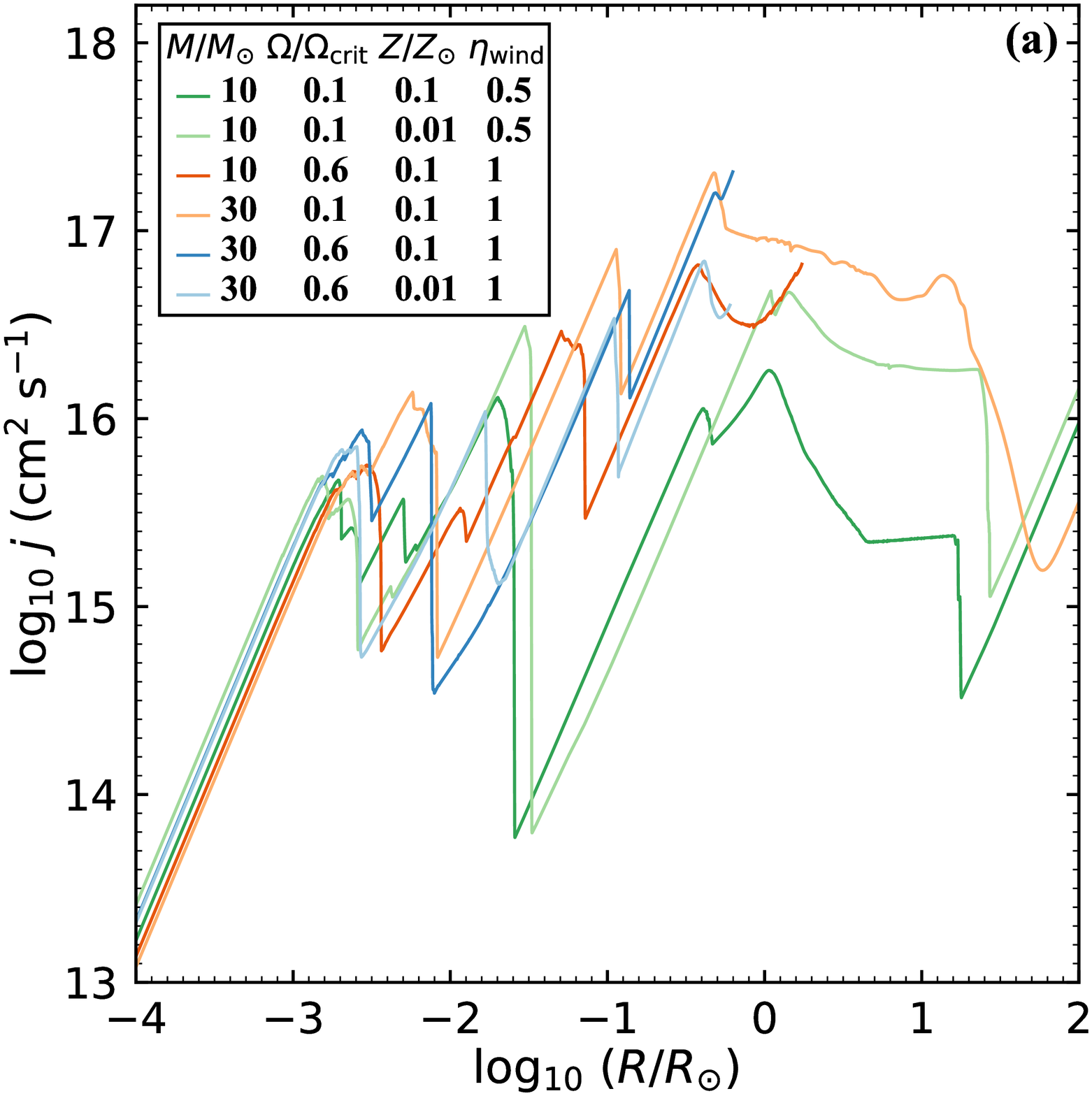}
\includegraphics[width=0.4\textwidth,height=0.4\textwidth]{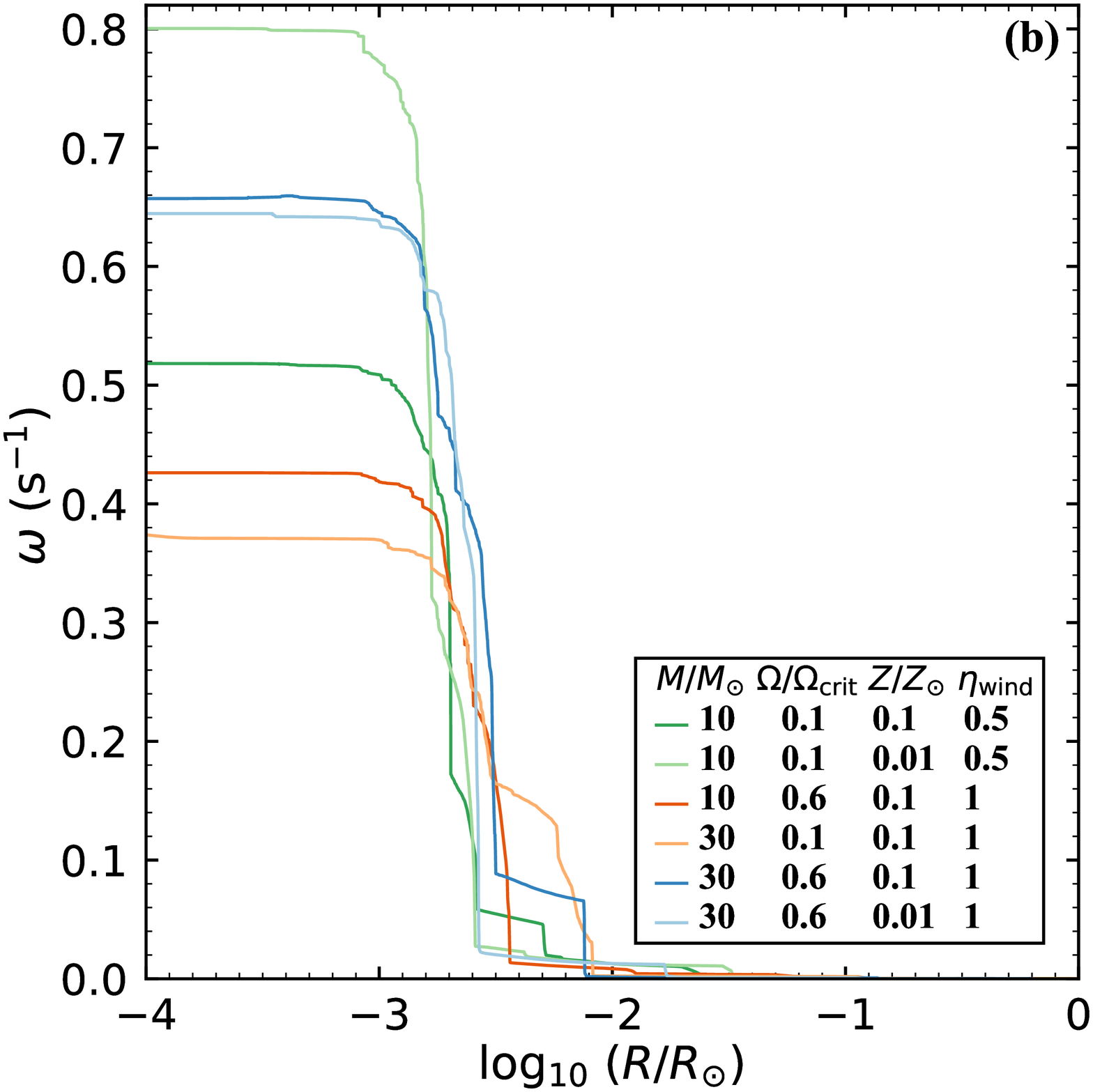}
\includegraphics[width=0.4\textwidth,height=0.4\textwidth]{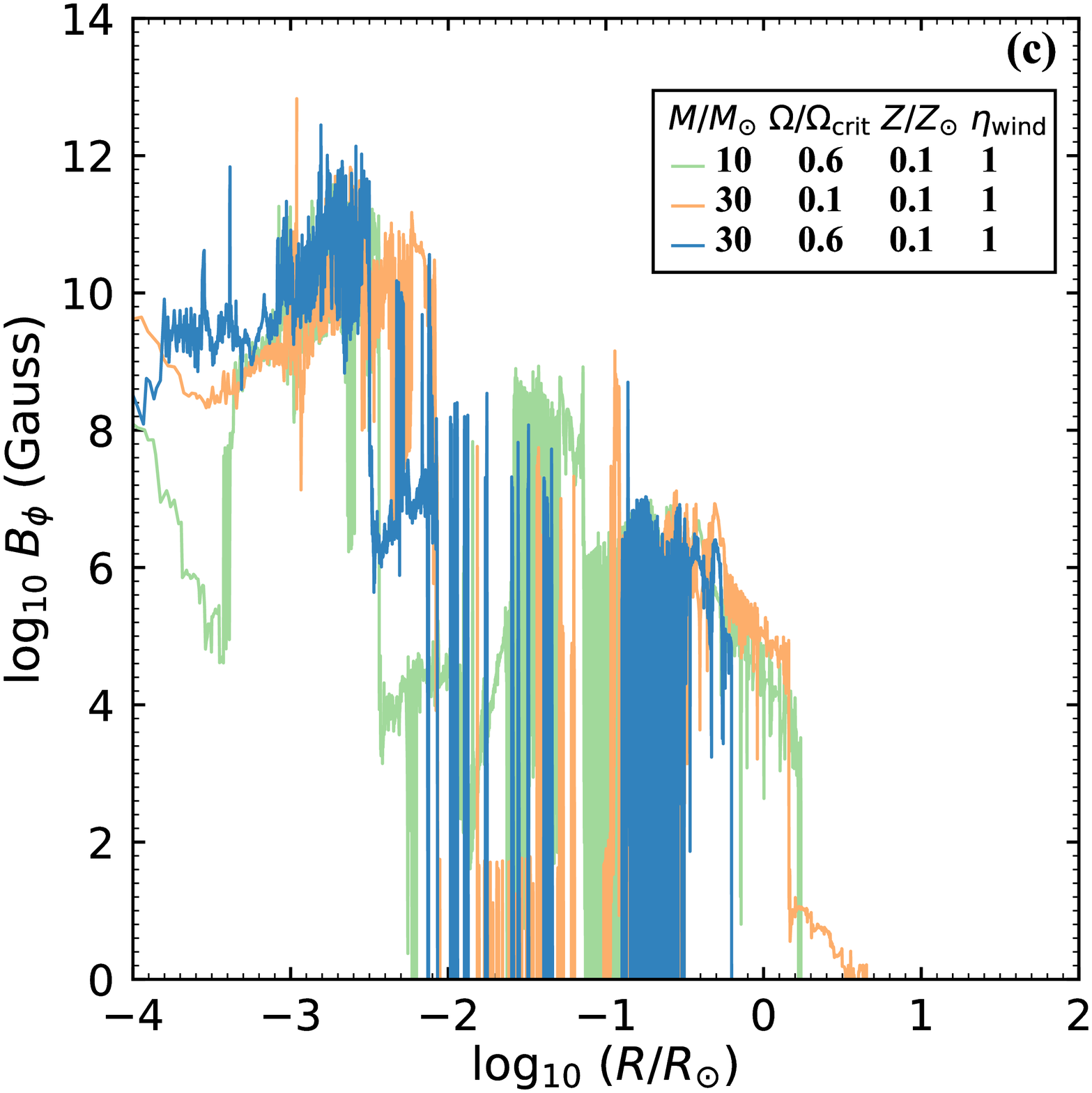}
\includegraphics[width=0.4\textwidth,height=0.4\textwidth]{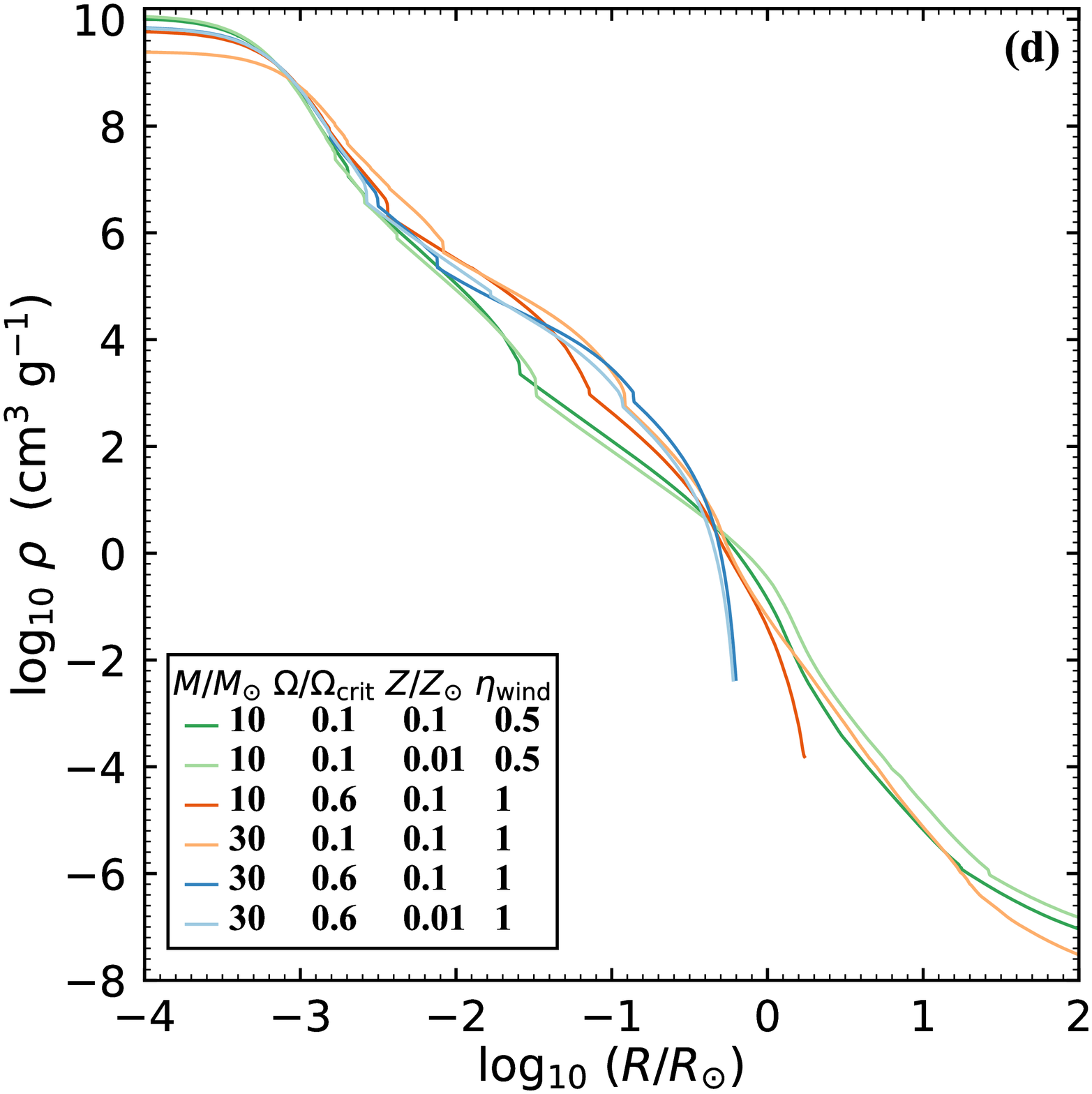}
\caption{Specific angular momentum (a), angular velocity (b), toroidal magnetic field strength (c), and density (d) variations as a function of the radius at the onset of core collapse.}
\label{struc_r}
\end{figure*}

\begin{figure*}
\centering
\includegraphics[width=0.4\textwidth,height=0.4\textwidth]{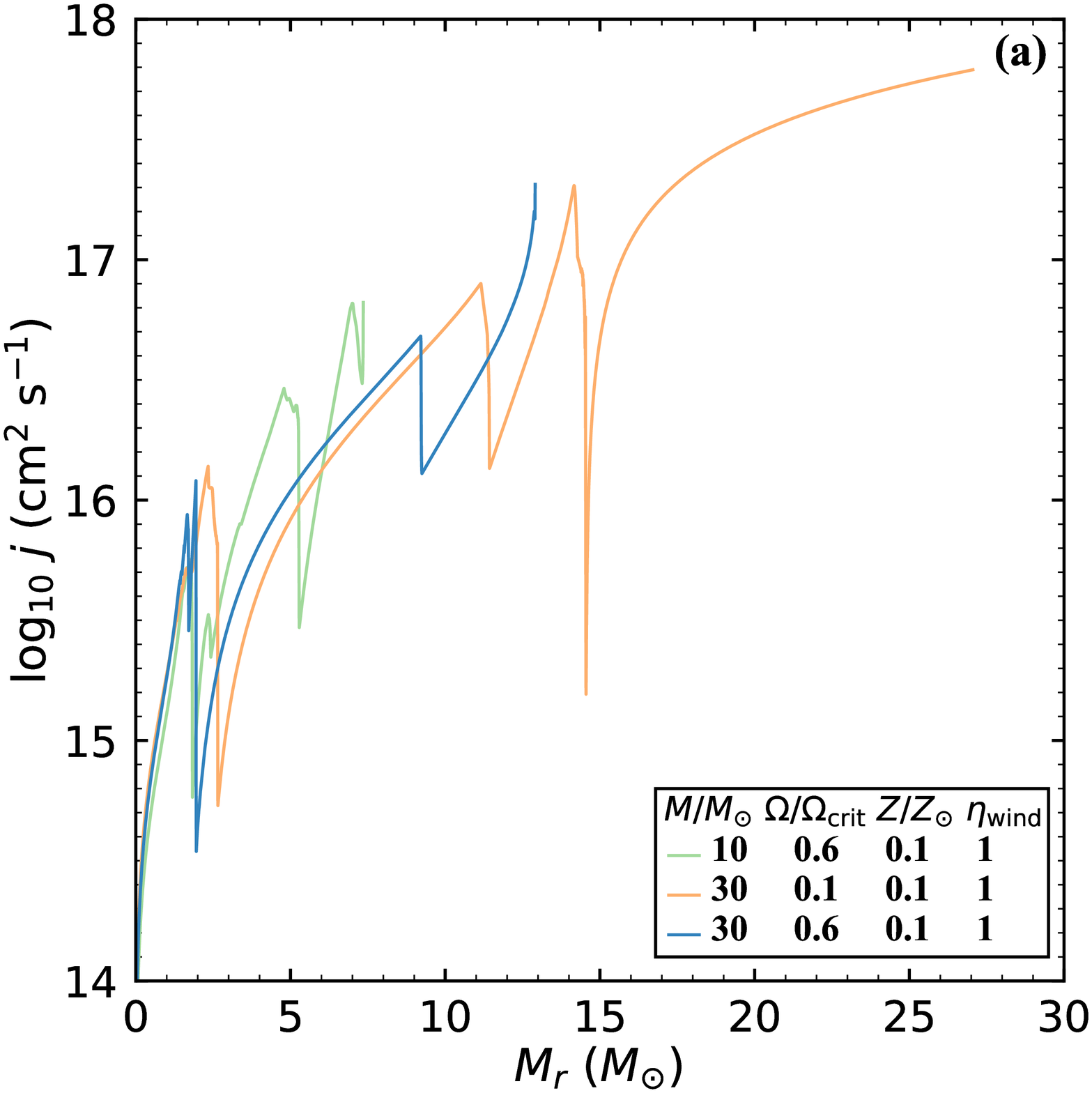}
\includegraphics[width=0.4\textwidth,height=0.4\textwidth]{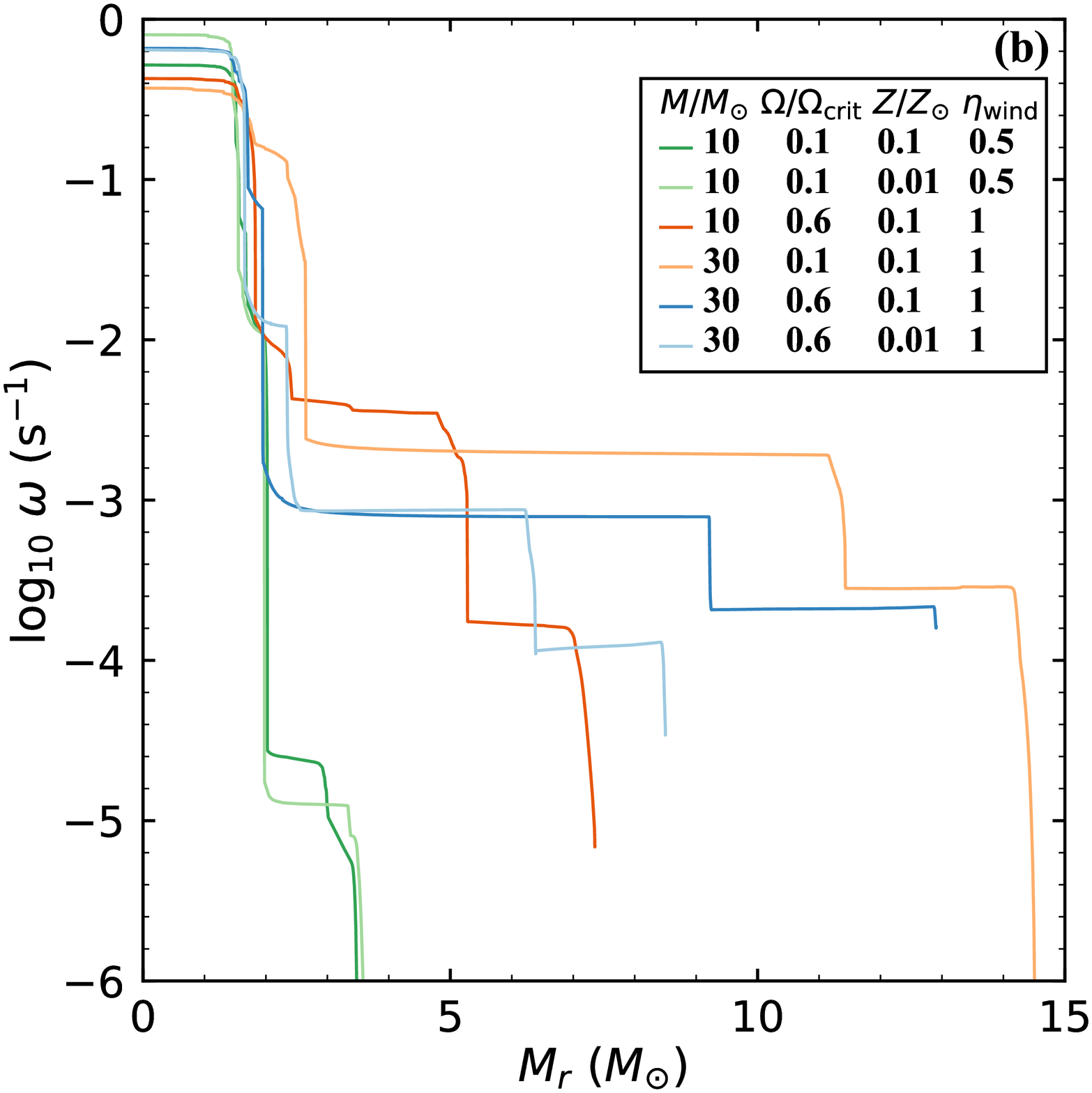}
\includegraphics[width=0.4\textwidth,height=0.4\textwidth]{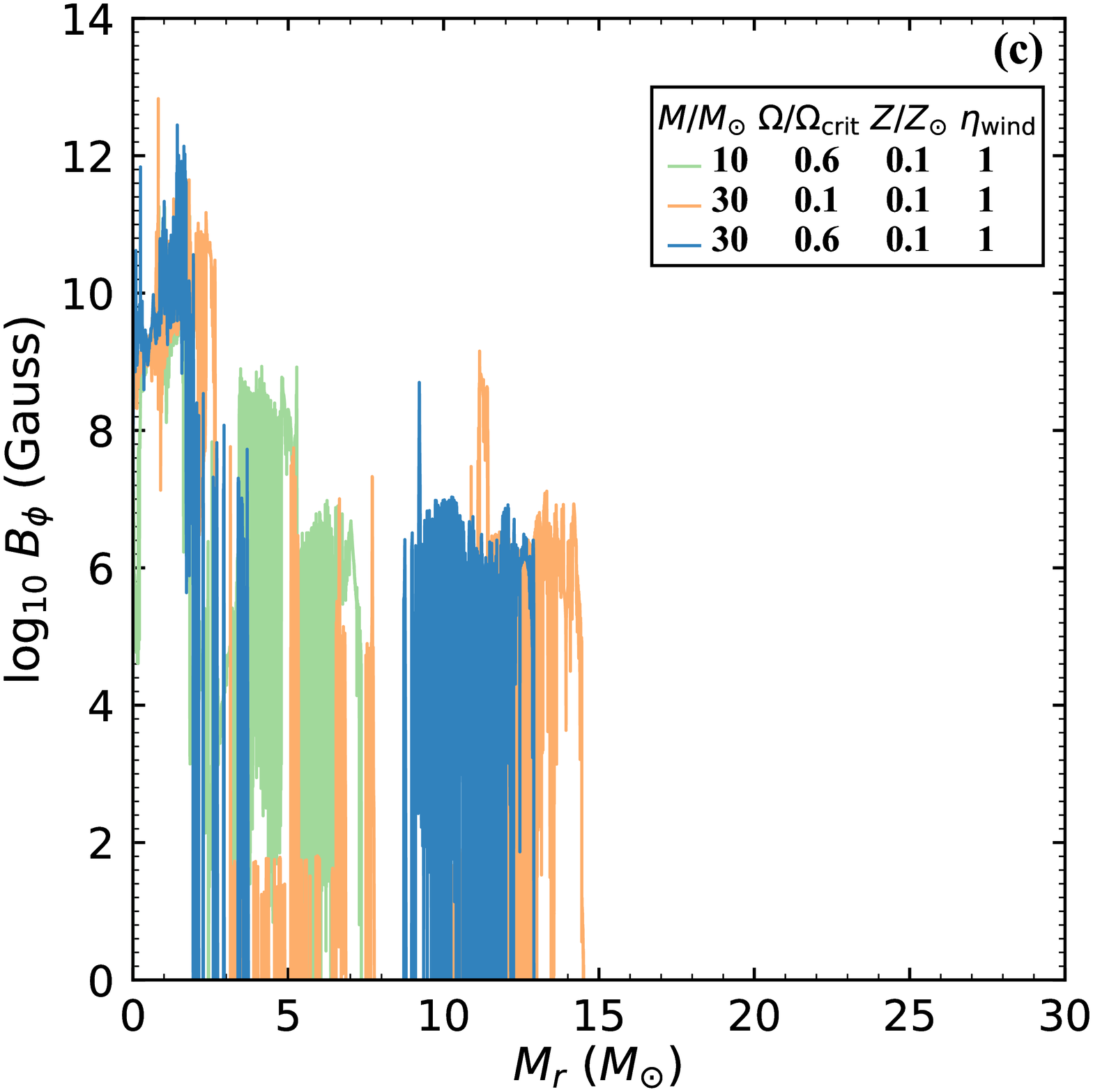}
\includegraphics[width=0.4\textwidth,height=0.4\textwidth]{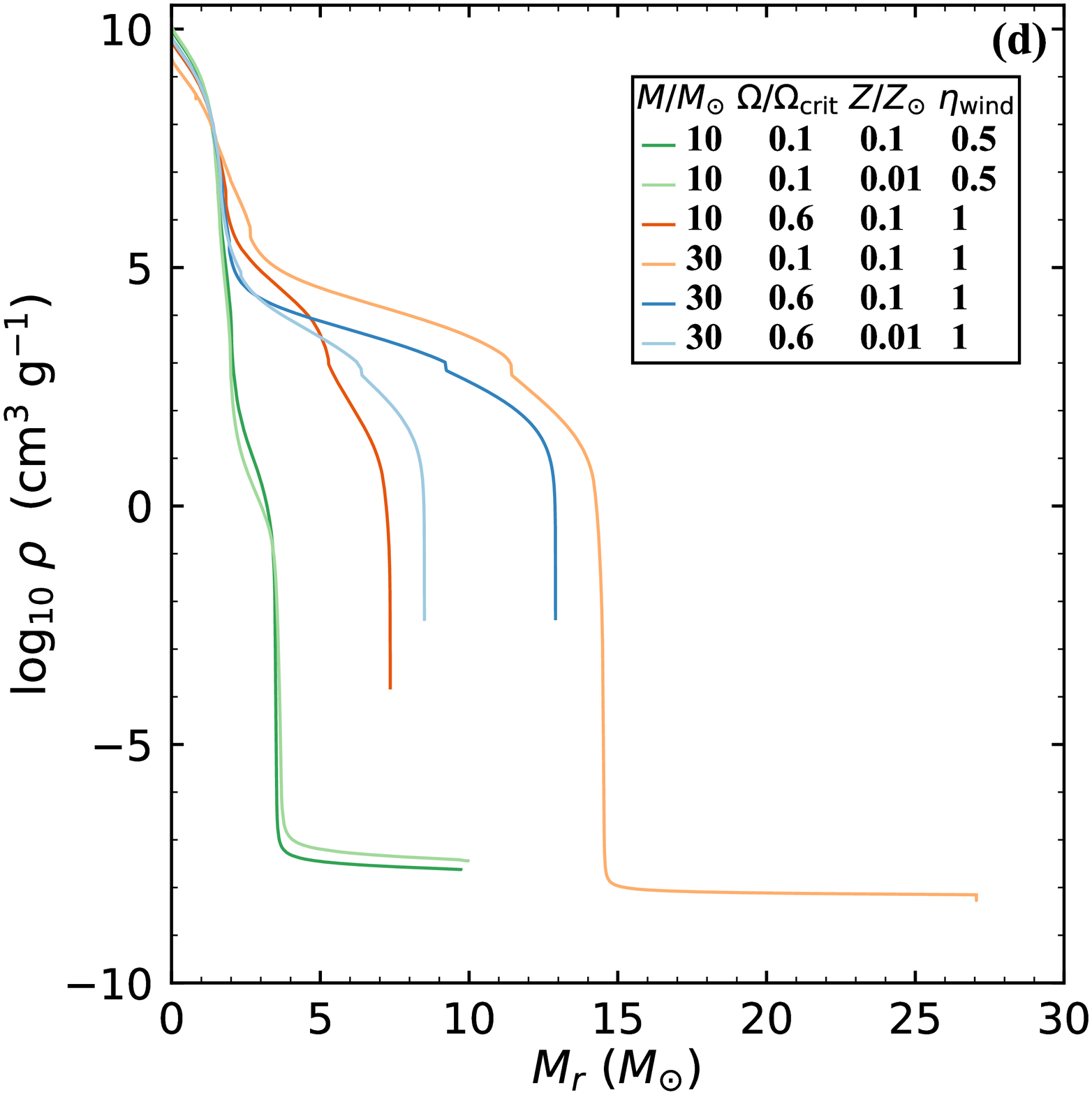}
\caption{Specific angular momentum (a), angular velocity (b), toroidal magnetic field strength  (c), and density (d) variations as a function of the mass coordinate at the onset of core collapse.}
\label{struc_m}
\end{figure*}

\begin{figure*}
\centering
\includegraphics[width=0.4\textwidth,height=0.4\textwidth]{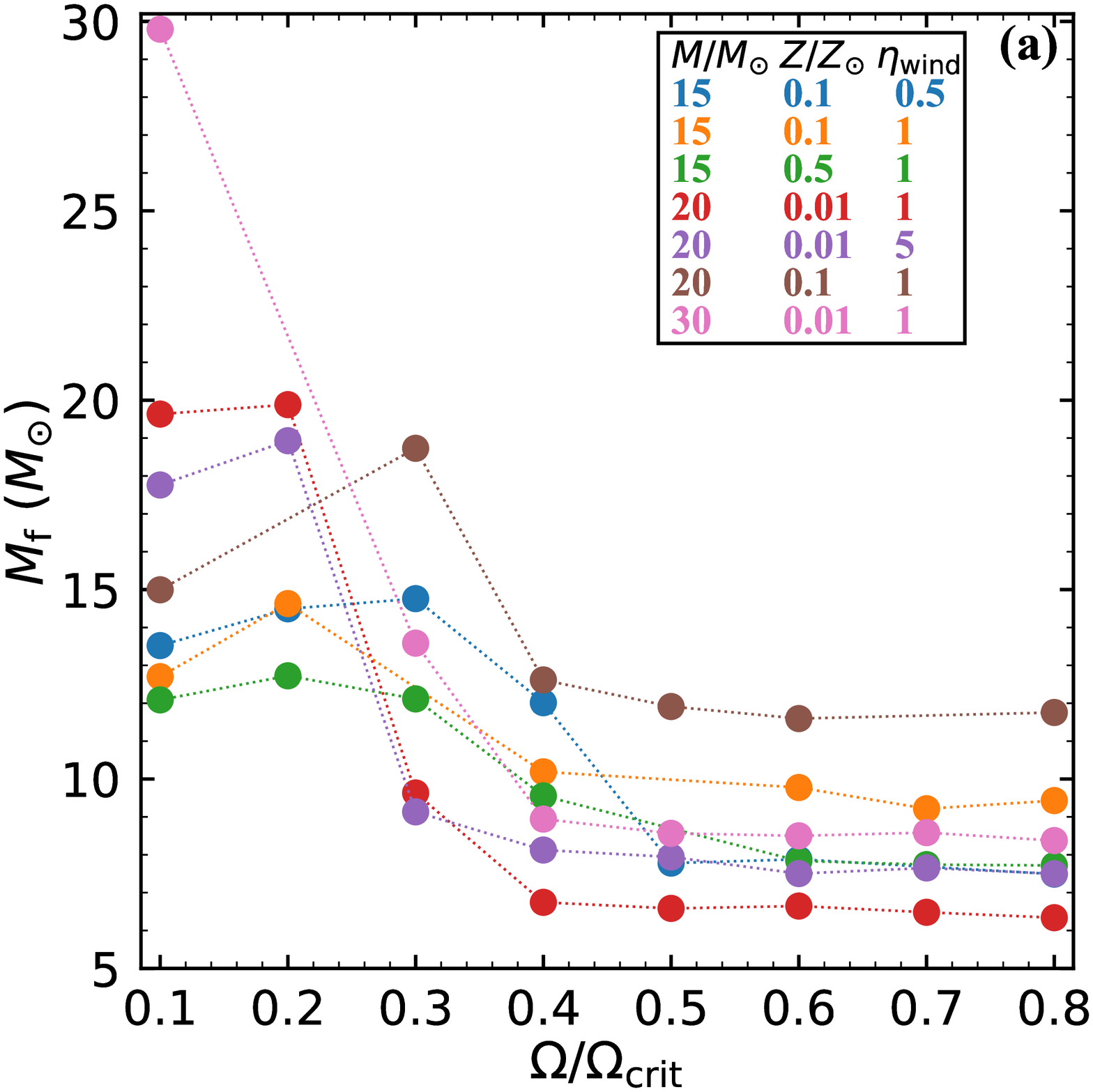}
\includegraphics[width=0.4\textwidth,height=0.4\textwidth]{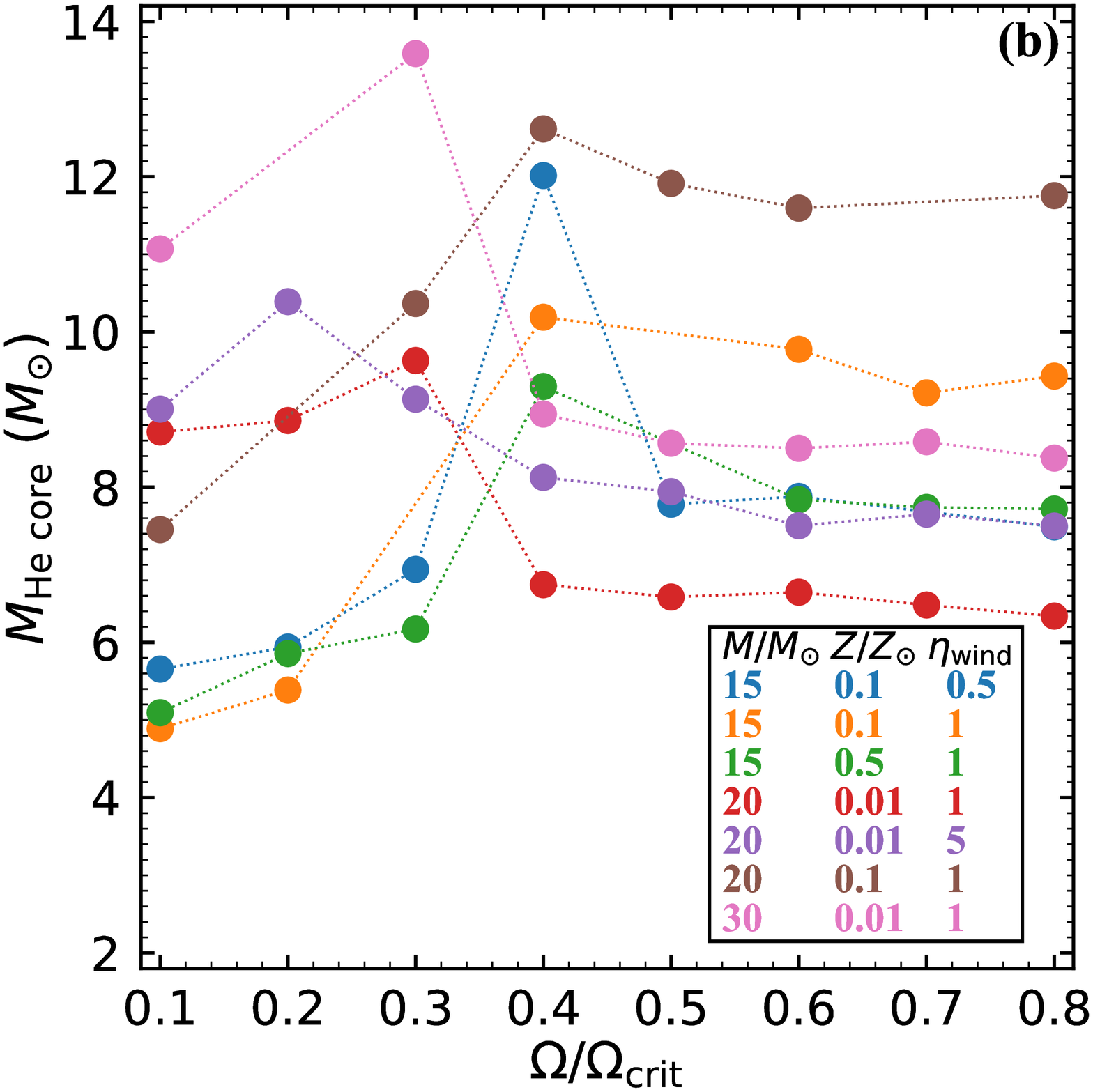}
\includegraphics[width=0.4\textwidth,height=0.4\textwidth]{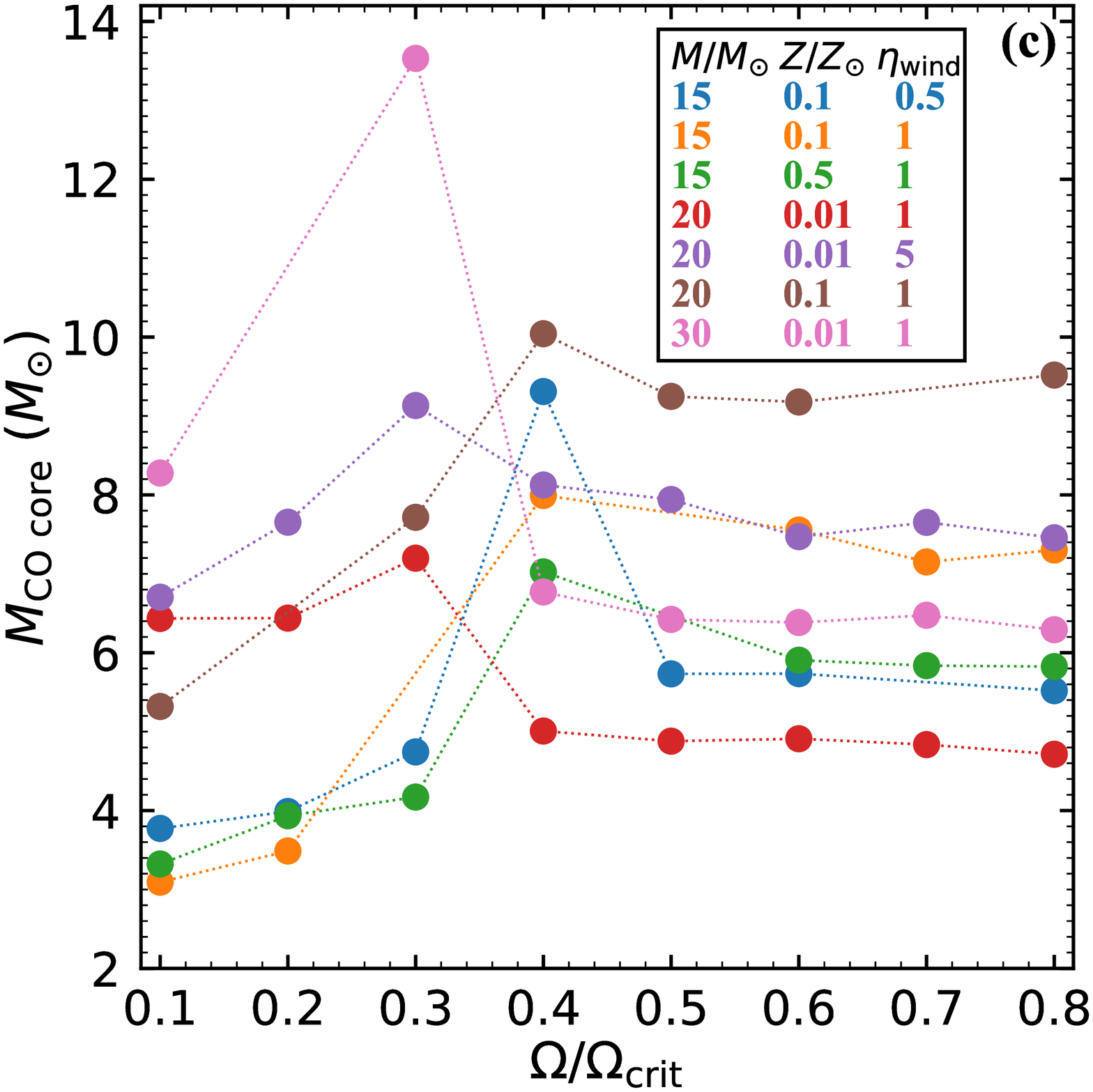}
\includegraphics[width=0.4\textwidth,height=0.4\textwidth]{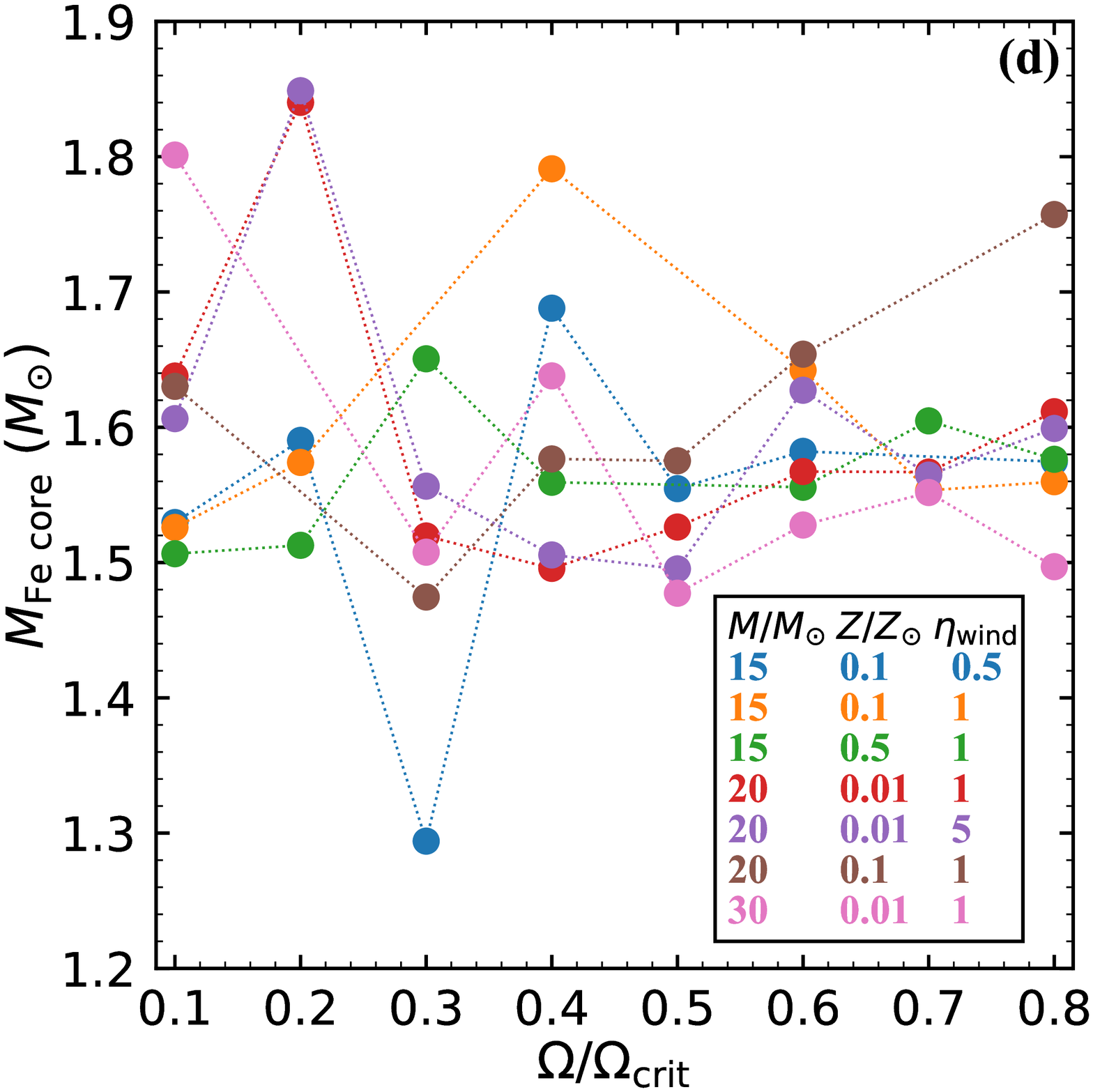}
\caption{Final stellar mass $M_{\rm f}$, He core mass $M_{\rm He~core}$, CO core mass $M_{\rm CO~core}$ and iron core mass $M_{\rm Fe~core}$ at the beginning of core collapse as a function of the initial rotational rate  $\Omega/\Omega_{\rm crit}$.}
\label{Momega}
\end{figure*}

\begin{figure*}
\centering
\includegraphics[width=0.4\textwidth,height=0.4\textwidth]{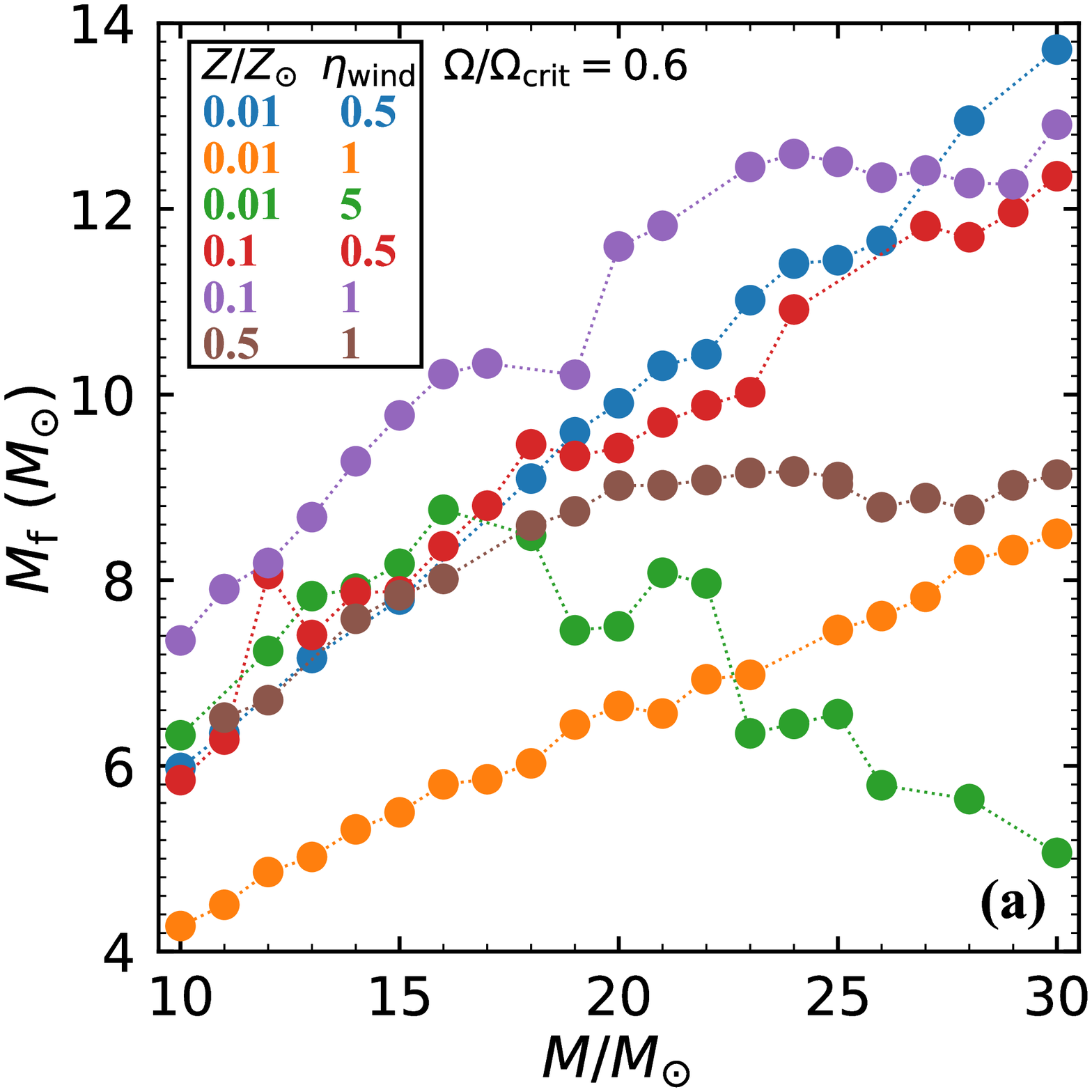}
\includegraphics[width=0.4\textwidth,height=0.4\textwidth]{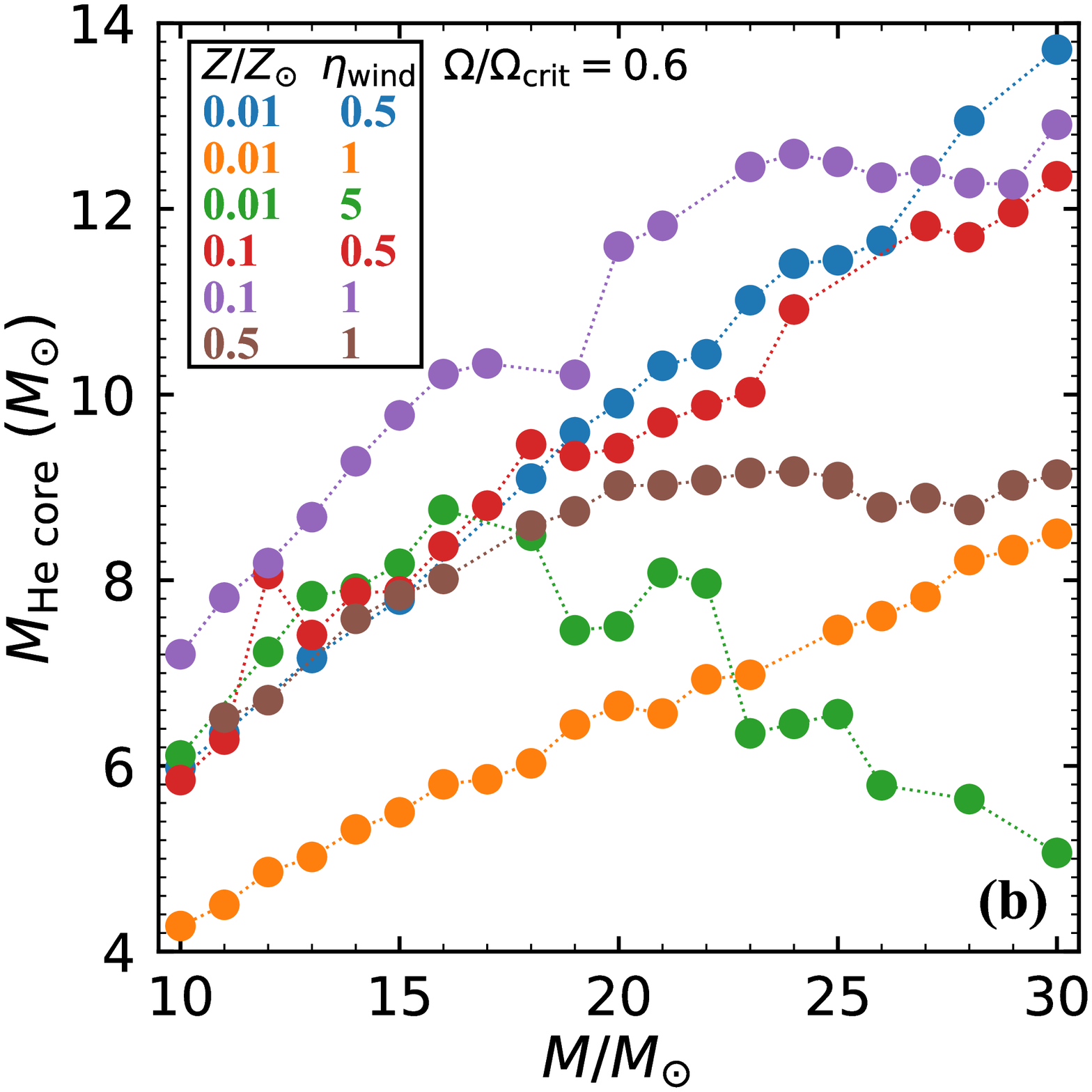}
\includegraphics[width=0.4\textwidth,height=0.4\textwidth]{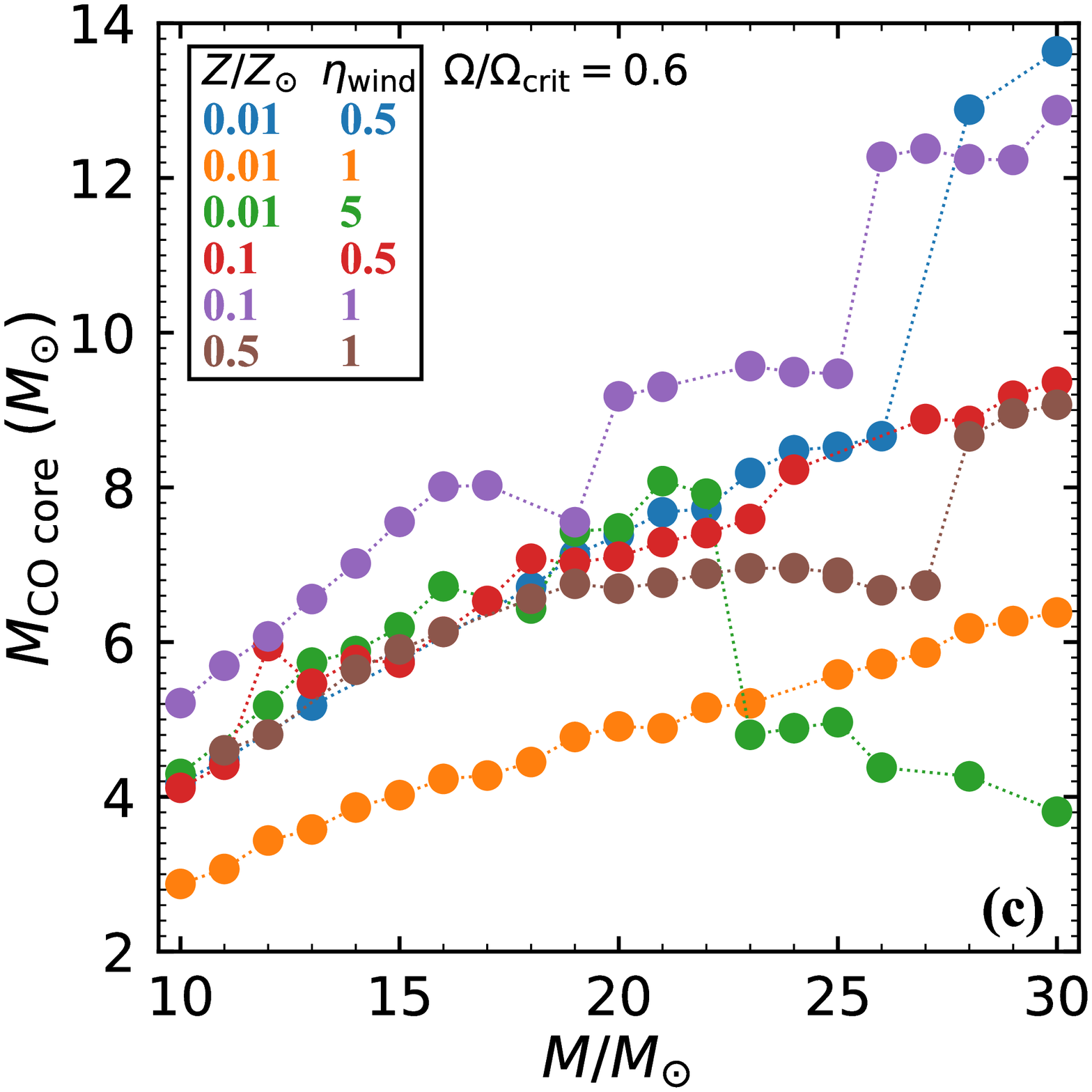}
\includegraphics[width=0.4\textwidth,height=0.4\textwidth]{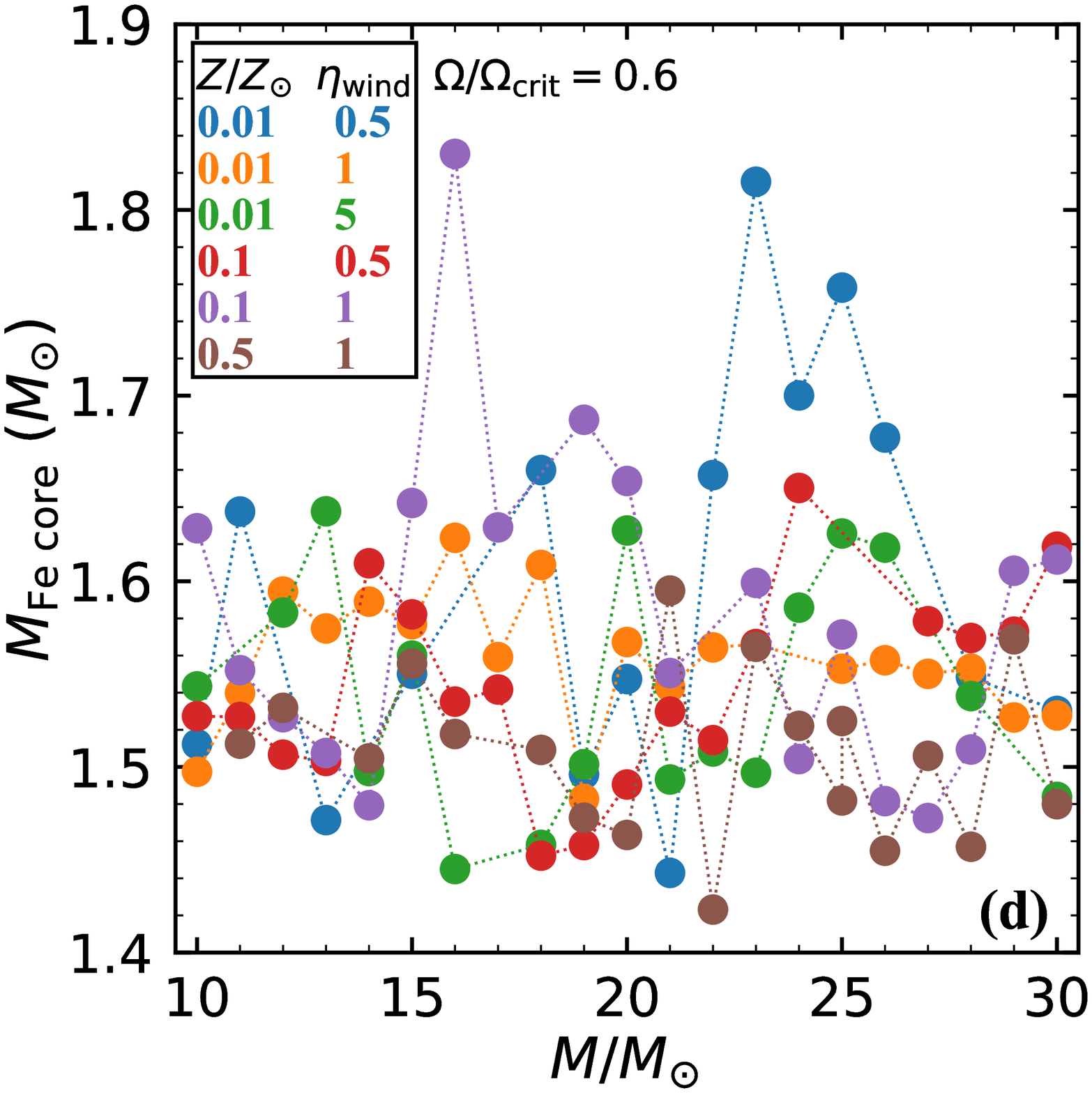}
\caption{Final stellar mass $M_{\rm f}$, He core mass $M_{\rm He~core}$, CO core mass $M_{\rm CO~core}$ and iron core mass $M_{\rm Fe~core}$ at the beginning of core collapse as a function of initial stellar mass when $\Omega/\Omega_{\rm crit}=0.6$.}
\label{MMi}
\end{figure*}

\begin{figure*}
\centering
\includegraphics[width=0.4\textwidth,height=0.4\textwidth]{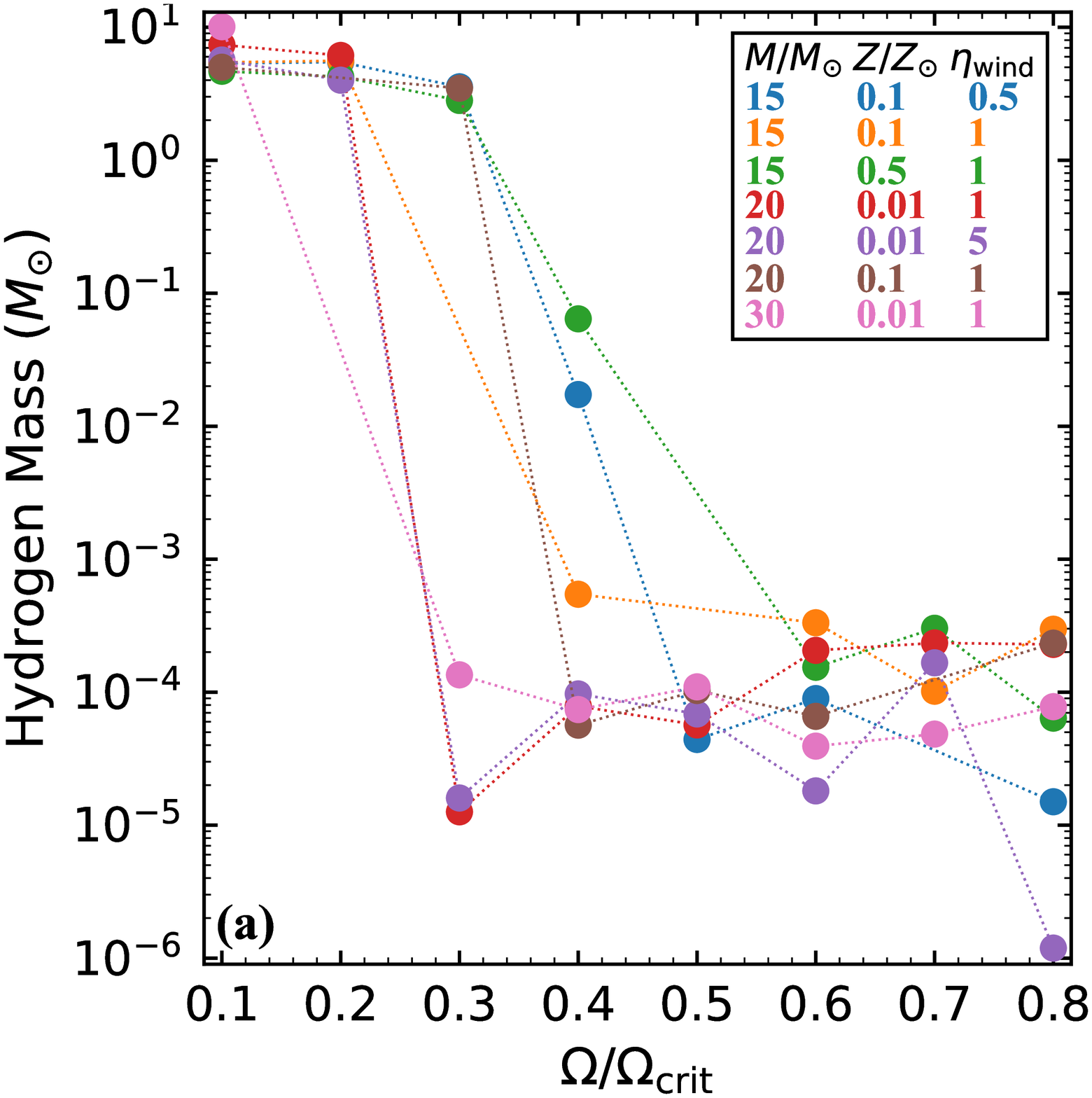}
\includegraphics[width=0.4\textwidth,height=0.4\textwidth]{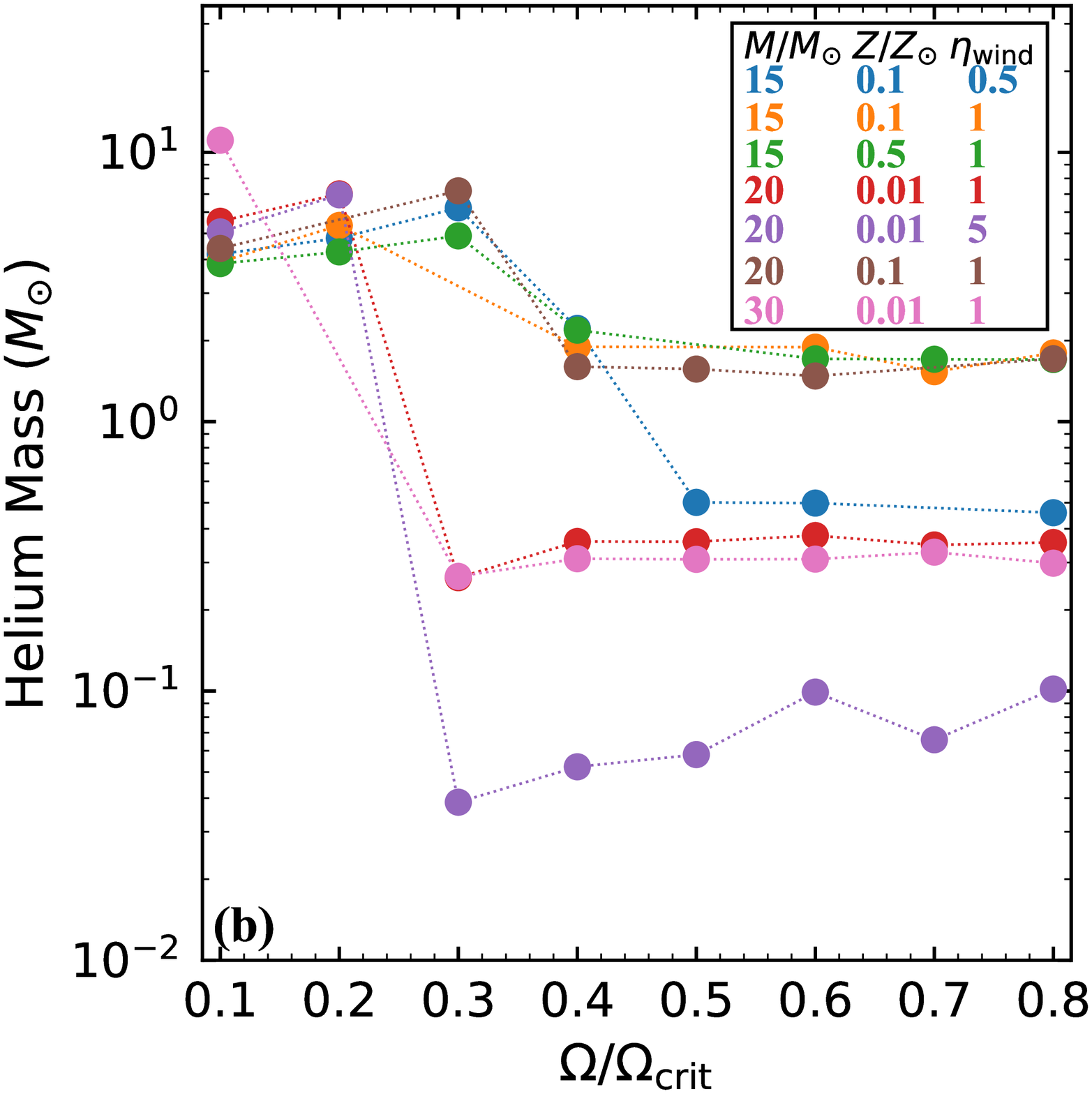}
\includegraphics[width=0.4\textwidth,height=0.4\textwidth]{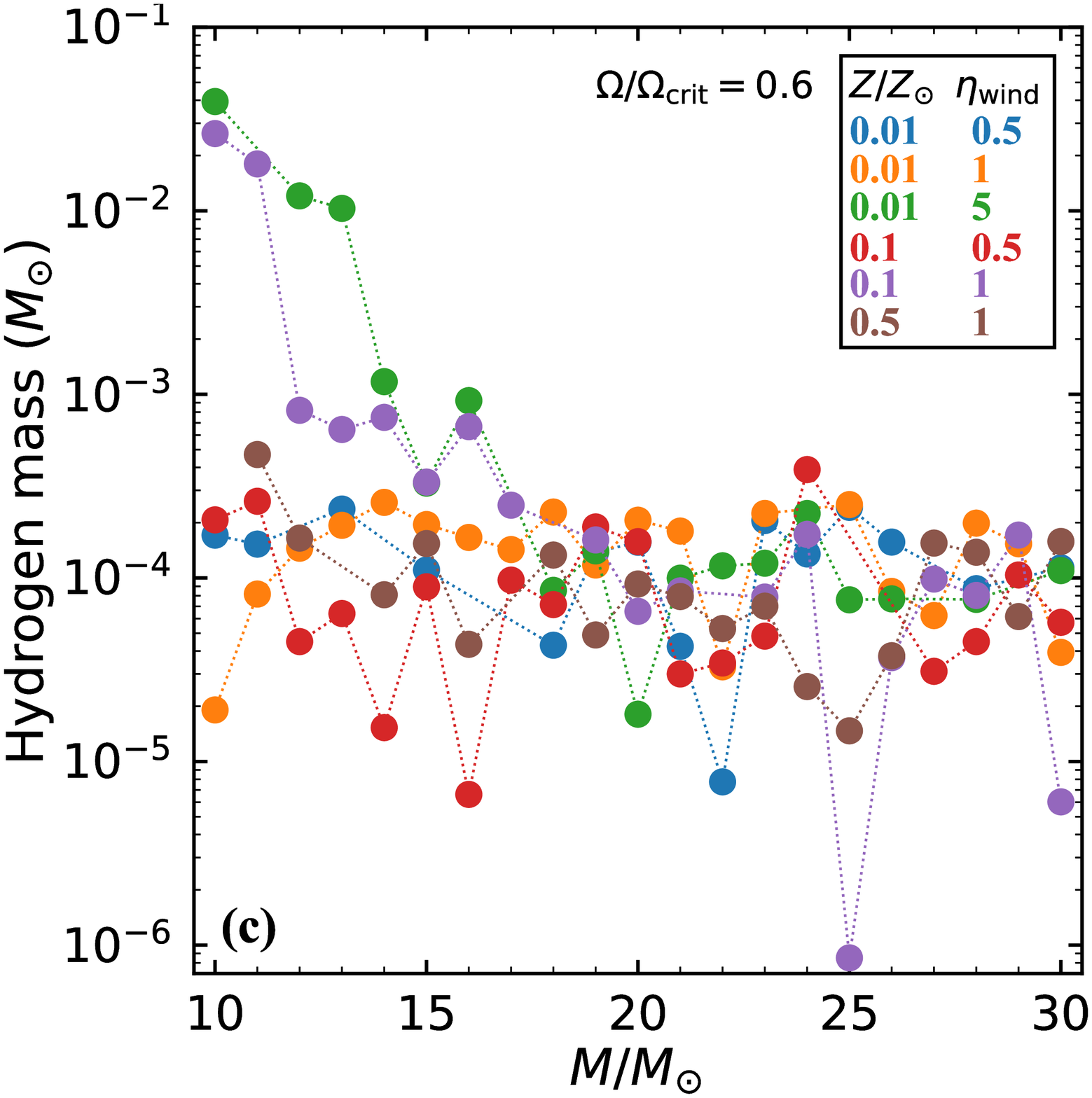}
\includegraphics[width=0.4\textwidth,height=0.4\textwidth]{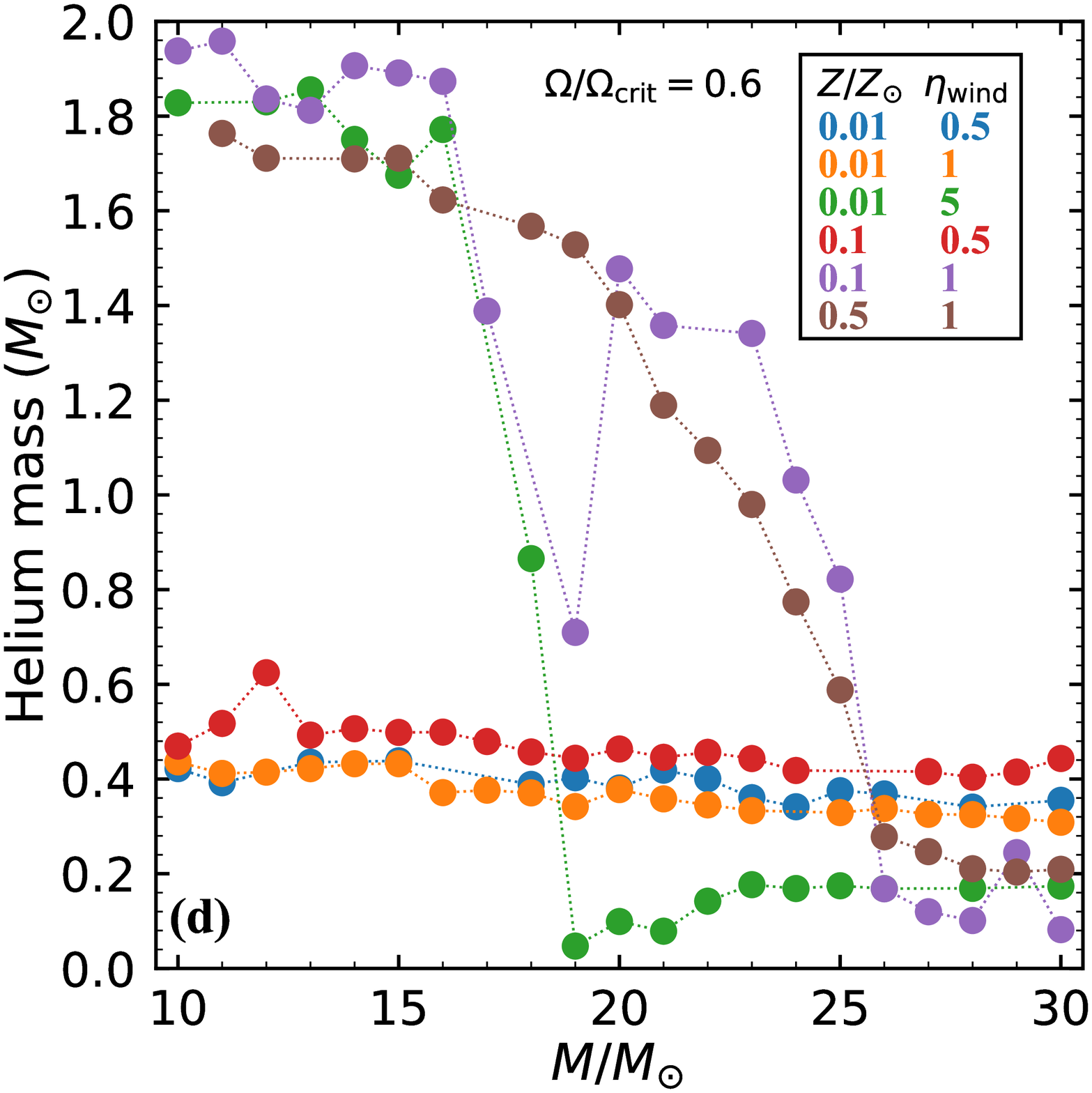}
\caption{Hydrogen and helium mass as a function of initial rotation rate (upper panels) and initial stellar mass (bottom panels) for pre-SN models. Different initial masses, metallicities, and wind loss scaling factors are indicated by different colors.}
\label{X_he_Omega}
\end{figure*}

Hertzsprung-Russell (H-R) diagram and evolution of the central density and temperature for some examples of our models are shown in Figure \ref{HR}. The different color lines correspond to different initial stellar masses $M=10, 15,$ and $30 M_{\odot}$,  initial rotation rates $\Omega / \Omega_{\rm crit} =0.1$ and $0.6$, initial metallicity $Z=0.01, 0.1$, and $0.5~Z_{\odot}$, and the ``Dutch'' wind scale factor $\eta_{\rm wind}=0.5, 1$, and $5$. The endpoint of lines is marked by a pentagram, corresponding to the time of the iron core collapse. In Figure \ref{HR}(b), locations of hydrogen, helium, carbon, neon, oxygen, and silicon burning are labeled. The grey dashed line indicates the $E_{F}/k_{B}T\approx 4$ electron degeneracy curve. The results show that the faster rotating model evolves toward the blue part of the H-R diagram and ends its life with a WR star.  The slower rotating model evolves toward the red part of the H-R diagram and becomes red supergiants (RSG). At the same initial rotation rate, the larger the initial mass, the higher the luminosity and effective temperature of the model evolving to the end. The luminosity generally increases with increasing initial metallicity when the initial mass is the same. Fast-rotating stars have higher central temperatures and densities than low-rotating stars. At the same initial rotation rate, the more massive the stars are, the higher the central density and temperature.

In Figure \ref{ages}, we present the evolution of stellar mass and total angular momentum. Similar to Figure \ref{HR}, different colored lines indicate vary initial parameters. Comparing the different colored lines in the diagram, we can see that the initial mass plays a decisive role in the lifetime of the star. More rapid rotation, lower metallicity, and a smaller scaling factor of wind can slightly extend the star lifetime. The final mass significantly decreases with increasing initial rotation rate. Stars with higher metallicity and larger ``Dutch'' wind scale factors have smaller stellar masses at the beginning of core-collapse for models with the same initial mass and fast rotation. This may be due to the fact that while larger stellar winds and high metallicity increase the mass loss on the one hand, they also reduce the stellar lifetime.  For models with the same initial rotation rate, the larger the initial mass, the larger the total initial angular momentum.  High rotation rate significantly increases angular momentum loss in stellar life.

\section{Results}

\subsection{Pre-SN Properties}

We summarize the properties of all models at the end of their evolution in Table \ref{TABpreSN} of the Appendix, listing the initial mass, metallicity, rotational rate, and corresponding velocity at the equator, the ``Dutch'' wind scale factor, the initial total angular momentum, final mass, ages, radius, effective temperature and luminosity, final angular momentum and masses of the He/CO/Fe core at the end of the calculation. The compactness parameter, mass, average magnetic field strength, and rotation period of the protomagnetar are also presented in the table. To show the final structure of pre-SN models, we present the specific angular momentum, angular velocity, toroidal magnetic field strength, and density variations as a function of the radius (Figure \ref{struc_r}) and the mass coordinate (Figure \ref{struc_m}).  It is clear to see that the radius of an RSG that evolved from a star with a lower initial rotation rate is two orders of magnitude higher than the radius of a WR star that evolved from a star with a larger initial rotation rate. The fast rotation causes the star to lose its envelope, which affects its surface parameters. The specific angular momentum, angular velocity, toroidal magnetic field strength, and density profiles are flatter in the interior of faster-rotating stellar models.  Convection transports angular momentum from the inside of the convection zone to the outside of the same zone, which produces a sawtooth shape of special angular momentum. During the evolution, $\Omega$ increases significantly in the central region when the core of the star contracts, and then the $\Omega$ curve flattens under the action of convection. In the pre-SN stage, the angular velocity  $\Omega$ reaches values of the order of 0.1 s$^{-1}$ in the center. The expansion of the outer layers of RSG  leads to the value of $\Omega$ decreasing sharply at the surface. Fast-rotating stars have higher magnetic field strengths in the inner regions and higher surface densities.

In Figure \ref{Momega}, we display the final stellar masses $M_{\rm f}$, helium core masses $M_{\rm He~core}$, carbon-oxygen core masses $M_{\rm CO~core}$, and iron core mass $M_{\rm Fe~core}$ of massive stars at the beginning of iron core collapse as a function of the initial rotational rate $\Omega$/$\Omega_{\rm crit}$. The He and CO core masses are measured using the mass coordinate, where the mass fraction of hydrogen is less than 0.1 for $M_{\rm He~core}$ and the mass fraction of helium decreases below 0.1 for $M_{\rm CO~core}$. Here, we define the iron core mass boundary as the outermost location where the Si mass fraction is less than 0.1. It should be noted that this definition is a bit arbitrary, as massive cores often have very shallow chemical gradients. The radius and composition of the stellar core might change with different core mass definitions \citep[e.g.,][]{Sukhbold2014,Laplace2021}. It is difficult to precisely define the stellar core boundary because there is a composition and density gradient at the edge. Moreover, the stellar core mass and boundary are also affected by semiconvection and overshoot \citep{Schootemeijer2019}.

It is obvious from Figures \ref{Momega}(a) and (b) that $M_{\rm He~core}$ values approximate $M_{\rm f}$ values when $\Omega$/$\Omega_{\rm crit} \geq$ 0.4. For the more massive and lower metallicity stars, the corresponding values $\Omega$/$\Omega_{\rm crit}$ $\geq$ 0.3. In other words, stars almost entirely lose their hydrogen envelopes and become naked He cores. Note that $M_{\rm f}$, $M_{\rm He~core}$, $M_{\rm CO~ core}$ and $M_{\rm Fe~core}$ show little change when $\Omega / \Omega_{\rm crit}\geq 0.4$. The values of $M_{\rm f}$, $M_{\rm He ~core} $, and $M_{\rm CO~ core}$ are positively correlated with the initial mass and negatively correlated with the initial metallicity $Z$ and a scaling factor of wind loss $\eta_{\rm wind}$ when the initial rotational rate is low. This is similar to models without rotation. Small fluctuations in the values of $M_{\rm f}$, $M_{\rm He ~core}$, and $M_{\rm CO~ core}$ when $\Omega$/$\Omega_{\rm crit}\geq$ 0.4. For high initial rotation rate, rotation plays an important role  in enhanced stellar mass loss \citep{Paxton2013} according to the prescription, i.e., $\dot{M} \propto 1/(1-\Omega/\Omega_{\rm crit})^{0.43}$.

Similar to Figure \ref{Momega}, $M_{\rm f}$, $M_{\rm He~core}$, $M_{\rm CO~core}$, and $M_{\rm Fe~core}$ of massive stars at the beginning of iron core collapse as a function of the initial stellar mass are presented in Figure \ref{MMi}. Here, we set $\Omega$/$\Omega_{\rm crit}=0.6$. For models with lower metallicity and smaller ``Dutch'' wind scale factor, $M_{\rm f}$, $M_{\rm He~core}$, and $M_{\rm CO~core}$ increase as the initial mass increases, except for some fluctuations. For models with higher metallicity and larger ``Dutch'' wind scale factor, the values of $M_{\rm f}$, $M_{\rm He~core}$, and $M_{\rm CO~core}$ climb up and then decline. $M_{\rm Fe~core}$ of massive stars with different initial $Z$ and $\eta_{\rm wind}$ fluctuates with $\Omega$ and stellar masses, and there is no obvious correlation. Iron core masses in our models range from 1.29 to 1.92 $M_{\odot}$, and most of them are concentrated between $1.4-1.6 ~M_{\odot}$. According to some previous studies \citep[e.g.][]{Sukhbold2014,Schneider2021}, $M_{\rm Fe~core}$ appears to peak in the range of $20-25~M_{\odot}$, and BH are produced at the corresponding peak. The position of the peak is influenced by the metallicity \citep{Woosley2007}. There is a peak of $M_{\rm Fe~core}$ in the mass range of $20-25~M_{\odot}$ for zero metallicity star. For stars with solar metallicity, this peak corresponds to masses around 40 $M_{\odot}$. However, there is no peak in our model and no BH formation, probably because of the large initial rotation rate $\Omega/\Omega_{\rm crit}$ of the model in Figure \ref{MMi}. All models in this diagram have gone through WR phase, and lost large amount of mass. This is similar to the fact that exceptionally massive stars ($\sim 100~ M_{\odot}$) can still produce NSs, probably both as a consequence of mass loss. Furthermore, the mapping from $M_{\rm CO~core}$ and $M_{\rm Fe~core}$ to initial mass could differ greatly in single stars compared to stripped stars \citep{Schneider2021}.

Figure \ref{X_he_Omega} demonstrates the variation of hydrogen and helium masses with initial rotation rate and initial stellar mass for the pre-SN model.  As shown in the upper panel of Figure \ref{X_he_Omega}, when $\Omega$/$\Omega_{\rm crit}$ $\geq$ 0.4, most of pre-SN models have hydrogen and helium mass below 0.001 and 0.5 $M_{\odot}$, respectively.  Only a small fraction of models with larger ``Dutch'' wind scale factor have helium mass below 0.2$M_{\odot}$. It should be noted that the hydrogen and helium mass show slight changes when $\Omega$/$\Omega_{\rm crit}$ $ \geq$ 0.4. In lower panel of Figure \ref{X_he_Omega}, we can find that  the hydrogen masses  fluctuate between 10$^{-5}$ and 10$^{-3}$ $M_{\odot}$ for the vast majority of models as the initial mass increases. For models with higher metallicity and larger ``Dutch'' wind scale factor, the helium mass decreases significantly with the growth of initial mass. Among them, a small percentage of models with initial masses greater than 20 $M_{\odot}$ have helium masses less than 0.2 $M_{\odot}$.

Massive stars without hydrogen envelopes could core-collapse and give rise to Type Ib or Ic SNe. Type Ib or Ic SNe are distinguished based on the presence or absence of helium lines in the spectrum. Sometimes, it is difficult to distinguish them from limited observational data. Most progenitor models still retain some helium after the core collapse. The different characteristics of the progenitor stars which lead to the difference between Type Ib and Ic SNe remain unknown \citep{Hachinger2012,Dessart2011,Dessart2012,Dessart2022}. Therefore, these stars are usually collectively referred to as Type Ibc SNe. The maximum H/He mass contained in progenitor that would give rise to a Type Ib/Ic SN appearance is uncertain.  It is proposed that the hydrogen mass of progenitors for Type Ib SN should less than 0.001 $M_{\odot}$ \citep{Yoon2010,Dessart2011}. \cite{Gilkis2022} estimate this threshold mass is 0.033 $M_{\odot}$. Estimates of the limit helium mass included in the progenitors of SNe Ic from previous work range from 0.06 to 0.6 $M_{\odot}$ \citep{Georgy2009,Yoon2010,Hachinger2012}. If the progenitor of Type Ib SN should contain hydrogen masses below 0.001 $M_{\odot}$ and the progenitors of Type Ic SN contain helium masses below 0.2 $M_{\odot}$, most of our models can produce Type Ib SNe  and only a small fraction produce Type Ic SNe. If the upper limit on the mass of helium hidden in the progenitor of type Ic SNe are relaxed a bit, more than half of the models can produce Type Ic SNe. Moreover, \cite{Eldridge2004} and \cite{Eldridge2005} used surface helium abundance $Y_{\rm surface}$ to estimate the SNe type produced by the progenitors. If $Y_{\rm surface}$ $\textless$ 0.3, a Type Ic SN occurs; if 0.3 $\textless$ $Y_{\rm surface}$ $\textless$ 0.7, they label the star as Type Ibc due to uncertain, and if $Y_{\rm surface}$ $\textgreater$ 0.7, a Type Ib SN results. The vast majority of the SN type created by our progenitor models can be designated as Type Ibc, with only a small fraction being SN Ic or Ib, based on the $Y_{\rm surface}$ value. The rates of SN Ib and Ic estimated by the rotating single star models is consistent with observations \citep[e.g.][]{Georgy2009,Groh2013}. Only a small fraction of Type Ic SN are observed in association with LGRBs \citep{Woosley2006,Cano2017}. It is uncertain what proportion of single and binary channels \citep[e.g.][]{Podsiadlowski1992,Eldridge2011} contribute to the production of Type Ic SNe.

The explodability of model is explored according to compactness, which was defined as \citep{Oconnor2011}
\begin{equation} \label{xi}
\xi_{\rm M_r}=\frac{M_r/M_{\odot}}{R(M_{\rm bary}=M_r)/1000 \rm ~km},
\end{equation}
where $M_r=2.5~ M_{\odot}$ was chosen to evaluate the compactness parameter \citep[e.g.,][]{Ugliano2012,Sukhbold2014}, which corresponds to the point where the collapse speed first reaches $1,000 ~\rm km~s^{-1}$. This has been pointed out as a rough indicator of whether the collapse of a nonrotating stellar core could lead to a successful neutrino-driven explosion ($\xi_{2.5}<0.45$) or form a BH ($\xi_{2.5}>0.45$). While this criterion is not sufficient to accurately predict the fate of rapidly spinning stars, it can still reveal structural features of the stellar core \citep[e.g.,][]{Ertl2016,Muller2016,Aguilera2020}.

\begin{figure*}
  \centering
  \includegraphics[width=0.4\textwidth,height=0.4\textwidth]{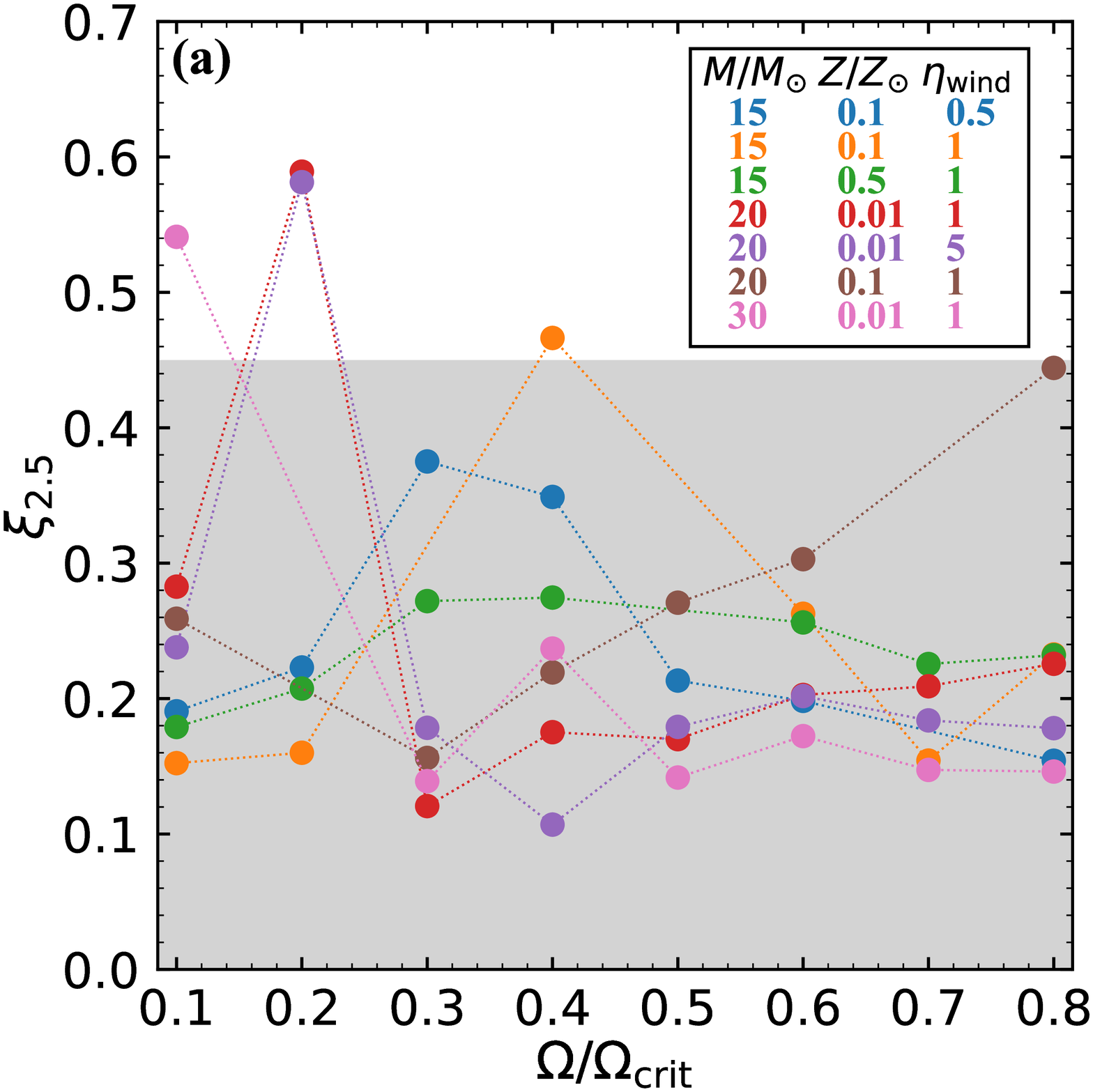}
  \includegraphics[width=0.4\textwidth,height=0.4\textwidth]{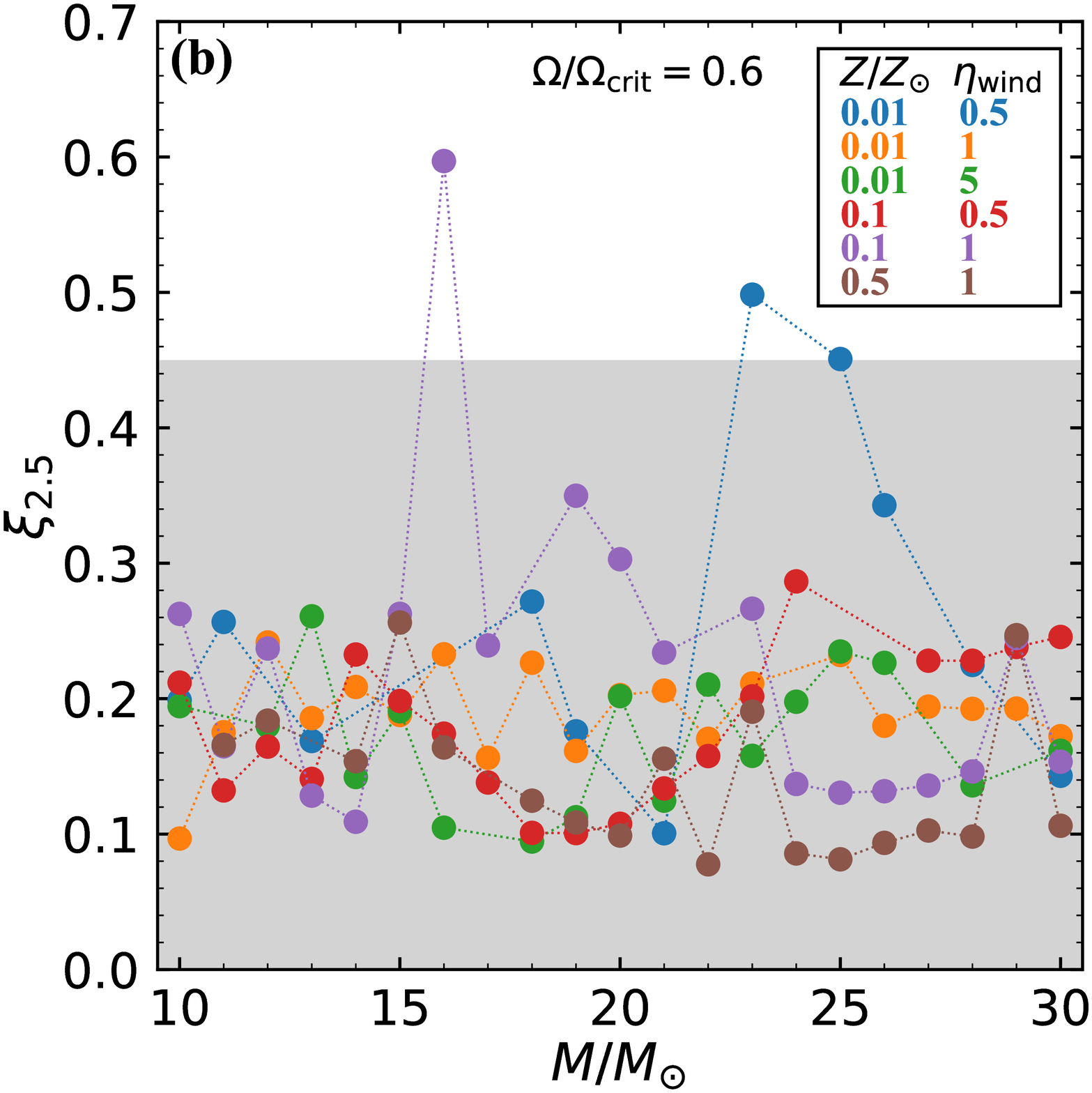}
  \caption{Compactness parameters $\xi_{2.5}$ of progenitor models as a function of initial rotation rate $\Omega/\Omega_{\rm crit}$ and stellar mass. The $\xi_{2.5}=0.45$ line separates models that could explode, which are driven by neutrino (gray) and disallowed (white) regions.  }
  \label{figComp}
\end{figure*}

The compactness parameter $\xi_{2.5}$ is calculated according to Equation (\ref{xi}). As shown in Figure \ref{figComp}, $\xi_{2.5}$ varies with the initial rotation rate $\Omega/\Omega_{\rm crit}$ and initial mass $M_{\rm ini}$ of progenitors. The $\xi_{2.5}$ = 0.45 line separates models that could explode, which are driven by neutrino (gray) and disallowed (white) regions. It is easy to find that most of our progenitor models could explode successfully. The models with $Z=0.01~Z_{\odot}$ and $\eta_{\rm wind}$ = 1 appear to have rather low compactness. The values of the compactness parameter are less affected by the initial velocity of stars, especially for $\eta_{\rm wind}$ = 1. The compactness parameter is positively correlated with the mass of the NSs in all cases. The models with the same initial rotation rate, the compactness parameter varies with the initial mass in a non-monotonic way. The non-monotonic behavior of the compactness for nonrotating massive stars has been investigated by some literatures. \cite{Sukhbold2014} found that the timing and location of several carbon and oxygen convective shells play a major role in generating this non-monotonic structure. The final compactness can be quite complicated because of the multiple episodes of shell burning. It is also sensitive to the initial mass, metallicity, convection way, and nuclear physics in code. The BHs are predicted to be produced by stars in the compactness peaks. When compared to a single star, the compactness in a stripped star shifts to a higher CO core mass \citep{Woosley2019,Ertl2020,Schneider2021}.

It is worth noting that the compactness parameter is really only meaningful for neutrino-driven explosions because these involve a threshold process. Neutrino heating mechanism could account for SNe with energy approximately 10$^{51}$ ergs \citep [see a review by][]{Janka2012}. For some more energetic SNe, magnetohydrodynamic mechanisms are required to drive explosions \citep[e.g.,][]{Akiyama2003, Burrows2007,Shankar2021}. The characteristics and mass distribution of the remnants are also influenced by the supernova explosion mechanism \citep[e.g.,][]{Fryer2012,Fryer2022a}.

\subsection{Protomagnetar signatures}

\begin{figure}
  \centering
  \includegraphics[width=0.4\textwidth,height=0.4\textwidth]{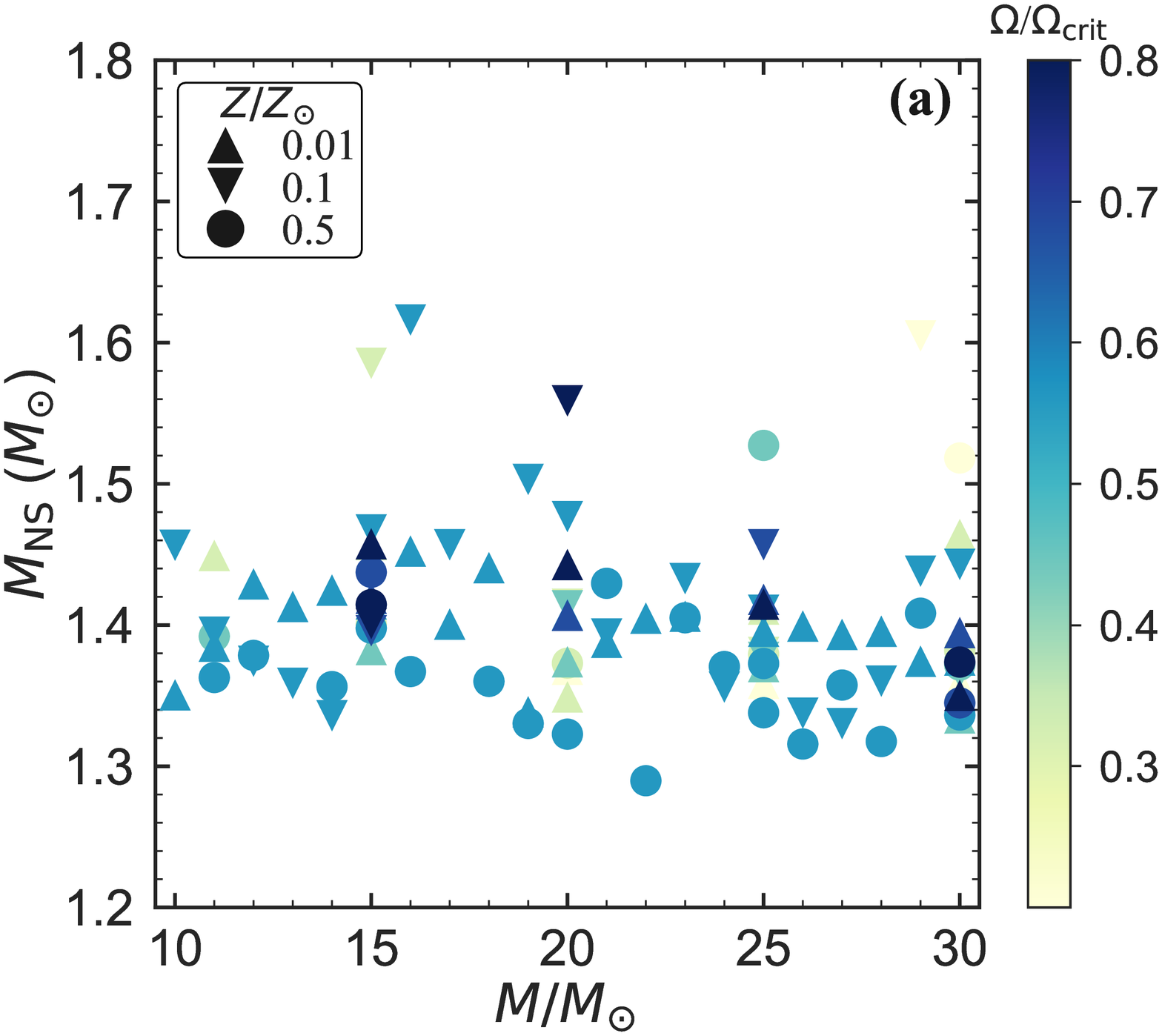}
  \includegraphics[width=0.4\textwidth,height=0.4\textwidth]{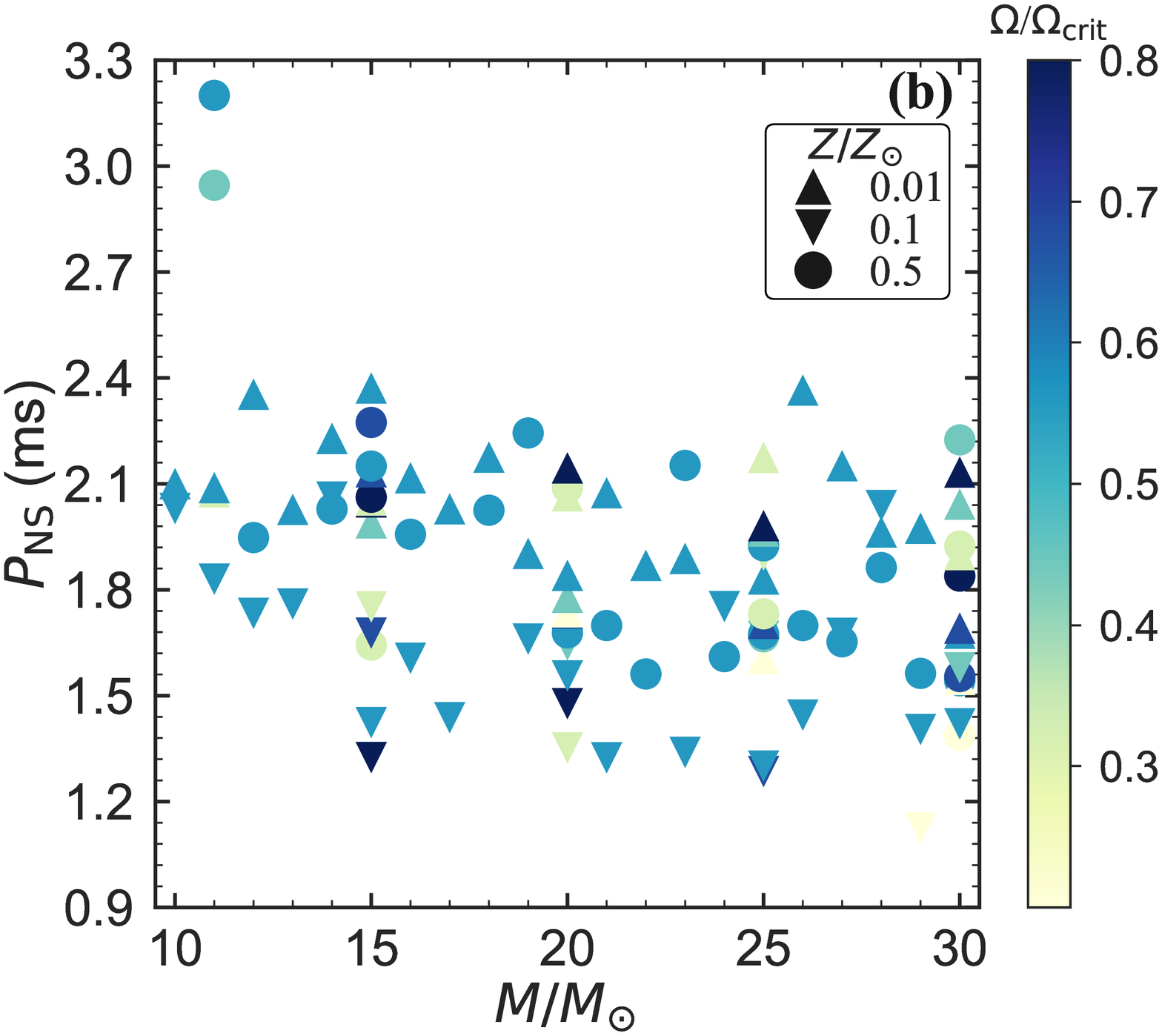}
  \includegraphics[width=0.4\textwidth,height=0.4\textwidth]{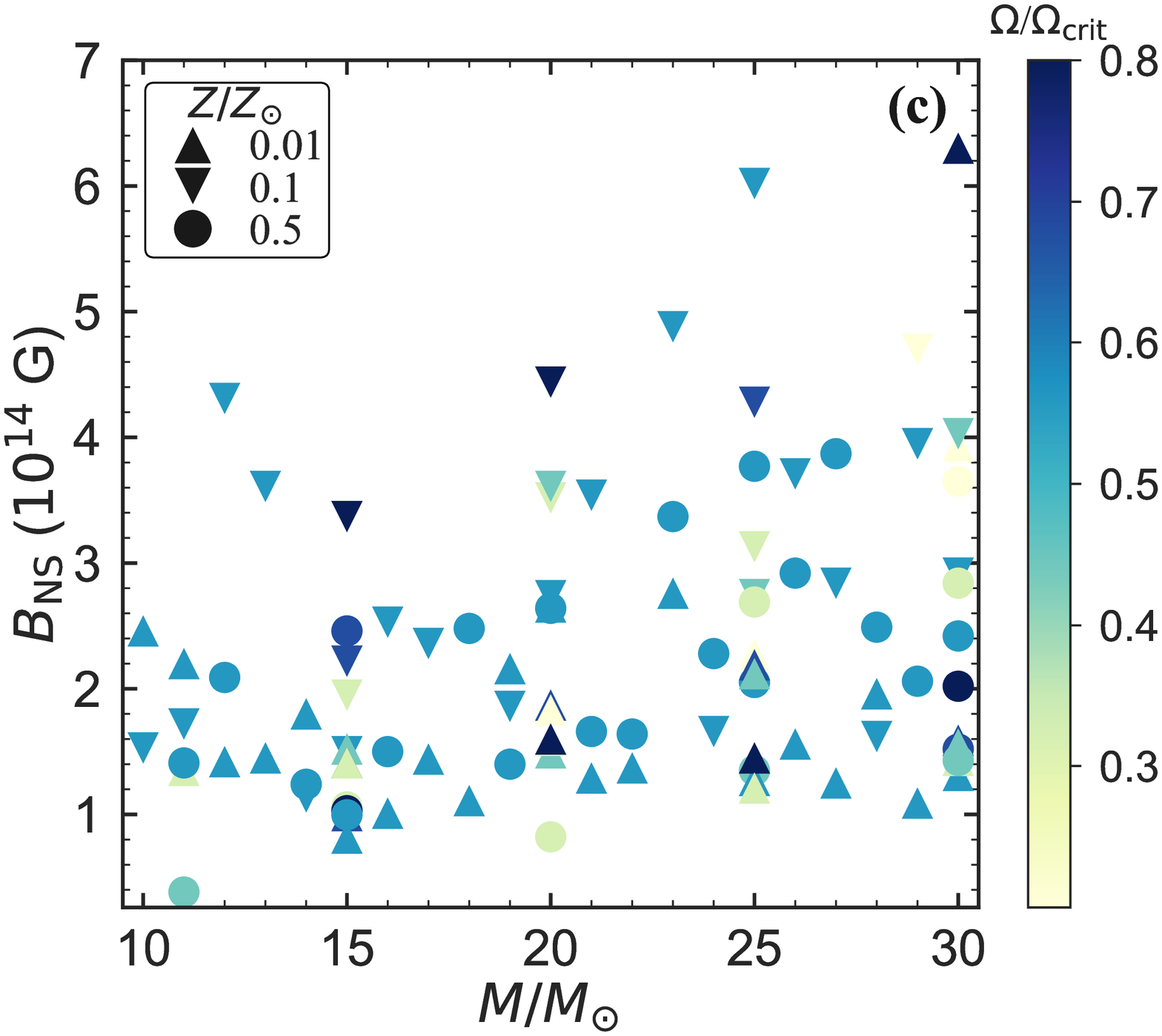}
   \caption{The mass $M_{\rm NS}$, periods $P_{\rm NS}$ and the average magnetic field strength $B_{\rm NS}$  of the protomagnetar as a function of initial stellar mass for models with different initial metallicity. All models involved in the diagram have the same ``Dutch'' wind scale factor $\eta_{\rm wind}=1$, and they eventually evolved into WR stars.}
  \label{figNSa}
\end{figure}

\begin{figure}
  \centering
  \includegraphics[width=0.4\textwidth,height=0.4\textwidth]{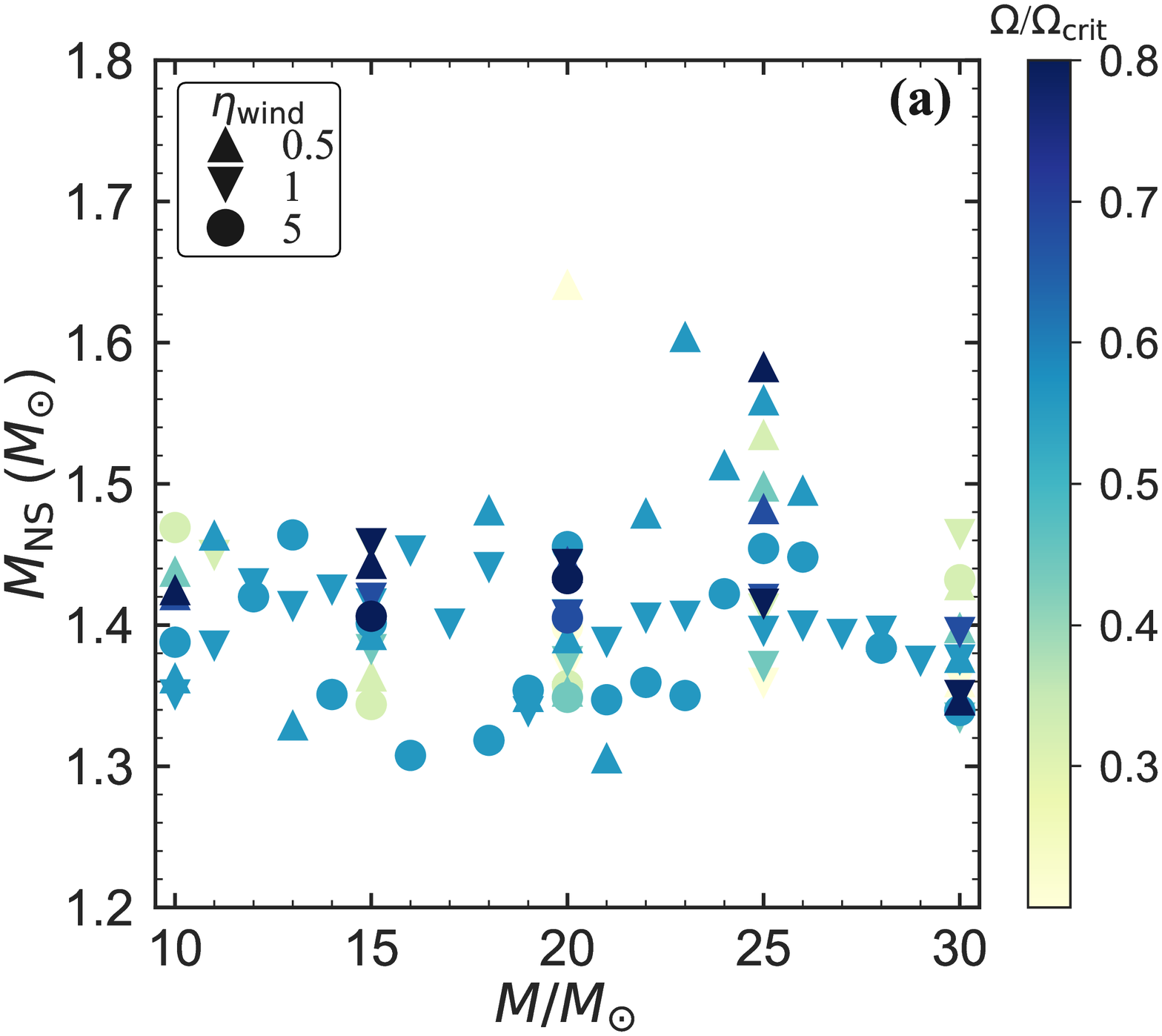}
  \includegraphics[width=0.4\textwidth,height=0.4\textwidth]{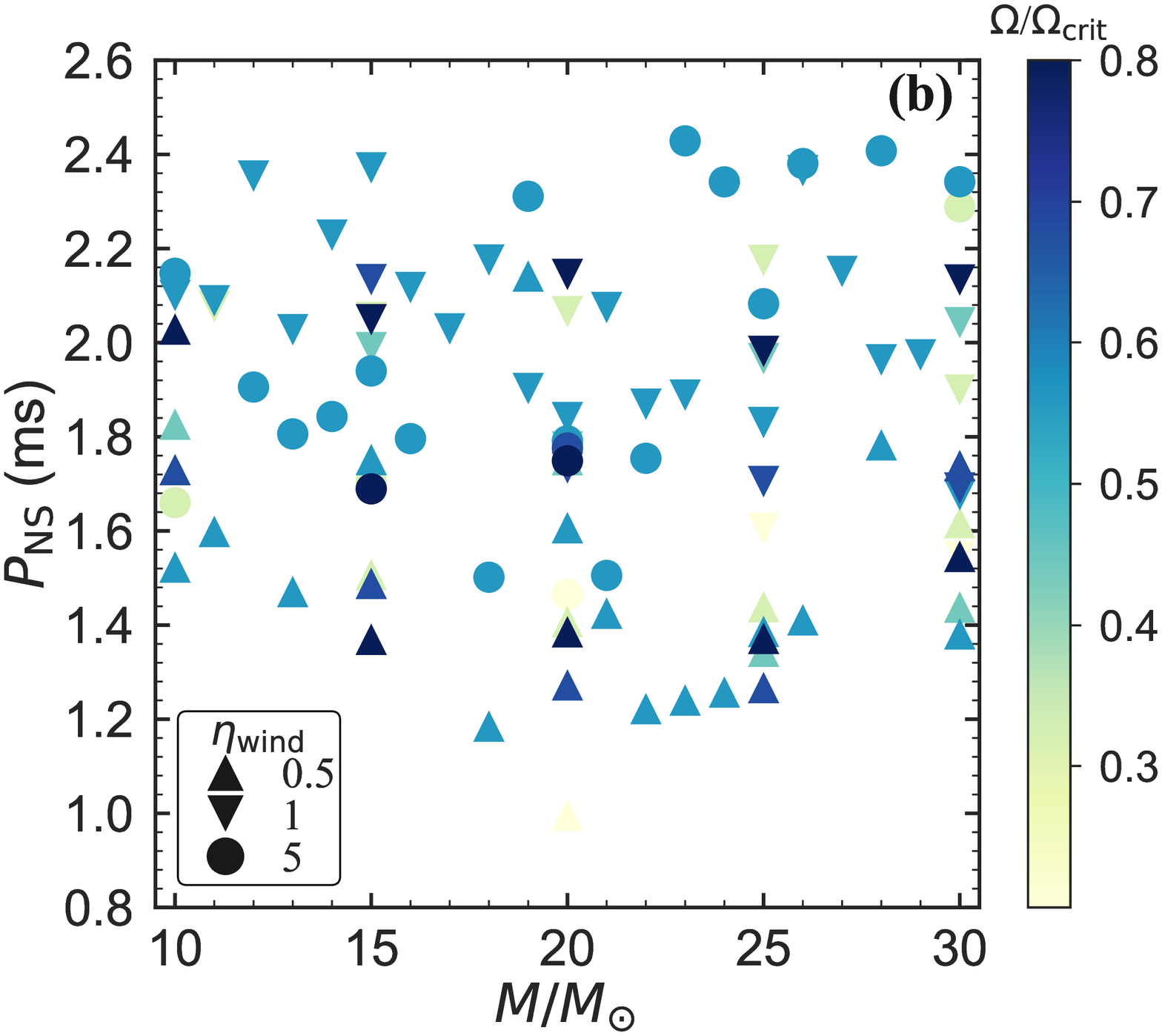}
  \includegraphics[width=0.4\textwidth,height=0.4\textwidth]{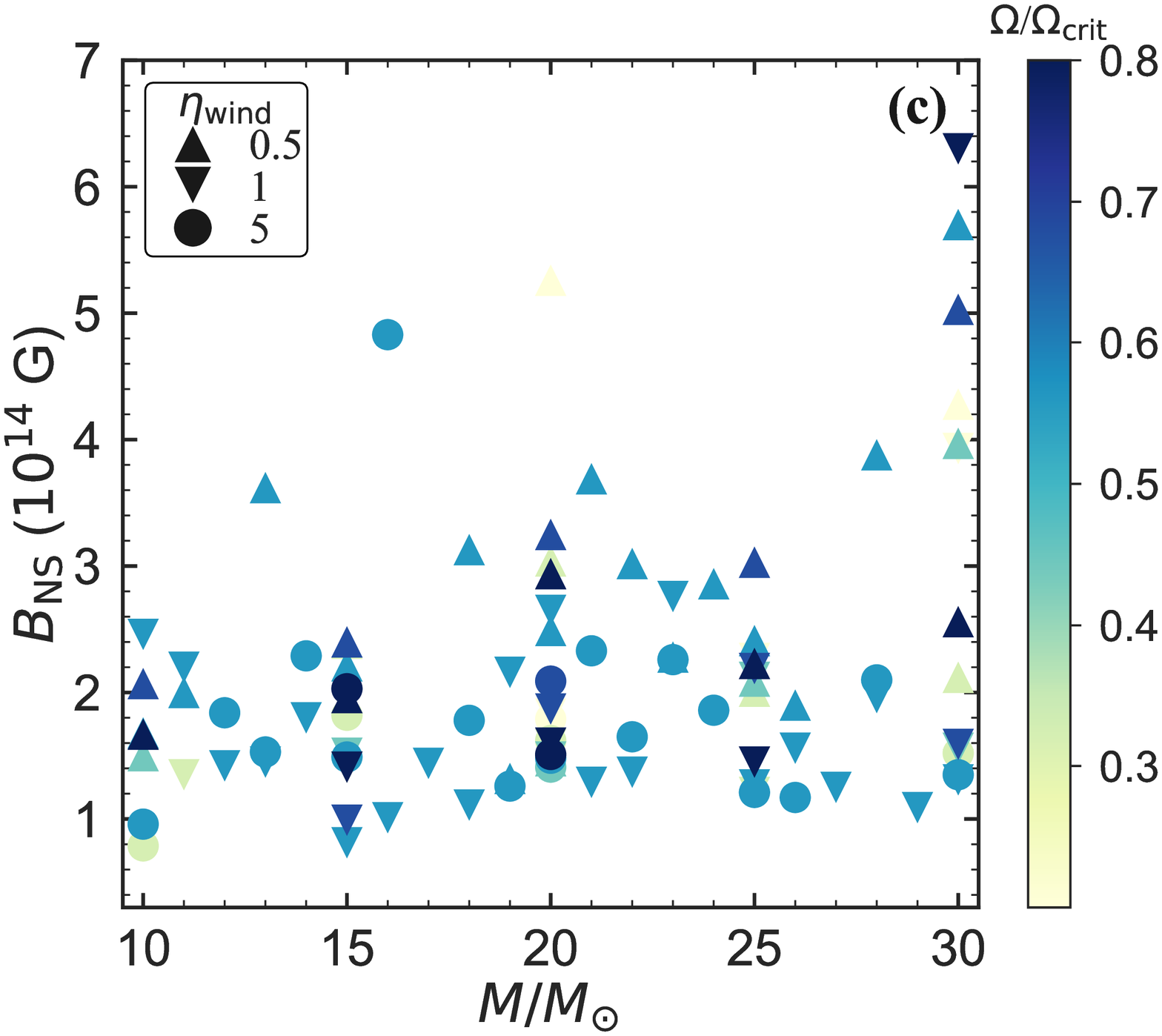}
   \caption{The mass $M_{\rm NS}$, periods $P_{\rm NS}$ and the average magnetic field strength $B_{\rm NS}$ of the protomagnetar as a function of initial stellar mass for models with different ``Dutch'' wind scale factor.  All models involved in the diagram have the same initial metallicity $Z=0.01Z_{\odot}$ , and they eventually evolved into WR stars.}
  \label{figNSb}
\end{figure}

\begin{figure*}
  \centering
  \includegraphics[width=0.4\textwidth,height=0.4\textwidth]{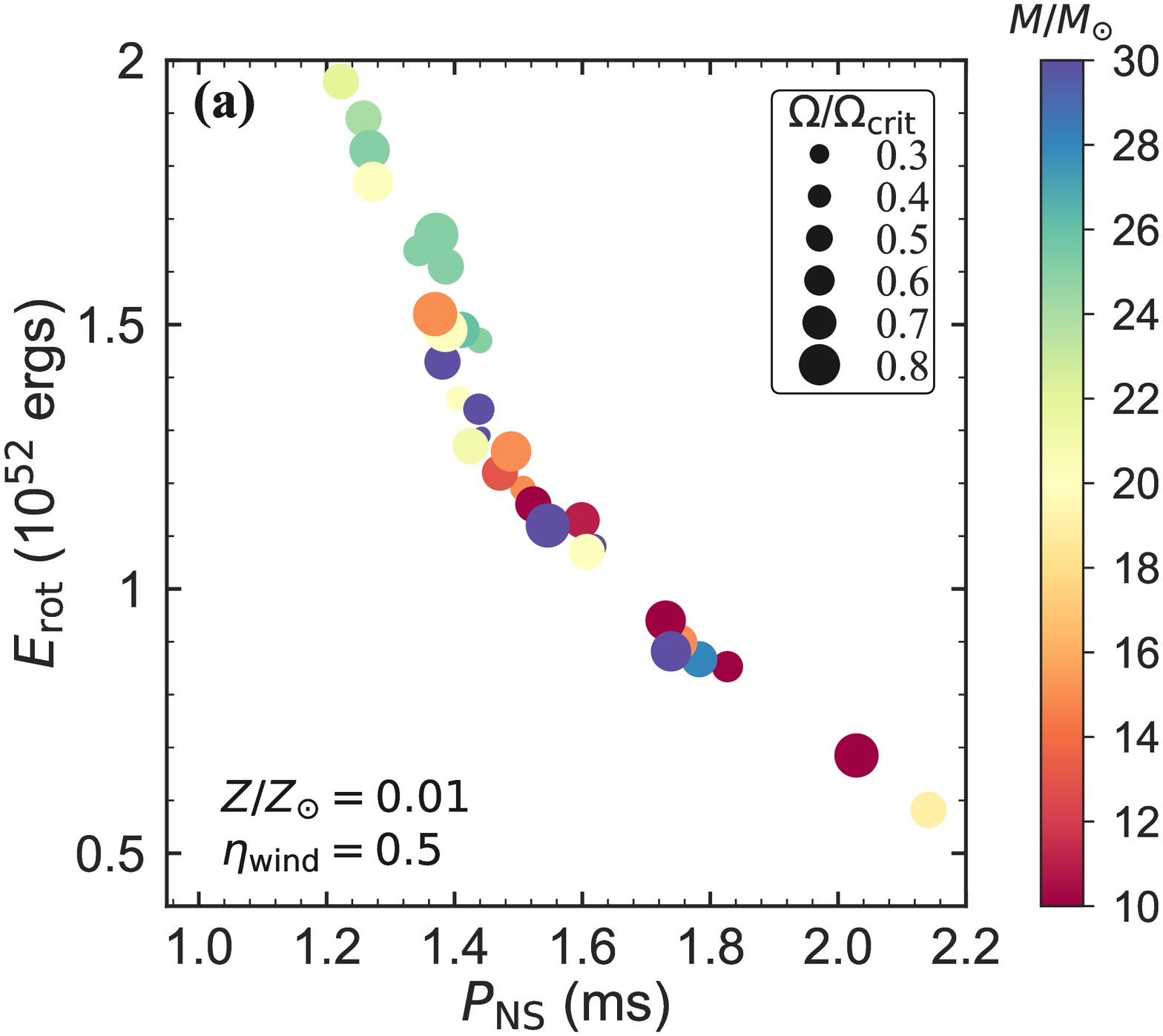}
  \includegraphics[width=0.4\textwidth,height=0.4\textwidth]{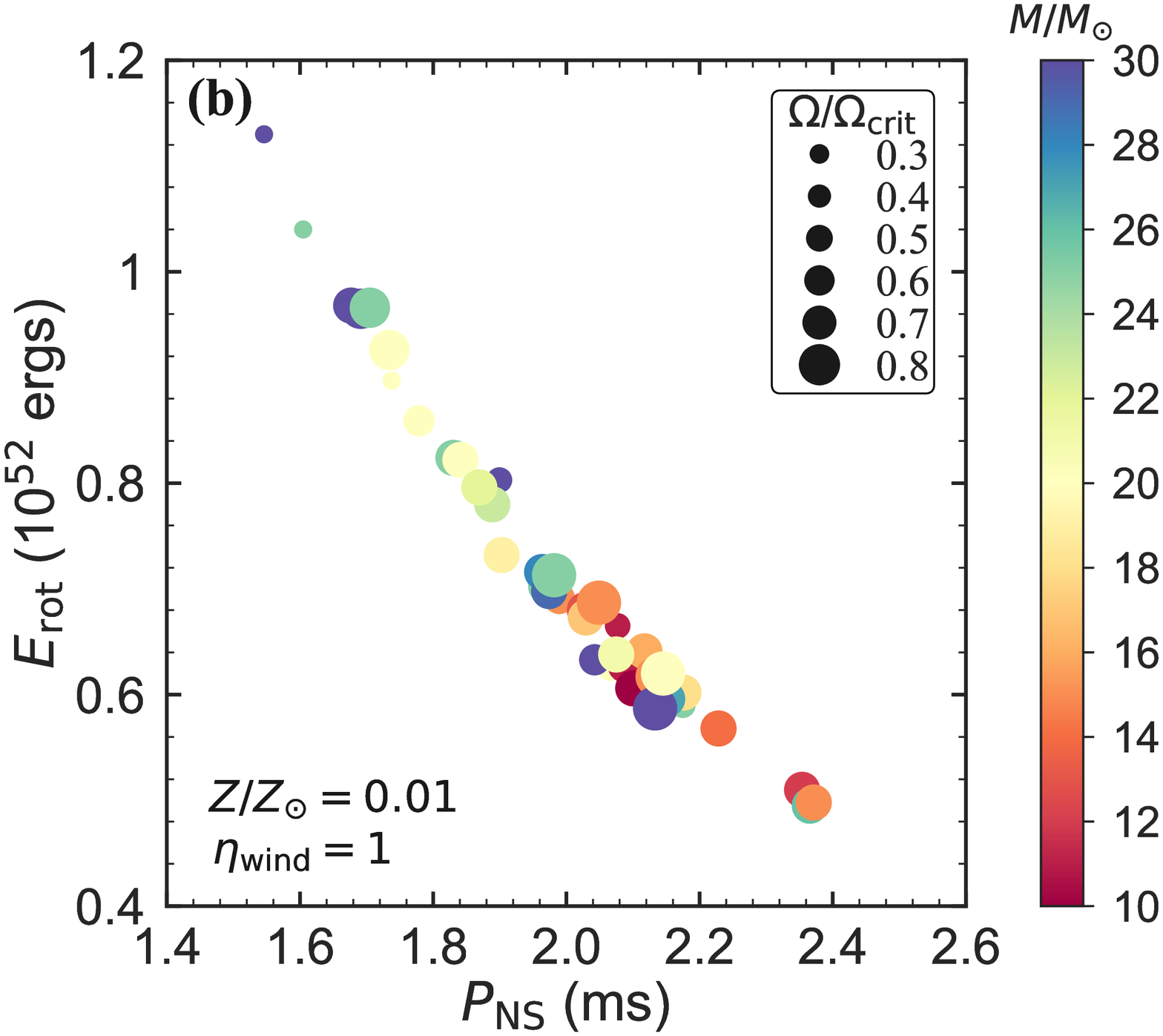}
  \includegraphics[width=0.4\textwidth,height=0.4\textwidth]{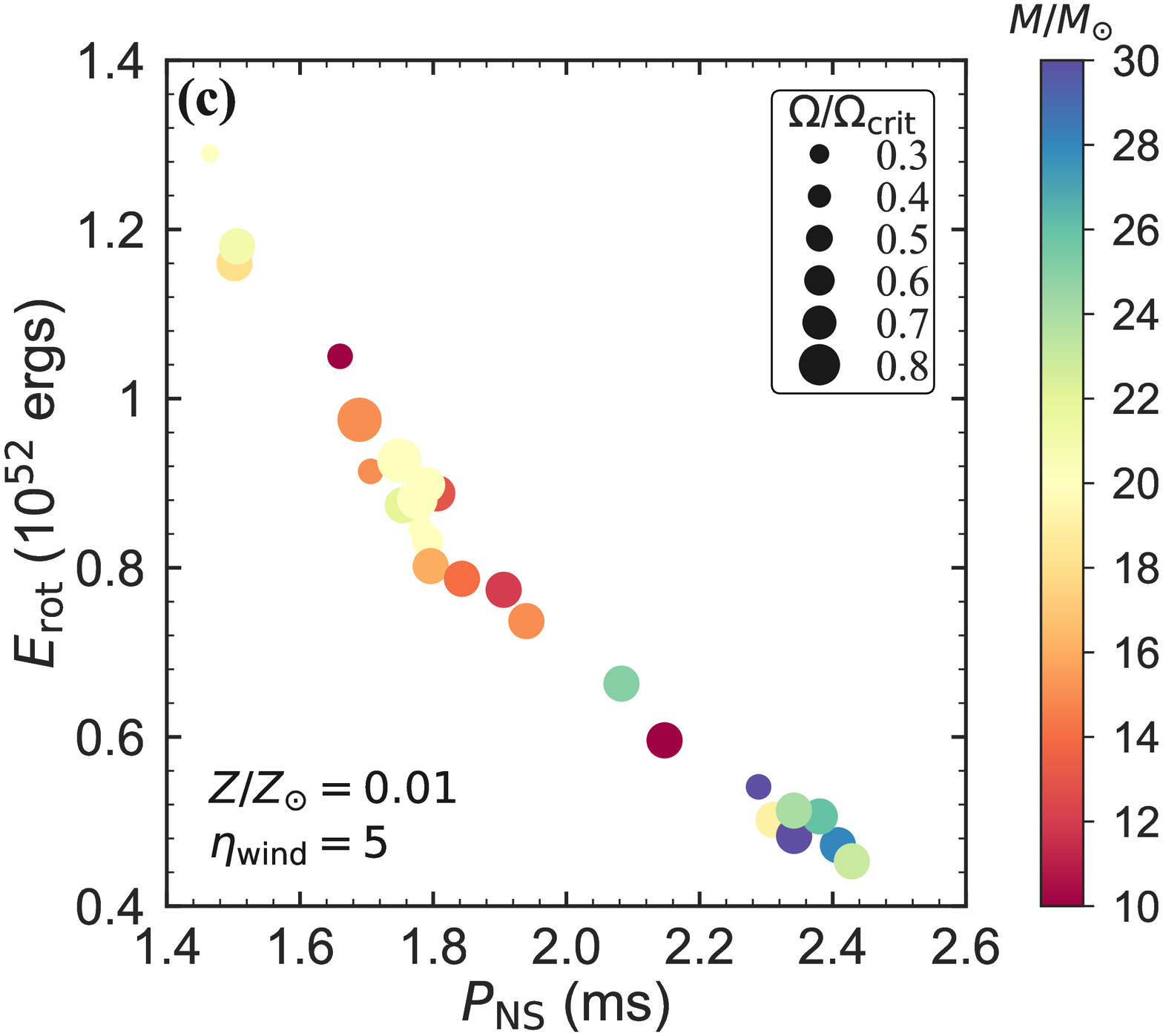}
  \includegraphics[width=0.4\textwidth,height=0.4\textwidth]{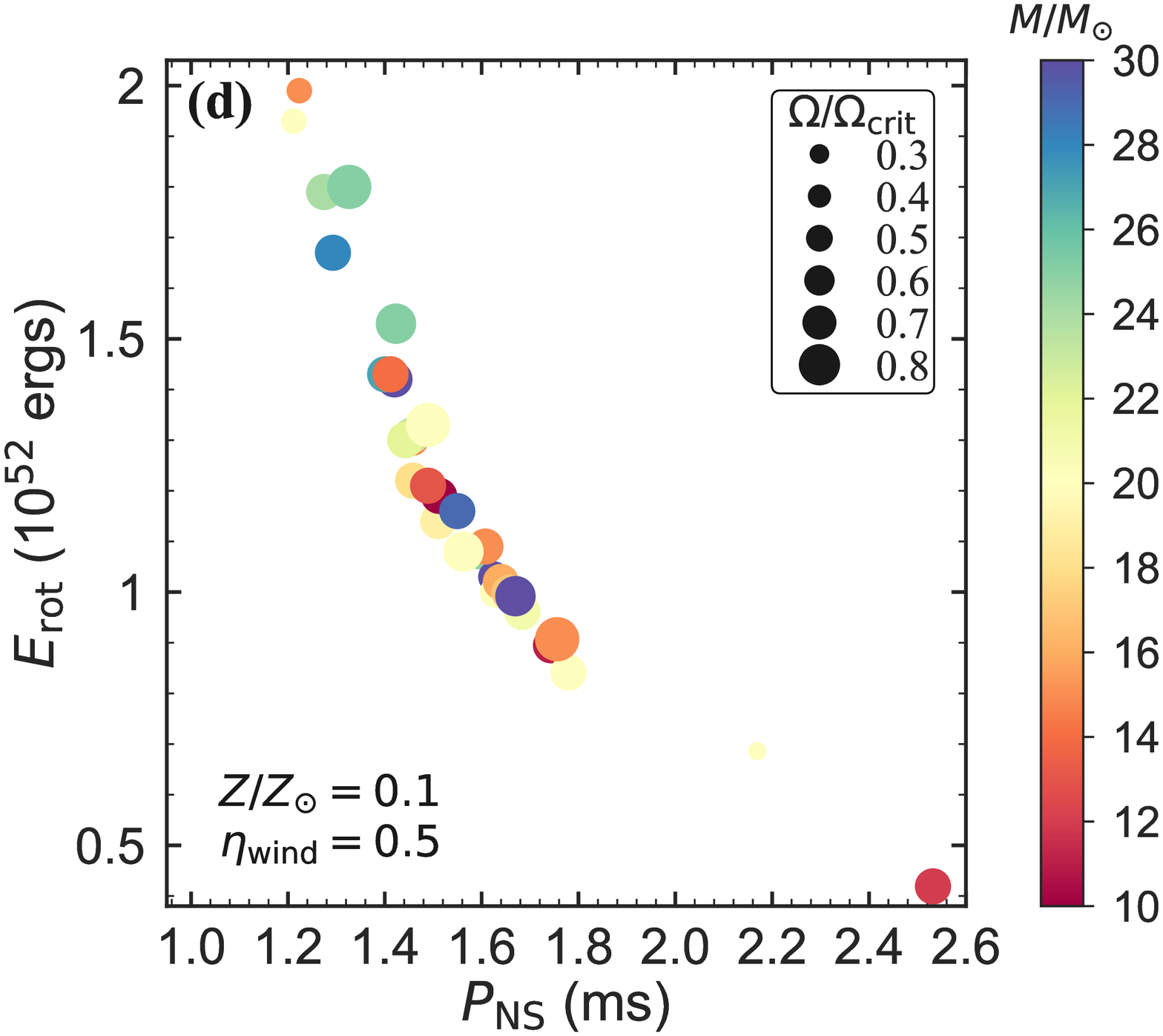}
  \includegraphics[width=0.4\textwidth,height=0.4\textwidth]{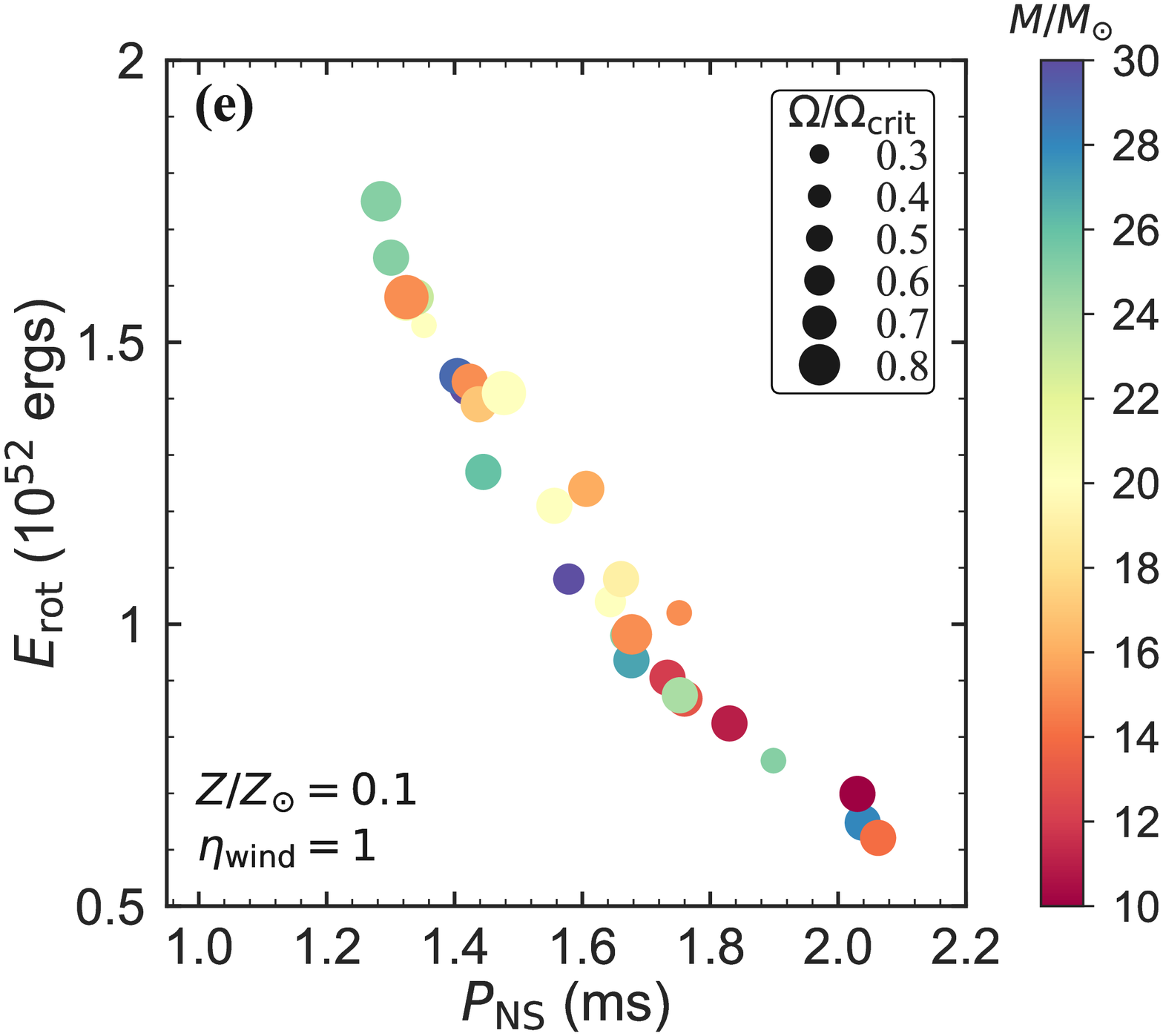}
  \includegraphics[width=0.4\textwidth,height=0.4\textwidth]{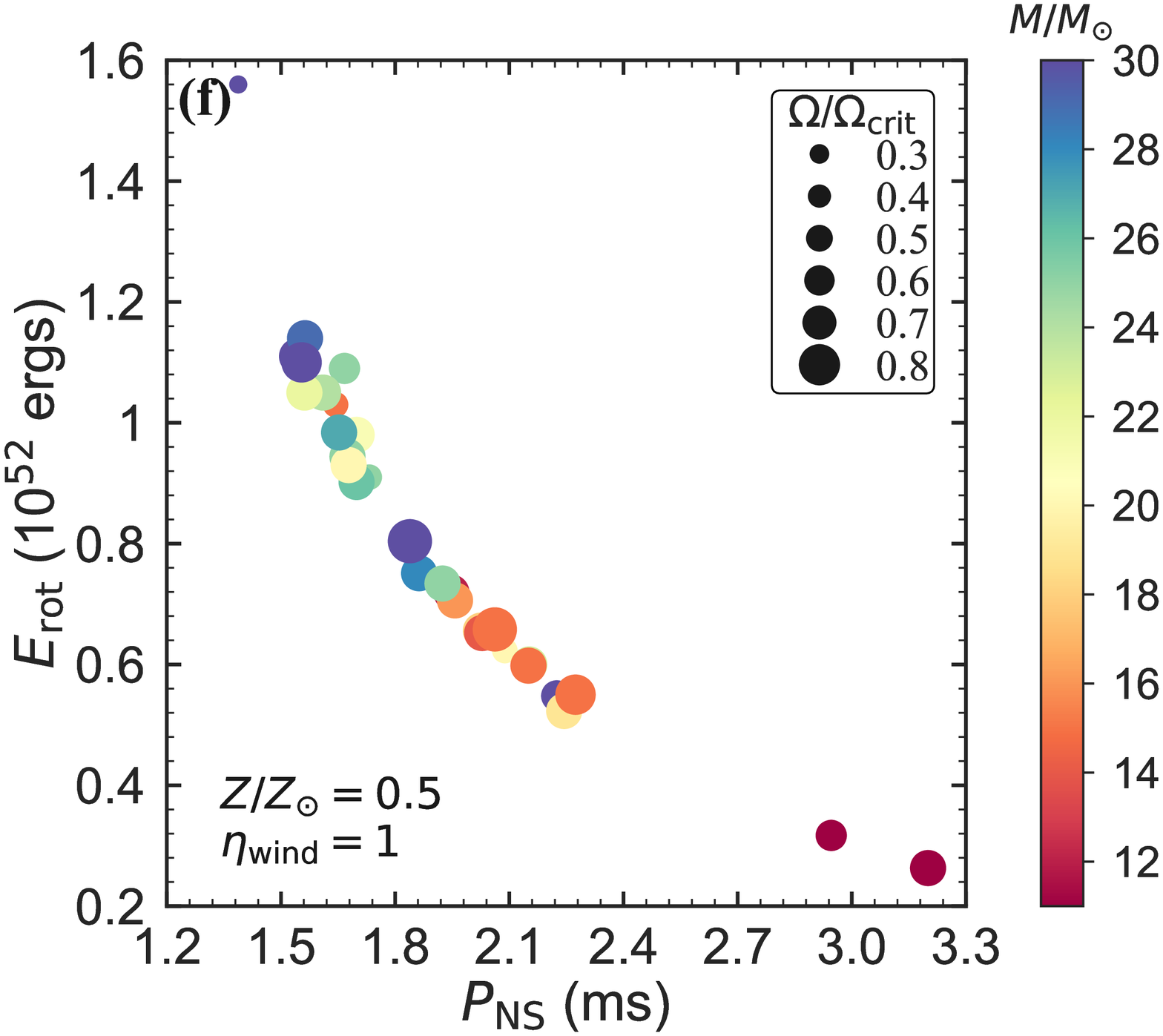}
    \caption{The total rotation energy $E_{\rm rot}$ as a function of periods $P_{\rm NS}$. All models involved in the diagram eventually evolved into WR stars. }
   \label{figErot}
\end{figure*}

\begin{figure*}
  \centering
  \includegraphics[width=0.4\textwidth,height=0.4\textwidth]{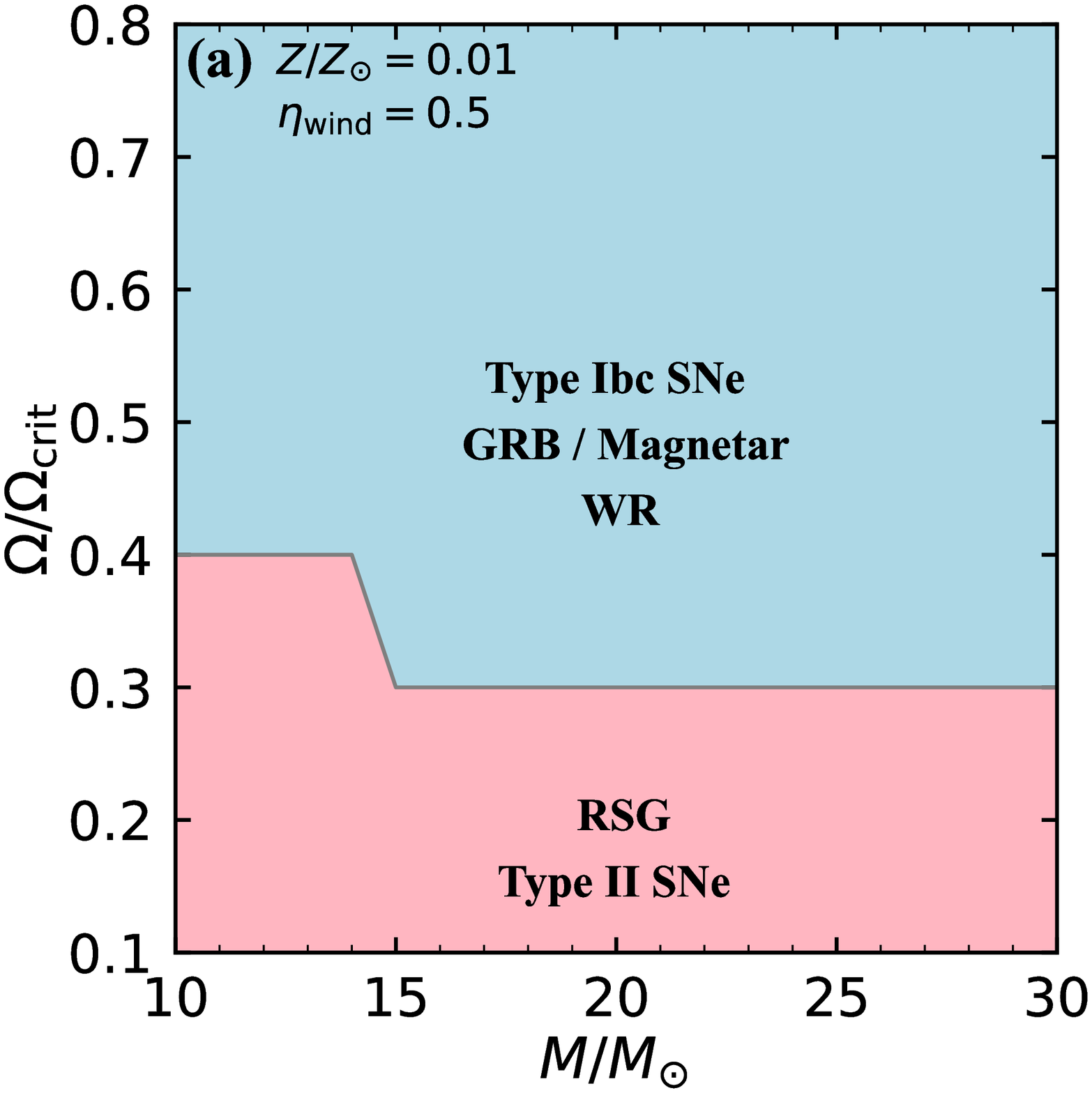}
  \includegraphics[width=0.4\textwidth,height=0.4\textwidth]{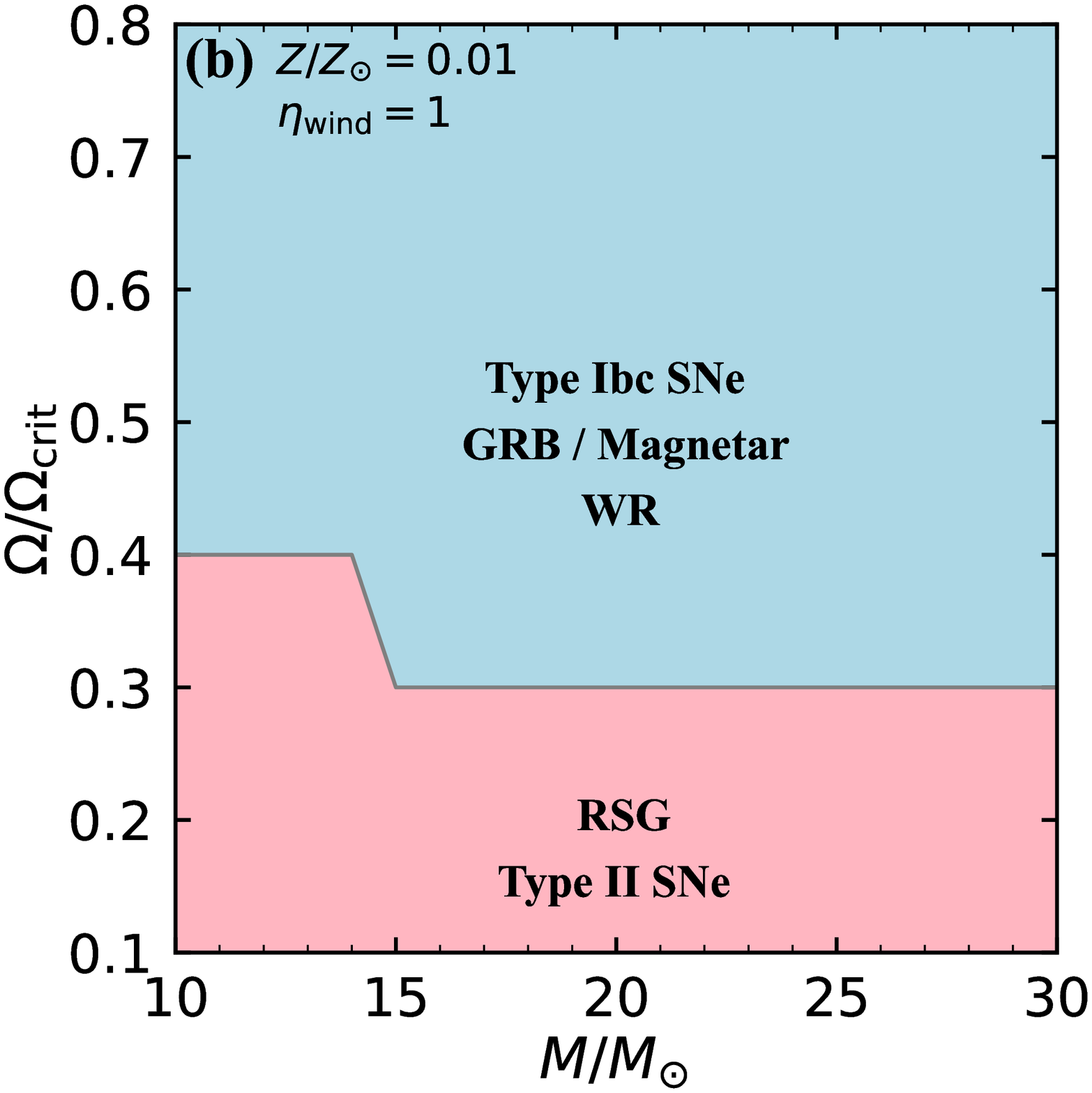}
  \includegraphics[width=0.4\textwidth,height=0.4\textwidth]{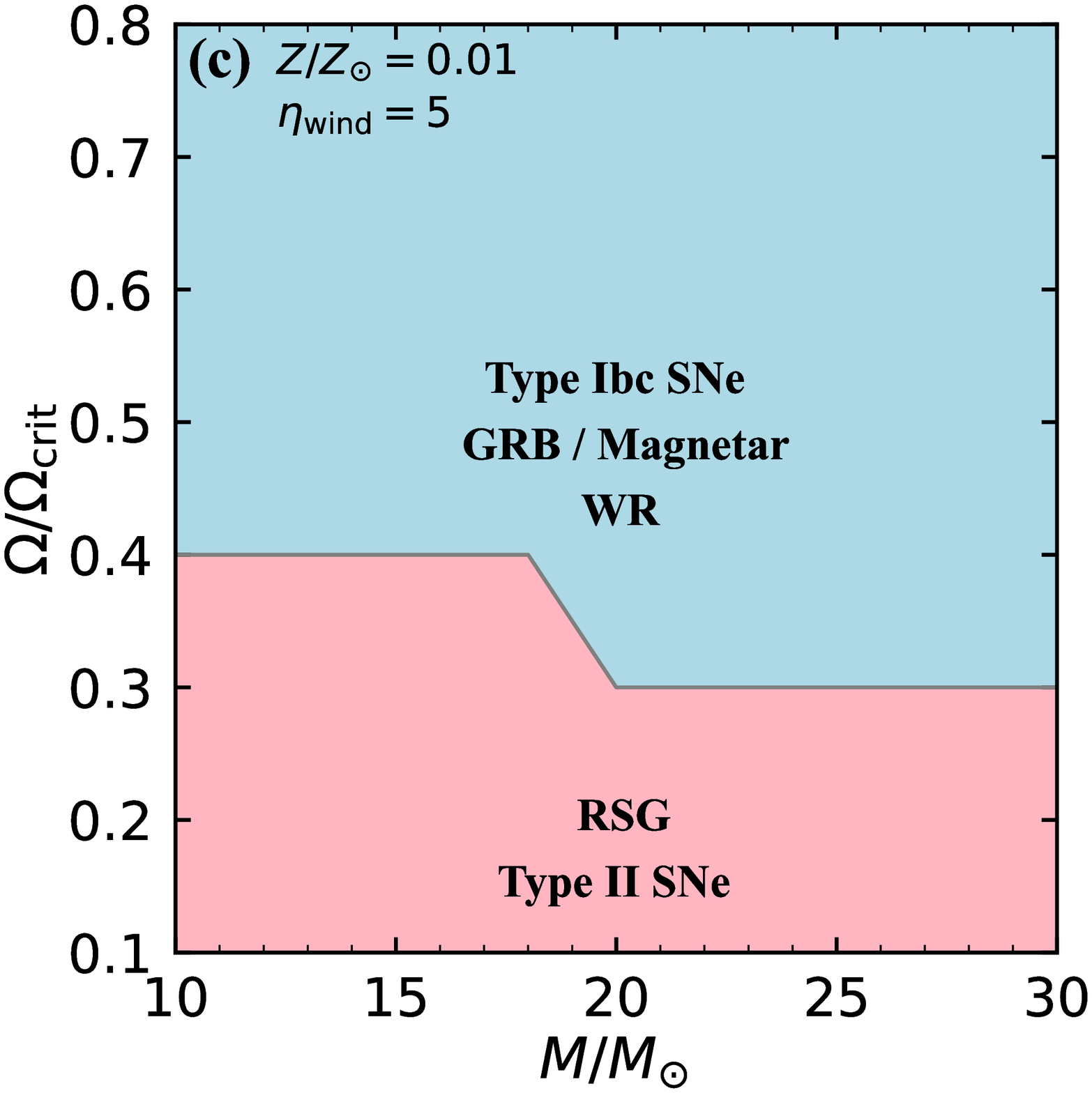}
  \includegraphics[width=0.4\textwidth,height=0.4\textwidth]{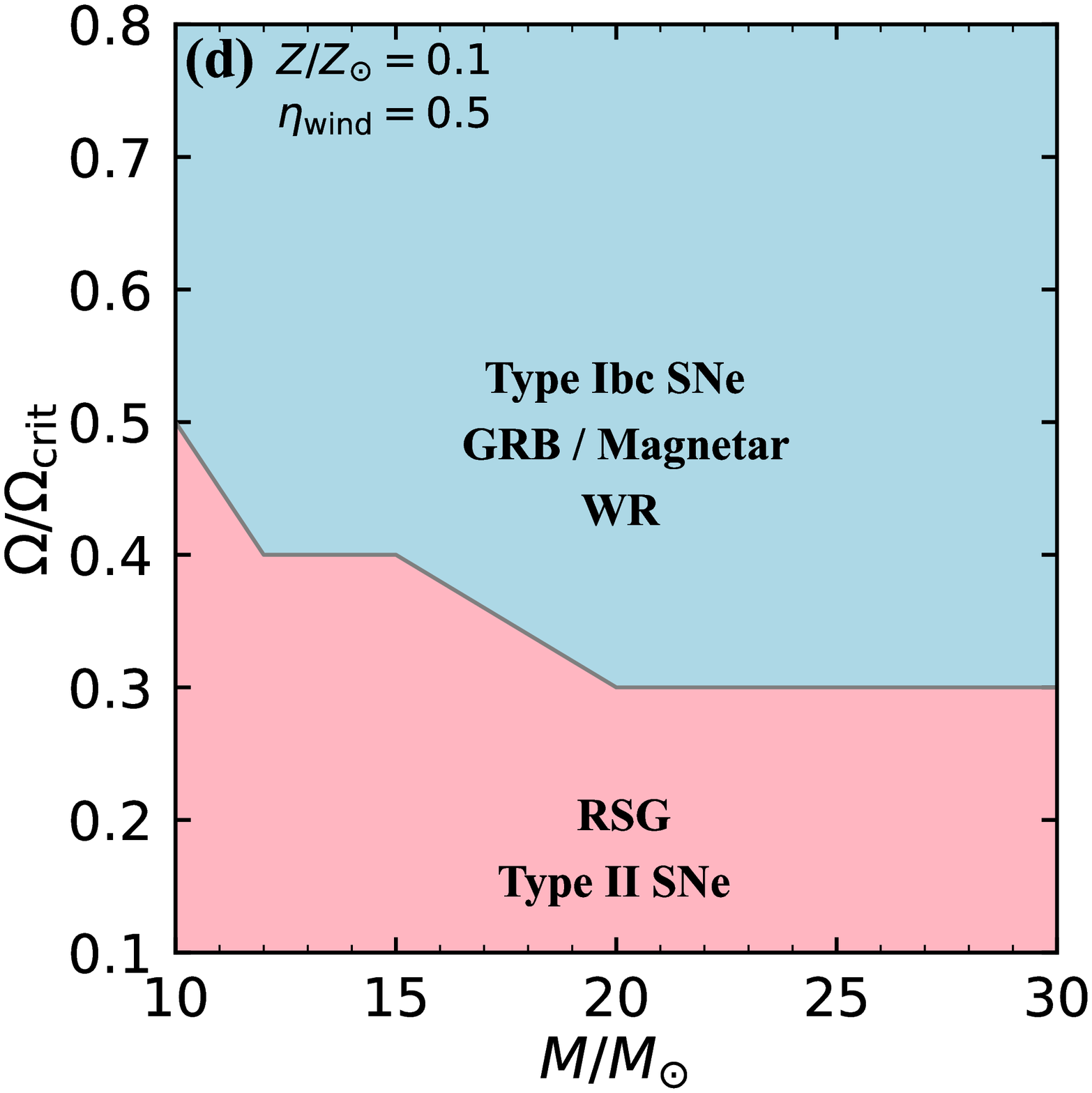}
  \includegraphics[width=0.4\textwidth,height=0.4\textwidth]{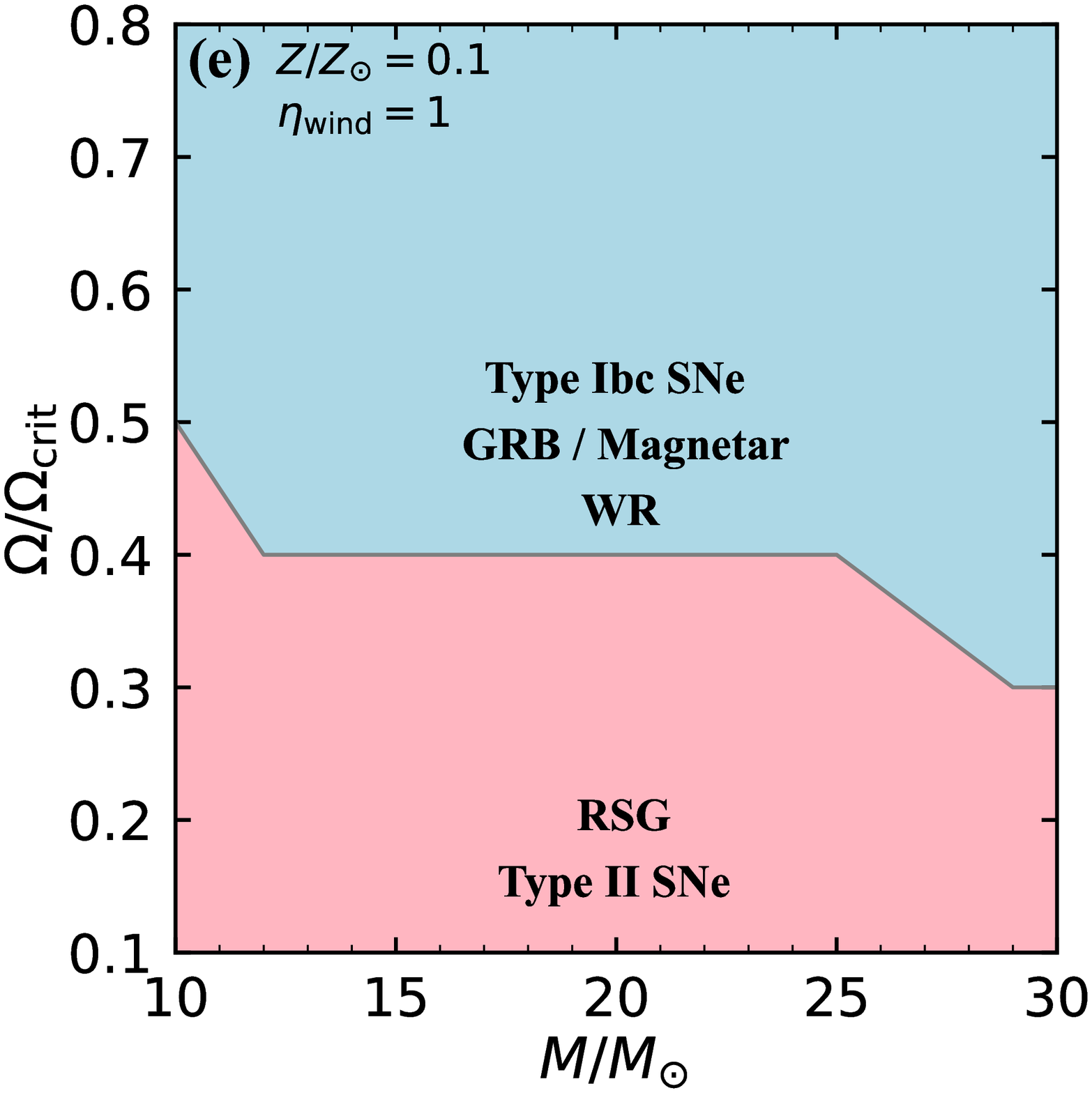}
  \includegraphics[width=0.4\textwidth,height=0.4\textwidth]{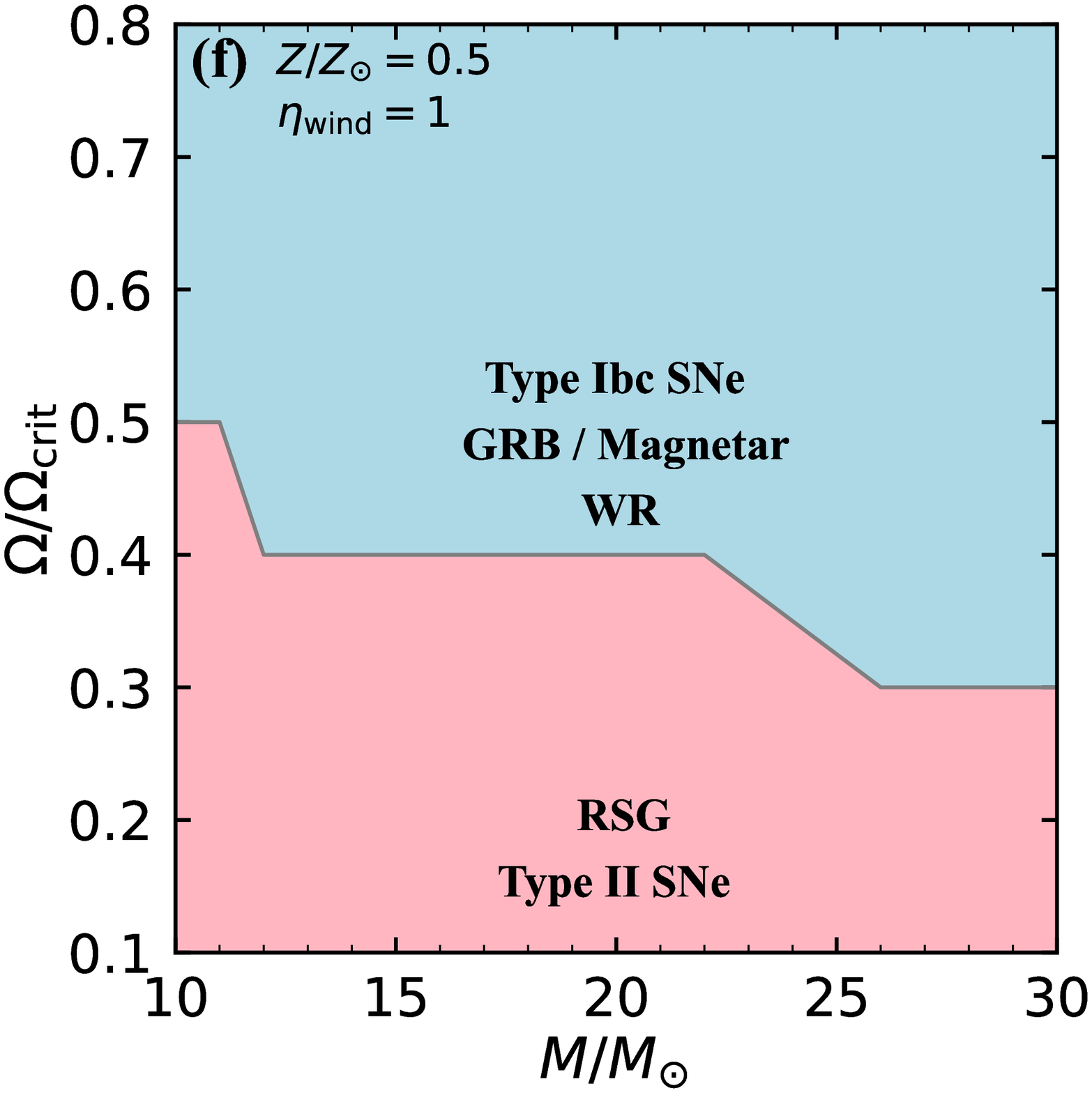}
   \caption{Ranges of masses and initial rotation rates required to produce magnetars from stellar models with different initial metallicity and ``Dutch'' wind scale factors. Only models with parameters for blue regions above the gray online can evolve into WR stars, which might eventually undergo SN explosions or GRBs when the core collapses, producing magnetars.}
  \label{figs_omega_m}
\end{figure*}

After a massive star collapses, a proto-NS might be born in the center. In this paper, we take the iron core masses as the NS baryonic masses $M_{\rm b}$. The binding energy will be released by emitting neutrinos during the iron core collapse to the NS. Following \citet{Lattimer2001}, the NS gravitational mass $M_{\rm NS}$ can be calculated by
\begin{equation}
\frac{M_{\rm b}-M_{\rm NS}}{M_{\rm NS}}  \approx \frac{0.6~GM_{\rm NS}/R_{\rm NS} c^2}{1-0.5~GM_{\rm NS}/R_{\rm NS} c^2}.
\end{equation}

The change in NS mass due to material fallback in accretion during the explosion is negligible, especially for stars with zero-age-main-sequence masses less than 30 $M_{\odot}$. The results of accurate calculations, including fallback, exhibit good agreement with this simple approach \citep[e.g.,][]{Sukhbold2016,Ertl2020}.

The Spruit-Tayler dynamo operates in radiative regions and taps into differential rotation to produce magnetic fields \citep[e.g.,][]{Spruit2002,Petrovic2005,Heger2005}. The corresponding toroidal and radial magnetic field strength  are \citep{Heger2005}
\begin{equation}
B_{\rm \phi}=2\pi^{1/2}\rho^{1/2}\nu_{\rm e}^{1/4}q^{1/2}\omega^{3/4}r^{1/2},
\end{equation}
\begin{equation}
B_{\rm r}=2\pi^{1/2}\rho^{1/2}\nu_{\rm e}^{3/4}q^{1/2}\omega^{1/4}r^{-1/2},
\end{equation}
where $\nu_{\rm e}$ is the effective viscosity,  $q=d$ ln $\omega/d$  ln $r$ is the shear. It is assumed that the initial magnetic field is sufficiently weak. The radial component $B_{\rm r}$ is wound up by differential rotation, which leads to the dominance of toroidal magnetic field $B_{\rm \phi}\gg B_{\rm r} $. A detailed description of the dynamo mechanism has been presented in \cite{Spruit2002}. It should be noted that MESA doesn’t calculate any Spruit-Tayler magnetic fields in convective regions \citep{Paxton2013}. The surface magnetic field produced by a dynamo \citep[e.g.,][]{Brandenburg2005} in the convective zone is also not included in star evolution.

The average strength of dynamo-generated magnetic fields inside the iron core is given by
\begin{equation}
<B_{\phi}>=\frac{\int_{0}^{M_{\rm Fe}} B_{\phi}(m)dm}{\int_{0}^{M_{\rm Fe}}dm}.
\end{equation}
Assuming the conservation of magnetic flux and iron core contracted homogeneously to the NS with radius $R_{\rm NS}=12$ km, we obtain the mean magnetic field strength of NSs.

The natal spin rates of NSs can be determined by angular momentum within the iron core:
\begin{equation}
P_{\rm NS}=\frac{2\pi I_{\rm NS}}{J_{\rm Fe}}.
\end{equation}
Here, the moment of inertia of the NSs is set as $I_{\rm NS}$ = 0.35$M_{\rm NS}R_{\rm NS}^2$ \citep[e.g.,][]{Lattimer2001}.

In Figure \ref{figNSa}, the mass $M_{\rm NS}$, periods $P_{\rm NS}$ and the mean magnetic field strength  of the protomagnetar as a function of initial stellar mass for models with different initial metallicity. All models involved in the diagram have the same ``Dutch'' wind scale factor $\eta_{\rm wind}=1$. The upward triangles, downward triangles, and circles  represent different metallicity $Z=0.01, 0.1,$ and $0.5~Z_{\odot}$, respectively. The initial rotation rate $\Omega/\Omega_{\rm crit}$ is indicated by the color bar, which varies from $0.3-0.8$. As mentioned before, it is only the models with initial rotation rates greater than 0.3 or 0.4 that can eventually evolve into WR stars, which could produce SNe or/and GRBs at core collapse. It is easy to see that the mass of most proto-NSs born in higher initial metallicity ($Z=0.5~Z_{\odot}$) progenitors are lower than those of progenitors produced by lower initial metallicity ($Z=0.01$ and $0.1~Z_{\odot}$). Almost all $P_{\rm NS}$ of proto-NSs born in progenitors with $Z=0.1~Z_{\odot}$ are shorter than those of progenitors with $Z=0.01$ and $0.5~Z_{\odot}$.  The higher the initial mass, the smaller the $P_{\rm NS}$, and this phenomenon is more evident in models with $Z=0.1$ and $0.5~Z_{\odot}$. The rotation rate of NSs does not increase monotonically with the initial rotation rate of the progenitor, which may be related to the fact that the angular momentum loss during star evolution is influenced by several factors.  The mean magnetic fields $B_{\rm NS}$ of most proto-NSs born from progenitors with $Z=0.1~Z_{\odot}$ are larger than those of progenitors with $Z=0.01$ and $0.5~Z_{\odot}$ when $\eta_{\rm wind}$ = 1.

Similarly, Figures \ref{figNSb}(a)-(c) correspond to $P_{\rm NS}$, $M_{\rm NS}$, and $B_{\rm NS}$ of proto-NSs as a function of initial stellar mass for models with different ``Dutch'' wind scale factor, respectively.  All models involved in the diagram have the same initial metallicity $Z$ = 0.01 $Z_{\odot}$, and they eventually evolved into WR stars. The upward triangles, downward triangles, and circles represent  the ``Dutch'' wind scale factor $\eta_{\rm wind}$ = 0.5, 1, and 5, respectively. We can find that the larger the ``Dutch'' wind scale factor, the smaller the mass of the final produced NSs for models with the same initial metallicity.  Models with $\eta_{\rm wind}$ = 0.5 produce proto-NSs with larger masses, shorter periods, and stronger mean magnetic fields than those with $\eta_{\rm wind}$ = 1 and 5. The masses $M_{\rm NS}$ span from 1.29 to 1.61 $M_{\odot}$, with a median value of 1.4 $M_{\odot}$. The $P_{\rm NS}$ range from $1.0-2.6$ ms. Even after taking into account the effects of initial rotation rate, stellar wind loss, and metallicity, progenitor stars with larger initial masses are still more likely to form fast-spinning NSs. The more massive NSs are more likely to have faster rotation rates and stronger magnetic field strengths. Regardless, these NSs are classified as millisecond magnetars.

Here we take the iron core mass as the baryonic mass of NS, then convert it to NS gravitational mass through an analytical, radius-dependent approach \citep{Lattimer2001} and do not consider detailed SNe explosion physics. The material fallback process during an SN explosion can affect the final properties of compact remnants \citep[e.g.,]{Zhang2008,Fryer2012,Ugliano2012,Janka2013,Muller2019,Ertl2020}. First, the mass of NS could grow by the accretion of fallback material. However, if a large amount of matter could fall on the newly born NS, a BH might form. Second, recent three-dimensional simulations of nonrotating progenitors indicate that the NS can be spun up to millisecond periods by fallback \citep{Chan2020}. \cite{Muller2016} obtained a multimodal NS mass distribution with peaks around 1.15, 1.45, and possibly $1.9 ~M_{\odot}$.  According to the simulation results of \cite{Sukhbold2016}, there is some bimodality in the distribution, with peaks around 1.25 and 1.4 $M_{\odot}$. Their maximum NS is 1.68 $M_{\odot}$, and the minimum is 1.23 $M_{\odot}$. Our result exhibits good agreement with these previous SNe explosion simulations.

To explore the ability of newborn magnetars to power GRBs, the total rotation energy $E_{\rm rot}$ and $P_{\rm NS}$ of magnetars born from different values of initial mass, metallicity,  ``Dutch'' wind scale factor, and initial rotation rate are shown in Figure \ref{figErot}. All of the progenitor stars in the diagram are WR stars that have lost their hydrogen envelope. $E_{\rm rot}$ ranges from $\sim 5\times 10^{51}$ to $\sim 2 \times 10^{52}$ ergs. We conclude that stars with larger initial masses tend to produce rapidly rotating magnetars with more rotational energy, even when different initial rotation rates and stellar wind losses are taken into account. This phenomenon is especially evident for stars with $Z=0.01~Z_{\odot}$. Considering that the envelopes are easily disrupted by jets in most cases, the above results indicate that our models could meet the energy requirements of most of the observed LGRBs, including some collapsar-origin SGRBs \citep[e.g.][]{Zhang2009,Levesque2010,Thone2011,Xin2011}. In other words, we present theoretical evidence on magnetar-origin GRBs by simulating the evolution of fast-rotating, low-metallicity, and massive stars.

In Figure \ref{figs_omega_m}, we present how the ranges of masses and initial rotation rates required to produce magnetars from stellar models vary with different initial metallicity and ``Dutch'' wind scale factors. The model of a star with initial mass and rotation rate in the red region eventually evolved into RSG to possibly produce Type II SNe. The model with parameters within the blue region evolved into a WR star, which might eventually undergo SN explosions or GRBs when the core collapses, producing magnetars. It can be found that the greater the metallicity and ``Dutch'' wind scale factor, the greater the minimum rotation rate required to produce a magnetar. For a star with a mass around 10 $M_{\odot}$, when $(Z/Z_{\odot}, \eta_{\rm wind})=(0.5, 1)$, the initial rotation rate at least 0.5 to evolve into a WR star. When $(Z/Z_{\odot}, \eta_{\rm wind})=(0.01, 0.5), (0.01, 1)$, this value of 0.4 is sufficient. As the initial mass increases, the minimum rotation rate required to form a magnetar gradually decreases.

\section{Conclusions and discussion}

We use MESA code to simulate a grid of stellar models with different initial rotation rates ($\Omega/\Omega_{\rm crit}=0.1-0.8$),  masses ($10-30~M_{\odot}$), metallicity ($Z=0.01, 0.1$, and $0.5~Z_{\odot}$), ``Dutch'' wind scaling factors ($\eta_{\rm wind}=0.5, 1$, and $5$). All models evolve from the pre-main-sequence stage until core collapse. The effects of the initial rotation, mass, metallicity, and mass loss on the formation and characteristics of the GRB progenitor and protomagnetar are explored. We find that the initial rotation rate and mass play a dominant role in whether a star can evolve into a GRB progenitor.  As the initial mass increases, the minimum rotation rate required to form a magnetar gradually decreases. The initial metallicity and the ``Dutch'' wind scaling factors affect the minimum initial rotation rate required to produce a magnetar. The greater the metallicity and ``Dutch'' wind scale factor, the greater the minimum rotation rate. In other words, massive stars with low metallicity are more likely to harbor magnetars.

The final stellar, helium, and carbon-oxygen core masses roughly increase with increasing initial mass and decrease moderately with increasing initial rotation rate for most cases. We also discuss the progenitors of the different types of supernovae. Most of our progenitor models would give rise to Type Ib SNe, and a small fraction of them will produce Type Ic. Furthermore, we find that the compactness parameter remains a non-monotonic function of the initial mass and initial rotation rate when the effects of different metallicity and wind mass loss are considered.

The period of a protomagnetar ranges from $1.0-3.2$ ms with a median value of 1.6 ms, and the mass ranges from $1.18-1.68~M_{\odot}$ with a median value of 1.4 $M_{\odot}$. The average magnetic field strength is concentrated on the order of $\sim 10^{14}$ G. The $E_{\rm rot}$ can be up to about $2 \times 10^{52}$ ergs. The rotational energy, magnetic field strength, and period of the protomagnetar are satisfied to power the typical LGRBs.

Observation and theory demonstrated that GRB rates and properties vary with redshift due to the different distributions of metallicity, star formation rate, and initial mass distributions of massive stars at different redshifts \citep[e.g.,][]{Yonetoku2004,Hirschi2005,Langer2006,Madau2014,Taggart2021}. The evolution of LGRB progenitors is also influenced by redshift \citep{Yoon2006,Fryer2022b}. However, the analysis of the metallicity of GRB host galaxies reveals that GRBs favor subsolar metallicity, even considering the redshift effect \citep{Levan2016}. Therefore, we only consider three subsolar metallicity values $Z=0.01, 0.1,$ and $0.5 ~Z_{\odot}$ in our stellar models.

It is unclear how much the evolutionary pathways of single and binary stars contributed to the production of Type Ibc SN and GRB progenitors \citep[e.g.,][]{Langer2012}. Here, we do not consider the interaction of binary stars and focus on exploring the effects of different parameters on single star evolution and the formation of LGRB central engines. The evolution of binary stars is more complicated than that of a single star. In addition to some usual uncertainties in single star evolution, as we mentioned above, the evolution path and outcome of a binary star system also depend on the initial mass ratio, orbital eccentricity, material, and angular momentum exchange. Binary channels might also be important for the formation of GRB progenitors \citep[e.g.,][]{Fryer1999,Fryer2005,Eldridge2011}. In a close binary system, the mass gainer can be spun up to break up the rotation velocity through mass and angular momentum transport \citep[e.g.,][]{Cantiello2007}. Recently, \cite{Hu2023} simulated binary evolution models with efficient angular momentum transport mechanisms and found that fast-spinning magnetars could form in these systems, powering the stripped-envelope SNe. Therefore, the binary progenitors for LGRBs do not require a fast initial rotation velocity to strip the stellar hydrogen envelope and retain sufficient angular momentum to produce a GRB jet \citep[e.g.,][]{Podsiadlowski2010,Chrimes2020}. Moreover, \cite{Laplace2021} studied the systematic differences in the core density and composition structure of solar metallicity nonrotating single and donor stars in binary systems with an initial mass range of $11-21 ~M_{\odot}$. They concluded that single stars systematically possessed more massive He cores than binary-striped stars with the same initial mass. \cite{Lloyd2022} proposed that radio-loud GRBs are produced by interacting binary systems, while radio-quiet GRBs originate from the collapse of single stars.

There is no firm conclusion regarding the upper and lower limits of NS mass born in nature \citep{Ozel2016}. It is difficult to form NSs with masses below 1.2 $M_{\odot}$ through iron-core collapse SN explosions. Some low-mass NSs candidates have been observed. \cite{Rawls2011} found NS mass of  $1.0\pm0.1~M_{\odot}$ for 4U 1538-52, and $1.04\pm0.09~M_{\odot}$ for SMC X-1. \cite{Ozel2012} reanalyzed these systems and estimate the mass are $1.18\pm0.25~M_{\odot}$ and $0.93\pm0.12~M_{\odot}$, respectively. Recently, \cite{Doroshenko2022} reported a strangely light NS with $0.77^{+0.20}_{-0.17}M_{\odot}$ in the central of SN remnant HESS J1731-347. Low-mass NSs may be born in an electron-capture SNe explosion with 8$-$10 $M_{\odot}$ progenitors \citep{Lattimer2012}. The well-measured massive pulsar masses, such as J1614+2230 (1.97$\pm$0.04 $M_{\odot}$), J0348+0432 (2.01$\pm$0.04 $M_{\odot}$), J2215+5135 (2.27$\pm$0.17 $M_{\odot}$), and J0740+6620 (2.14$\pm$0.1 $M_{\odot}$), indicate that the maximum NS mass can be at least 2 $M_{\odot}$ \citep{Demorest2010, Antoniadis2013,Linares2018,Cromartie2020}. The observation of a compact object in a binary merger GW190814 \citep{Abbott2020} could be an indication of the most massive NS mass reaching $2.6-2.9 ~M_{\odot}$ \citep{Godzieba2021}. The NS mass distributions in binary systems have been investigated by many authors \citep{Thorsett1999,Fryer2012}. \cite{Woosley2019}, \cite{Ertl2020}, and \cite{Woosley2020} calculated the evolution and explosion of a grid of nonrotating helium stars and studied the remnant mass distributions. They simplify binary star evolution and assume that the entire hydrogen envelope is promptly removed by binary interaction at the moment of helium core ignition. Their results show that the median NS masses in binary systems are in the range of $1.32-1.37 ~M_{\odot}$ and vary slightly with mass loss and metallicity \citep{Woosley2020}. They set the lightest and heaviest NS gravitational mass at 1.24 and 2.3 $M_{\odot}$, respectively. Previous observations have shown overwhelming evidence that the mass distribution of NSs is bimodal \citep[e.g.,][]{Ozel2012,Antoniadis2016,Alsing2018}, with two peaks at 1.3 $M_{\odot}$ and 1.5$-$1.7 $M_{\odot}$.

The distribution of the compactness parameter for nonrotating stars has been studied by some authors \citep{Oconnor2011,Ugliano2012,Sukhbold2014,Muller2016}. The approximate critical value between black hole and NS formation varies in different studies. \cite{Muller2016} proposed a semianalytic method to determine explodability and found that $\xi_{2.5}=0.278$ is the best value for discriminating successful or failed explosions. \cite{Aguilera2020} calculated the compactness of 42 rotating massive star models with initial equatorial rotation velocities of $600 ~\rm km ~s^{-1}$ and $Z=0.02~Z_{\odot}$. They show that semianalytic exploding models are compatible with the BH and NS thresholds $\eta_{2.5}=0.45$. They suggested that the compactness is a nonmonotonic function of the initial mass for rapidly rotating stars. In this paper, we explore the effects of the initial rotation rate, initial mass, metallicity and mass loss on compactness. The results show that the compactness parameter remains a nonmonotonic function of the initial mass and initial velocity when the effects of different metallicity and wind mass loss are considered. In the future, we will test the accuracy of the compactness parameter as a criterion for rotating massive star explosions by simulating the process of SN explosion. We will also investigate the effects of the initial mass, metallicity and rotational rate on the SN explosion energy, ejecta mass and remnant characteristics.

\software
{MESA \citep{Paxton2011,Paxton2013,Paxton2015,Paxton2018,Paxton2019,Jermyn2023}, MESA SDK 21.4.1 \citep{Townsend2021}, MatPlotLib \citep{Hunter2007}, NASA ADS, py$\_$mesa$\_$reader, NumPy \citep{Van2011}, and Pandas \citep{Reback2022}.}

\acknowledgments
We appreciate Prof. Alexander Heger and Prof. Bernhard M\"{u}ller for their constructive suggestions and comments, and thank Prof. Dong Lai, Dr. Tuan Yi, Dr. Zhenyu Zhu, and Shuai Zha for helpful discussion. The computations in this paper were run on the $\pi$2.0 cluster supported by the Center for High Performance Computing at Shanghai Jiao Tong University. This work was supported by the National Natural Science Foundation of China under grants 12103033, 12173031, and 12221003.

\appendix
The properties of the star models at the end of their evolution are shown in Table \ref{TABpreSN}.

\begin{longrotatetable}
\begin{deluxetable*}{ccccccccccccccccccccc}
\tablecaption{Properties of the models at the end of their evolution. \label{chartable}}
\tablewidth{700pt}
\movetabledown=20mm
\setlength\tabcolsep{0.3mm}%
\tabletypesize{\scriptsize}

\tablehead{
\multicolumn{6}{c}{Progenitor Parameters}  & \multicolumn{11}{c}{Stellar Properties at Core Collapse}
&
\multicolumn{4}{c}{Protomagnetar} \\
\cmidrule(r){1-6} \cmidrule(r){7-18} \cmidrule(r){19-21}
\colhead{\emph{$M$}} & \colhead{\emph{Z}} & \colhead{\emph{$\Omega$}} &
\colhead{\emph{$V_{\rm equ}$}} & \colhead{\emph{$\eta_{\rm wind}$}} &
\colhead{\emph{$J_{\rm ini}$}} & \colhead{\emph{M$_{\rm f}$}} &
\colhead{\emph{Ages}} &
\colhead{\emph{R$_{\rm f}$}} & \colhead{log \emph{T$_{\rm eff}$}} &
\colhead{log\emph{L}} & \colhead{\emph{J$_{\rm He~core}$}} &\colhead{\emph{M$_{\rm He~core}$}} & \colhead{\emph{J$_{\rm CO~core}$}} &\colhead{\emph{M$_{\rm CO~core}$}}  & \colhead{\emph{J$_{\rm Fe~core}$}} & \colhead{\emph{M$_{\rm Fe~core}$}} &
\colhead{\emph{$\xi_{2.5}$}} &
\colhead{\emph{M$_{\rm NS}$}} & \colhead{\emph{B$_{\rm NS}$}} & \colhead{\emph{P$_{\rm NS}$}}
\\
\colhead{(\emph{M$_{\odot}$)}}  & \colhead{(\emph{Z$_{\odot}$)}} & \colhead{($\Omega_{\rm crit}$)} & \colhead{(km s$^{-1}$)} & \colhead{} & \colhead{(10$^{52}$ergs s)} & \colhead{(\emph{M$_{\odot}$)}} & \colhead{(Myr)} & \colhead{(\emph{R$_{\odot}$)}} &
\colhead{(K)} & \colhead{(\emph{L$_{\odot}$)}} & \colhead{( 10$^{50}$ ergs s)} & \colhead{(\emph{M$_{\odot}$)}} & \colhead{(10$^{49}$ ergs s)} & \colhead{(\emph{M$_{\odot}$)}} & \colhead{(10$^{48}$ ergs s)} & \colhead{(\emph{M$_{\odot}$)}}  &
\colhead{} &
\colhead{(\emph{M$_{\odot}$)}} & \colhead{(10$^{14}$ G)} & \colhead{(ms)}  \\ }
\startdata
        10 & 0.01 & 0.1 & 89 & 0.5 & 0.31 & 9.97 & 28.31 & 558.77 & 3.59 & 4.81 & 0.61 & 3.51 & 1.15 & 1.98 & 4.19 & 1.47 & 0.01 & 1.33 & 1.69 & 2.00 \\
        10 & 0.01 & 0.3 & 265 & 1 & 0.91 & 9.66 & 39.34 & 25.73 & 4.35 & 5.17 & 1.90 & 6.20 & 6.91 & 4.12 & 7.03 & 1.65 & 0.25 & 1.48 & 2.84 & 1.32 \\
        10 & 0.01 & 0.3 & 265 & 0.5 & 0.91 & 9.97 & 38.49 & 29.79 & 4.28 & 5.04 & 1.41 & 5.38 & 4.54 & 3.43 & 5.15 & 1.47 & 0.11 & 1.33 & 2.04 & 1.62 \\
        10 & 0.01 & 0.4 & 348 & 5 & 1.17 & 7.37 & 41.50 & 6.58 & 4.66 & 5.24 & 1.73 & 6.56 & 7.51 & 4.58 & 5.57 & 1.64 & 0.27 & 1.47 & 0.79 & 1.66 \\
        10 & 0.01 & 0.5 & 428 & 0.5 & 1.40 & 5.87 & 43.95 & 0.92 & 5.12 & 5.36 & 3.17 & 5.87 & 3.62 & 3.44 & 4.96 & 1.61 & 0.23 & 1.44 & 1.50 & 1.83 \\
        10 & 0.01 & 0.6 & 502 & 1 & 1.58 & 4.28 & 45.02 & 0.95 & 5.04 & 5.05 & 0.76 & 4.28 & 2.48 & 2.87 & 4.05 & 1.50 & 0.10 & 1.35 & 2.46 & 2.10 \\
        10 & 0.01 & 0.6 & 502 & 5 & 1.58 & 6.33 & 44.51 & 2.05 & 4.91 & 5.20 & 1.31 & 6.11 & 5.81 & 4.30 & 4.07 & 1.54 & 0.19 & 1.39 & 0.96 & 2.15 \\
        10 & 0.01 & 0.6 & 502 & 0.5 & 1.58 & 5.98 & 44.93 & 0.68 & 5.15 & 5.24 & 3.66 & 5.98 & 8.64 & 4.18 & 5.63 & 1.51 & 0.20 & 1.36 & 1.68 & 1.52 \\
        10 & 0.01 & 0.7 & 571 & 0.5 & 1.71 & 5.93 & 45.77 & 0.68 & 5.15 & 5.23 & 3.61 & 5.93 & 8.43 & 4.13 & 5.18 & 1.59 & 0.21 & 1.42 & 2.07 & 1.73 \\
        10 & 0.01 & 0.8 & 593 & 0.5 & 1.56 & 5.82 & 44.95 & 0.70 & 5.16 & 5.27 & 3.13 & 5.82 & 6.20 & 4.00 & 4.42 & 1.59 & 0.24 & 1.43 & 1.67 & 2.03 \\
        10 & 0.1 & 0.1 & 82 & 0.5 & 0.31 & 9.73 & 28.48 & 657.59 & 3.55 & 4.81 & 0.27 & 3.47 & 0.82 & 2.02 & 2.99 & 1.47 & 0.02 & 1.33 & 0.80 & 2.80 \\
        10 & 0.1 & 0.6 & 461 & 1 & 1.58 & 7.35 & 44.70 & 1.73 & 5.03 & 5.55 & 1.87 & 7.20 & 8.51 & 5.21 & 4.52 & 1.63 & 0.26 & 1.46 & 1.53 & 2.03 \\
        10 & 0.1 & 0.6 & 461 & 0.5 & 1.58 & 5.85 & 45.62 & 0.78 & 5.11 & 5.17 & 2.87 & 5.85 & 5.55 & 4.12 & 5.73 & 1.53 & 0.21 & 1.37 & 2.95 & 1.51 \\
        11 & 0.01 & 0.4 & 354 & 1 & 1.40 & 5.21 & 36.60 & 0.67 & 5.21 & 5.44 & 1.08 & 5.21 & 4.96 & 3.76 & 4.40 & 1.62 & 0.20 & 1.45 & 1.35 & 2.08 \\
        11 & 0.01 & 0.6 & 510 & 1 & 1.90 & 4.50 & 38.43 & 0.81 & 5.09 & 5.11 & 0.81 & 4.50 & 2.93 & 3.07 & 4.18 & 1.54 & 0.18 & 1.39 & 2.20 & 2.09 \\
        11 & 0.01 & 0.6 & 510 & 0.5 & 1.90 & 6.35 & 38.36 & 0.68 & 5.25 & 5.61 & 3.74 & 6.35 & 8.90 & 4.48 & 5.77 & 1.64 & 0.26 & 1.46 & 2.00 & 1.60 \\
        11 & 0.1 & 0.6 & 469 & 1 & 1.90 & 7.91 & 38.02 & 1.51 & 5.01 & 5.34 & 2.33 & 7.81 & 11.00 & 5.70 & 4.80 & 1.55 & 0.16 & 1.39 & 1.72 & 1.83 \\
        11 & 0.1 & 0.6 & 469 & 0.5 & 1.90 & 6.28 & 38.65 & 0.81 & 5.11 & 5.21 & 2.57 & 6.28 & 8.06 & 4.42 & 4.97 & 1.53 & 0.13 & 1.37 & 1.80 & 1.74 \\
        11 & 0.5 & 0.4 & 272 & 1 & 1.48 & 9.08 & 33.67 & 757.09 & 3.59 & 5.08 & 1.16 & 5.31 & 4.08 & 3.43 & 5.28 & 1.52 & 0.15 & 1.37 & 1.79 & 1.64 \\
        11 & 0.5 & 0.5 & 334 & 1 & 1.76 & 6.69 & 37.09 & 1.11 & 5.08 & 5.35 & 1.39 & 6.69 & 6.15 & 4.70 & 2.98 & 1.55 & 0.19 & 1.39 & 0.38 & 2.95 \\
        11 & 0.5 & 0.6 & 392 & 1 & 1.99 & 6.52 & 37.90 & 1.11 & 5.08 & 5.36 & 1.04 & 6.52 & 4.70 & 4.60 & 2.68 & 1.51 & 0.17 & 1.36 & 1.41 & 3.20 \\
        12 & 0.01 & 0.6 & 518 & 1 & 2.24 & 4.86 & 33.26 & 0.62 & 5.17 & 5.21 & 0.94 & 4.86 & 3.87 & 3.44 & 3.82 & 1.59 & 0.24 & 1.43 & 1.42 & 2.35 \\
        12 & 0.01 & 0.6 & 518 & 5 & 2.24 & 7.24 & 32.79 & 1.25 & 5.10 & 5.57 & 2.09 & 7.23 & 9.38 & 5.18 & 4.69 & 1.58 & 0.18 & 1.42 & 1.84 & 1.91 \\
        12 & 0.1 & 0.1 & 85 & 0.5 & 0.45 & 11.72 & 20.79 & 688.92 & 3.57 & 4.92 & 0.74 & 4.10 & 1.96 & 2.43 & 3.43 & 1.43 & 0.04 & 1.30 & 1.04 & 2.38 \\
        12 & 0.1 & 0.6 & 476 & 1 & 2.26 & 8.19 & 33.00 & 0.80 & 5.15 & 5.37 & 2.56 & 8.19 & 13.30 & 6.07 & 5.00 & 1.53 & 0.24 & 1.37 & 4.31 & 1.73 \\
        12 & 0.1 & 0.6 & 476 & 0.5 & 2.26 & 8.07 & 33.12 & 0.70 & 5.18 & 5.35 & 4.22 & 8.07 & 18.10 & 5.95 & 3.38 & 1.51 & 0.16 & 1.36 & 1.05 & 2.53 \\
        12 & 0.5 & 0.6 & 398 & 1 & 2.38 & 6.71 & 32.89 & 0.97 & 5.09 & 5.28 & 1.41 & 6.71 & 7.08 & 4.81 & 4.46 & 1.53 & 0.18 & 1.38 & 2.09 & 1.95 \\
        13 & 0.01 & 0.6 & 525 & 1 & 2.61 & 5.02 & 29.34 & 0.87 & 5.17 & 5.52 & 0.96 & 5.02 & 4.23 & 3.58 & 4.39 & 1.57 & 0.19 & 1.41 & 1.45 & 2.03 \\
        13 & 0.01 & 0.6 & 525 & 5 & 2.61 & 7.83 & 28.70 & 1.06 & 5.11 & 5.44 & 2.39 & 7.83 & 11.70 & 5.74 & 5.11 & 1.64 & 0.26 & 1.46 & 1.53 & 1.81 \\
        13 & 0.01 & 0.6 & 525 & 0.5 & 2.61 & 7.16 & 29.20 & 0.67 & 5.18 & 5.30 & 4.03 & 7.16 & 13.50 & 5.18 & 5.69 & 1.47 & 0.17 & 1.33 & 3.62 & 1.47 \\
        13 & 0.1 & 0.6 & 483 & 1 & 2.64 & 8.68 & 29.06 & 0.91 & 5.13 & 5.38 & 2.74 & 8.68 & 14.90 & 6.56 & 4.86 & 1.51 & 0.13 & 1.36 & 3.61 & 1.76 \\
        13 & 0.1 & 0.6 & 483 & 0.5 & 2.64 & 7.41 & 29.33 & 0.70 & 5.25 & 5.65 & 3.39 & 7.41 & 13.80 & 5.46 & 5.73 & 1.50 & 0.14 & 1.36 & 1.97 & 1.49 \\
        14 & 0.01 & 0.3 & 280 & 1 & 1.71 & 13.91 & 21.79 & 67.45 & 4.16 & 5.25 & 2.34 & 7.25 & 9.31 & 5.01 & 6.95 & 1.76 & 0.48 & 1.56 & 2.10 & 1.41 \\
        14 & 0.01 & 0.6 & 531 & 1 & 3.00 & 5.32 & 26.14 & 0.61 & 5.18 & 5.25 & 1.05 & 5.32 & 5.00 & 3.86 & 4.03 & 1.59 & 0.21 & 1.42 & 1.80 & 2.23 \\
        14 & 0.01 & 0.6 & 531 & 5 & 3.00 & 7.91 & 25.81 & 0.93 & 5.11 & 5.34 & 2.42 & 7.91 & 12.10 & 5.88 & 4.62 & 1.50 & 0.14 & 1.35 & 2.29 & 1.84 \\
        14 & 0.1 & 0.6 & 490 & 1 & 3.05 & 9.28 & 25.90 & 1.03 & 5.12 & 5.45 & 3.59 & 9.28 & 18.70 & 7.02 & 4.08 & 1.48 & 0.11 & 1.34 & 1.14 & 2.06 \\
        14 & 0.1 & 0.6 & 490 & 0.5 & 3.05 & 7.87 & 26.12 & 0.61 & 5.21 & 5.36 & 3.60 & 7.87 & 15.40 & 5.77 & 6.43 & 1.61 & 0.23 & 1.44 & 2.50 & 1.41 \\
        14 & 0.5 & 0.6 & 412 & 1 & 3.22 & 7.58 & 25.32 & 0.91 & 5.11 & 5.31 & 1.87 & 7.58 & 10.30 & 5.64 & 4.21 & 1.50 & 0.15 & 1.36 & 1.24 & 2.03 \\
        15 & 0.01 & 0.2 & 190 & 1 & 1.33 & 14.95 & 17.20 & 570.08 & 3.68 & 5.18 & 1.82 & 6.06 & 6.65 & 4.07 & 6.31 & 1.64 & 0.26 & 1.46 & 2.41 & 1.46 \\
        15 & 0.01 & 0.2 & 190 & 5 & 1.33 & 14.84 & 16.68 & 109.82 & 3.98 & 4.97 & 1.30 & 5.53 & 4.45 & 3.59 & 5.57 & 1.57 & 0.14 & 1.41 & 1.75 & 1.59 \\
        15 & 0.01 & 0.4 & 372 & 1 & 2.52 & 5.68 & 22.57 & 0.59 & 5.18 & 5.22 & 1.13 & 5.68 & 5.26 & 4.18 & 4.47 & 1.63 & 0.22 & 1.46 & 1.41 & 2.05 \\
        15 & 0.01 & 0.4 & 372 & 5 & 2.52 & 8.61 & 22.33 & 0.99 & 5.11 & 5.37 & 2.90 & 8.61 & 16.00 & 6.55 & 4.96 & 1.49 & 0.11 & 1.34 & 1.82 & 1.71 \\
        15 & 0.01 & 0.4 & 372 & 0.5 & 2.52 & 8.09 & 22.48 & 0.58 & 5.22 & 5.37 & 4.23 & 8.09 & 17.20 & 6.00 & 5.70 & 1.51 & 0.20 & 1.36 & 2.39 & 1.51 \\
        15 & 0.01 & 0.5 & 458 & 1 & 3.02 & 5.55 & 23.07 & 0.60 & 5.17 & 5.18 & 1.10 & 5.55 & 4.50 & 4.05 & 4.38 & 1.54 & 0.19 & 1.38 & 1.52 & 1.99 \\
        15 & 0.01 & 0.6 & 537 & 1 & 3.42 & 5.50 & 23.56 & 0.64 & 5.15 & 5.18 & 1.08 & 5.50 & 4.97 & 4.02 & 3.76 & 1.58 & 0.19 & 1.41 & 0.81 & 2.37 \\
        15 & 0.01 & 0.6 & 537 & 5 & 3.42 & 8.18 & 23.34 & 0.79 & 5.16 & 5.37 & 2.52 & 8.18 & 13.50 & 6.19 & 4.55 & 1.56 & 0.19 & 1.40 & 1.49 & 1.94 \\
        15 & 0.01 & 0.6 & 537 & 0.5 & 3.42 & 7.79 & 23.48 & 0.58 & 5.22 & 5.37 & 4.00 & 7.79 & 15.50 & 5.75 & 5.01 & 1.55 & 0.20 & 1.39 & 2.21 & 1.75 \\
        15 & 0.01 & 0.7 & 612 & 1 & 3.72 & 5.34 & 23.94 & 0.64 & 5.16 & 5.19 & 1.01 & 5.34 & 4.84 & 3.85 & 4.19 & 1.58 & 0.18 & 1.42 & 0.99 & 2.13 \\
        15 & 0.01 & 0.7 & 612 & 0.5 & 3.72 & 7.75 & 23.88 & 0.54 & 5.24 & 5.38 & 3.98 & 7.75 & 16.10 & 5.69 & 5.99 & 1.58 & 0.23 & 1.41 & 2.40 & 1.49 \\
        15 & 0.01 & 0.8 & 658 & 1 & 3.69 & 5.44 & 23.91 & 0.58 & 5.19 & 5.25 & 1.07 & 5.44 & 4.78 & 3.98 & 4.48 & 1.63 & 0.24 & 1.46 & 1.41 & 2.05 \\
        15 & 0.01 & 0.8 & 658 & 5 & 3.69 & 8.15 & 23.68 & 0.83 & 5.14 & 5.36 & 2.55 & 8.15 & 13.40 & 6.14 & 5.24 & 1.57 & 0.19 & 1.41 & 2.03 & 1.69 \\
        15 & 0.01 & 0.8 & 658 & 0.5 & 3.69 & 7.95 & 23.83 & 0.60 & 5.21 & 5.36 & 4.52 & 7.95 & 16.90 & 5.80 & 6.64 & 1.61 & 0.22 & 1.44 & 1.96 & 1.37 \\
        15 & 0.1 & 0.1 & 88 & 1 & 0.69 & 12.70 & 14.51 & 719.67 & 3.59 & 5.04 & 1.19 & 4.89 & 3.93 & 3.09 & 4.13 & 1.53 & 0.15 & 1.37 & 2.70 & 2.09 \\
        15 & 0.1 & 0.1 & 88 & 0.5 & 0.69 & 13.52 & 15.55 & 748.97 & 3.61 & 5.13 & 1.46 & 5.66 & 5.27 & 3.77 & 4.59 & 1.53 & 0.19 & 1.38 & 1.56 & 1.89 \\
        15 & 0.1 & 0.2 & 175 & 1 & 1.36 & 14.63 & 16.44 & 697.36 & 3.61 & 5.10 & 0.98 & 5.39 & 3.38 & 3.49 & 4.51 & 1.57 & 0.16 & 1.41 & 1.83 & 1.97 \\
        15 & 0.1 & 0.2 & 175 & 0.5 & 1.36 & 14.49 & 16.71 & 750.76 & 3.62 & 5.17 & 1.81 & 5.94 & 6.18 & 3.99 & 5.48 & 1.59 & 0.22 & 1.43 & 1.73 & 1.64 \\
        15 & 0.1 & 0.3 & 261 & 0.5 & 1.99 & 14.76 & 19.35 & 670.84 & 3.66 & 5.24 & 1.55 & 6.94 & 5.85 & 4.74 & 3.44 & 1.29 & 0.38 & 1.18 & 0.04 & 2.16 \\
        15 & 0.1 & 0.4 & 343 & 1 & 2.57 & 10.19 & 22.24 & 0.70 & 5.22 & 5.51 & 3.34 & 10.19 & 19.40 & 7.99 & 5.70 & 1.79 & 0.47 & 1.59 & 1.95 & 1.75 \\
        15 & 0.1 & 0.4 & 343 & 0.5 & 2.57 & 12.01 & 22.24 & 1.15 & 5.16 & 5.71 & 8.89 & 12.01 & 42.40 & 9.31 & 7.74 & 1.69 & 0.35 & 1.50 & 2.88 & 1.22 \\
        15 & 0.1 & 0.5 & 422 & 0.5 & 3.08 & 7.78 & 23.11 & 0.61 & 5.21 & 5.35 & 3.02 & 7.78 & 14.00 & 5.73 & 6.03 & 1.55 & 0.21 & 1.40 & 3.10 & 1.46 \\
        15 & 0.1 & 0.6 & 495 & 1 & 3.48 & 9.78 & 23.33 & 0.70 & 5.21 & 5.47 & 3.94 & 9.78 & 21.70 & 7.55 & 6.49 & 1.64 & 0.26 & 1.47 & 1.50 & 1.42 \\
        15 & 0.1 & 0.6 & 496 & 0.5 & 3.48 & 7.88 & 23.55 & 0.67 & 5.19 & 5.35 & 3.71 & 7.88 & 14.70 & 5.74 & 5.56 & 1.58 & 0.20 & 1.42 & 2.08 & 1.61 \\
        15 & 0.1 & 0.7 & 565 & 1 & 3.79 & 9.21 & 23.69 & 0.81 & 5.16 & 5.39 & 3.12 & 9.21 & 18.30 & 7.15 & 5.24 & 1.55 & 0.15 & 1.40 & 2.22 & 1.68 \\
        15 & 0.1 & 0.8 & 622 & 1 & 3.90 & 9.43 & 23.79 & 0.79 & 5.17 & 5.45 & 3.37 & 9.43 & 19.40 & 7.30 & 6.66 & 1.56 & 0.23 & 1.40 & 3.37 & 1.32 \\
        15 & 0.1 & 0.8 & 623 & 0.5 & 3.90 & 7.49 & 24.06 & 0.66 & 5.18 & 5.33 & 3.32 & 7.49 & 13.50 & 5.52 & 5.07 & 1.57 & 0.15 & 1.41 & 3.91 & 1.76 \\
        15 & 0.5 & 0.1 & 74 & 1 & 0.73 & 12.09 & 14.79 & 901.88 & 3.55 & 5.07 & 0.97 & 5.09 & 3.47 & 3.32 & 3.93 & 1.51 & 0.18 & 1.36 & 0.79 & 2.17 \\
        15 & 0.5 & 0.2 & 148 & 1 & 1.44 & 12.72 & 16.57 & 938.26 & 3.56 & 5.16 & 1.62 & 5.86 & 5.84 & 3.93 & 4.52 & 1.51 & 0.21 & 1.36 & 2.09 & 1.90 \\
        15 & 0.5 & 0.3 & 219 & 1 & 2.12 & 12.11 & 18.44 & 877.22 & 3.59 & 5.19 & 0.95 & 6.17 & 3.62 & 4.17 & 3.81 & 1.65 & 0.27 & 1.47 & 0.78 & 2.43 \\
        15 & 0.5 & 0.4 & 291 & 1 & 2.70 & 9.55 & 21.29 & 3.49 & 4.86 & 5.47 & 2.96 & 9.30 & 14.70 & 7.02 & 5.37 & 1.56 & 0.27 & 1.40 & 1.06 & 1.64 \\
        15 & 0.5 & 0.6 & 421 & 1 & 3.66 & 7.84 & 22.85 & 0.77 & 5.16 & 5.35 & 1.63 & 7.84 & 9.37 & 5.91 & 4.09 & 1.56 & 0.26 & 1.40 & 1.00 & 2.15 \\
        15 & 0.5 & 0.7 & 481 & 1 & 3.99 & 7.74 & 23.16 & 0.82 & 5.14 & 5.34 & 1.70 & 7.74 & 9.88 & 5.84 & 3.98 & 1.60 & 0.23 & 1.44 & 2.46 & 2.27 \\
        15 & 0.5 & 0.8 & 536 & 1 & 4.18 & 7.72 & 23.39 & 0.79 & 5.15 & 5.34 & 1.63 & 7.72 & 9.50 & 5.82 & 4.32 & 1.58 & 0.23 & 1.41 & 1.03 & 2.06 \\
        16 & 0.01 & 0.6 & 543 & 1 & 3.86 & 5.80 & 21.45 & 0.58 & 5.19 & 5.24 & 1.19 & 5.80 & 5.10 & 4.23 & 4.32 & 1.62 & 0.23 & 1.45 & 1.01 & 2.12 \\
        16 & 0.01 & 0.6 & 543 & 5 & 3.87 & 8.76 & 21.26 & 0.97 & 5.11 & 5.38 & 2.95 & 8.76 & 16.60 & 6.72 & 4.59 & 1.44 & 0.10 & 1.31 & 4.83 & 1.80 \\
        16 & 0.1 & 0.6 & 501 & 1 & 3.94 & 10.22 & 21.17 & 0.68 & 5.22 & 5.50 & 3.84 & 10.22 & 22.50 & 8.01 & 6.34 & 1.83 & 0.60 & 1.62 & 2.53 & 1.61 \\
        16 & 0.1 & 0.6 & 501 & 0.5 & 3.94 & 8.37 & 21.42 & 0.64 & 5.21 & 5.40 & 4.14 & 8.37 & 18.40 & 6.13 & 5.31 & 1.54 & 0.17 & 1.38 & 2.73 & 1.64 \\
        16 & 0.5 & 0.6 & 422 & 1 & 4.20 & 8.01 & 20.83 & 0.76 & 5.17 & 5.38 & 2.11 & 8.01 & 12.90 & 6.13 & 4.40 & 1.52 & 0.16 & 1.37 & 1.50 & 1.96 \\
        17 & 0.01 & 0.6 & 549 & 1 & 4.32 & 5.86 & 19.73 & 0.65 & 5.19 & 5.32 & 1.16 & 5.86 & 5.44 & 4.28 & 4.35 & 1.56 & 0.16 & 1.40 & 1.44 & 2.03 \\
        17 & 0.1 & 0.6 & 507 & 1 & 4.42 & 10.34 & 19.44 & 0.78 & 5.18 & 5.45 & 3.78 & 10.34 & 22.70 & 8.03 & 6.38 & 1.63 & 0.24 & 1.46 & 2.36 & 1.44 \\
        17 & 0.1 & 0.6 & 507 & 0.5 & 4.42 & 8.80 & 19.61 & 0.73 & 5.17 & 5.37 & 4.12 & 8.80 & 17.80 & 6.53 & 5.27 & 1.54 & 0.14 & 1.39 & 4.83 & 1.66 \\
        18 & 0.01 & 0.6 & 554 & 1 & 4.81 & 6.03 & 18.22 & 0.56 & 5.20 & 5.26 & 1.18 & 6.03 & 5.90 & 4.45 & 4.17 & 1.61 & 0.23 & 1.44 & 1.11 & 2.18 \\
        18 & 0.01 & 0.6 & 554 & 5 & 4.81 & 8.48 & 18.06 & 0.92 & 5.11 & 5.31 & 2.56 & 8.48 & 14.70 & 6.43 & 5.53 & 1.46 & 0.09 & 1.32 & 1.78 & 1.50 \\
        18 & 0.01 & 0.6 & 554 & 0.5 & 4.81 & 9.10 & 18.14 & 0.61 & 5.22 & 5.41 & 4.92 & 9.10 & 21.90 & 6.71 & 7.87 & 1.66 & 0.27 & 1.48 & 3.13 & 1.19 \\
        18 & 0.1 & 0.6 & 512 & 0.5 & 4.93 & 9.46 & 18.14 & 0.78 & 5.17 & 5.40 & 4.83 & 9.46 & 22.00 & 7.08 & 5.67 & 1.45 & 0.10 & 1.31 & 3.25 & 1.46 \\
        18 & 0.5 & 0.6 & 431 & 1 & 5.28 & 8.59 & 17.56 & 0.80 & 5.21 & 5.61 & 2.44 & 8.59 & 14.80 & 6.56 & 4.23 & 1.51 & 0.12 & 1.36 & 2.48 & 2.03 \\
        19 & 0.01 & 0.6 & 559 & 1 & 5.32 & 6.45 & 16.93 & 0.59 & 5.19 & 5.25 & 1.38 & 6.45 & 7.04 & 4.77 & 4.43 & 1.48 & 0.16 & 1.34 & 2.16 & 1.90 \\
        19 & 0.01 & 0.6 & 559 & 5 & 5.32 & 7.46 & 16.84 & 0.50 & 5.22 & 5.22 & 1.98 & 7.46 & 19.40 & 7.43 & 3.69 & 1.50 & 0.11 & 1.35 & 1.26 & 2.31 \\
        19 & 0.01 & 0.6 & 559 & 0.5 & 5.32 & 9.59 & 16.87 & 0.59 & 5.24 & 5.46 & 5.31 & 9.59 & 23.90 & 7.13 & 3.97 & 1.50 & 0.18 & 1.35 & 1.32 & 2.14 \\
        19 & 0.1 & 0.6 & 516 & 1 & 5.46 & 10.21 & 16.71 & 0.64 & 5.23 & 5.49 & 3.64 & 10.21 & 19.30 & 7.55 & 5.70 & 1.69 & 0.35 & 1.50 & 1.86 & 1.66 \\
        19 & 0.1 & 0.6 & 516 & 0.5 & 5.46 & 9.34 & 16.87 & 0.75 & 5.20 & 5.51 & 4.33 & 9.34 & 21.00 & 7.03 & 5.50 & 1.46 & 0.10 & 1.32 & 2.51 & 1.51 \\
        19 & 0.5 & 0.6 & 436 & 1 & 5.86 & 8.74 & 16.36 & 0.79 & 5.15 & 5.36 & 2.55 & 8.74 & 16.00 & 6.75 & 3.73 & 1.47 & 0.11 & 1.33 & 1.40 & 2.24 \\
        20 & 0.01 & 0.1 & 100 & 1 & 1.14 & 19.63 & 10.95 & 657.71 & 3.70 & 5.41 & 3.70 & 8.71 & 18.00 & 6.43 & 6.71 & 1.64 & 0.28 & 1.46 & 1.92 & 1.37 \\
        20 & 0.01 & 0.1 & 100 & 5 & 1.14 & 17.76 & 11.31 & 644.03 & 3.71 & 5.42 & 4.09 & 9.01 & 19.60 & 6.70 & 7.24 & 1.61 & 0.24 & 1.44 & 2.70 & 1.25 \\
        20 & 0.01 & 0.2 & 199 & 1 & 2.26 & 19.88 & 12.02 & 429.99 & 3.78 & 5.33 & 3.54 & 8.86 & 15.90 & 6.44 & 7.41 & 1.84 & 0.59 & 1.62 & 2.49 & 1.38 \\
        20 & 0.01 & 0.2 & 199 & 5 & 2.26 & 18.93 & 12.93 & 516.81 & 3.79 & 5.53 & 3.53 & 10.39 & 21.40 & 7.65 & 8.07 & 1.85 & 0.58 & 1.63 & 2.50 & 1.27 \\
        20 & 0.01 & 0.3 & 296 & 1 & 3.32 & 9.63 & 14.74 & 0.64 & 5.21 & 5.41 & 3.11 & 9.63 & 16.40 & 7.20 & 4.96 & 1.52 & 0.12 & 1.37 & 1.83 & 1.74 \\
        20 & 0.01 & 0.3 & 296 & 5 & 3.32 & 9.13 & 14.54 & 0.55 & 5.23 & 5.35 & 2.97& 9.13 & 29.70 & 9.13 & 6.01 & 1.56 & 0.18 & 1.40 & 1.79 & 1.46 \\
        20 & 0.01 & 0.3 & 296 & 0.5 & 3.32 & 18.01 & 14.58 & 2.24 & 5.06 & 5.89 & 27.30 & 18.01 & 42.20 & 8.93 & 10.40 & 1.86 & 0.61 & 1.64 & 5.26 & 1.00 \\
        20 & 0.01 & 0.4 & 390 & 1 & 4.29 & 6.74 & 15.21 & 0.57 & 5.20 & 5.28 & 1.41 & 6.74 & 7.40 & 5.01 & 4.11 & 1.50 & 0.18 & 1.35 & 1.59 & 2.07 \\
        20 & 0.01 & 0.4 & 390 & 5 & 4.30 & 8.13 & 15.07 & 0.53 & 5.22 & 5.28 & 2.42 & 8.13 & 24.20 & 8.13 & 4.80 & 1.51 & 0.11 & 1.36 & 1.62 & 1.78 \\
        20 & 0.01 & 0.4 & 390 & 0.5 & 4.29 & 10.27 & 15.15 & 0.67 & 5.21 & 5.44 & 5.79 & 10.27 & 27.90 & 7.77 & 6.08 & 1.51 & 0.13 & 1.36 & 3.04 & 1.41 \\
        20 & 0.01 & 0.5 & 480 & 1 & 5.15 & 6.58 & 15.56 & 0.58 & 5.20 & 5.27 & 1.40 & 6.58 & 7.45 & 4.88 & 4.87 & 1.53 & 0.17 & 1.37 & 1.49 & 1.78 \\
        20 & 0.01 & 0.5 & 479 & 5 & 5.16 & 7.94 & 15.47 & 0.40 & 5.29 & 5.33 & 2.31 & 7.94 & 23.10 & 7.94 & 4.74 & 1.50 & 0.18 & 1.35 & 1.41 & 1.79 \\
        20 & 0.01 & 0.5 & 480 & 0.5 & 5.15 & 9.72 & 15.48 & 0.74 & 5.18 & 5.43 & 4.84 & 9.72 & 22.80 & 7.22 & 4.87 & 1.50 & 0.11 & 1.35 & 1.46 & 1.75 \\
        20 & 0.01 & 0.6 & 564 & 1 & 5.84 & 6.65 & 15.83 & 0.55 & 5.21 & 5.28 & 1.47 & 6.65 & 7.17 & 4.91 & 4.82 & 1.57 & 0.20 & 1.41 & 2.65 & 1.84 \\
        20 & 0.01 & 0.6 & 563 & 5 & 5.86 & 7.50 & 15.76 & 0.49 & 5.28 & 5.47 & 1.91 & 7.50 & 18.80 & 7.47 & 5.12 & 1.63 & 0.20 & 1.46 & 1.48 & 1.79 \\
        20 & 0.01 & 0.6 & 564 & 0.5 & 5.84 & 9.91 & 15.76 & 0.67 & 5.21 & 5.45 & 5.45 & 9.91 & 25.40 & 7.38 & 5.45 & 1.55 & 0.14 & 1.39 & 2.49 & 1.61 \\
        20 & 0.01 & 0.7 & 644 & 1 & 6.36 & 6.48 & 16.07 & 0.53 & 5.21 & 5.27 & 1.38 & 6.48 & 7.12 & 4.84 & 5.11 & 1.57 & 0.21 & 1.41 & 1.87 & 1.73 \\
        20 & 0.01 & 0.7 & 643 & 5 & 6.38 & 7.65 & 15.97 & 0.41 & 5.28 & 5.30 & 2.07 & 7.65 & 20.70 & 7.65 & 4.98 & 1.56 & 0.18 & 1.41 & 2.09 & 1.78 \\
        20 & 0.01 & 0.7 & 644 & 0.5 & 6.36 & 9.82 & 16.00 & 0.55 & 5.26 & 5.47 & 5.39 & 9.82 & 25.00 & 7.35 & 7.15 & 1.61 & 0.22 & 1.45 & 3.25 & 1.27 \\
        20 & 0.01 & 0.8 & 716 & 1 & 6.67 & 6.34 & 16.22 & 0.55 & 5.23 & 5.35 & 1.29 & 6.34 & 6.59 & 4.71 & 4.24 & 1.61 & 0.23 & 1.44 & 1.60 & 2.15 \\
        20 & 0.01 & 0.8 & 715 & 5 & 6.68 & 7.50 & 16.12 & 0.45 & 5.29 & 5.40 & 1.93 & 7.50 & 18.80 & 7.46 & 5.16 & 1.60 & 0.18 & 1.43 & 1.51 & 1.75 \\
        20 & 0.01 & 0.8 & 716 & 0.5 & 6.67 & 9.76 & 16.14 & 0.50 & 5.28 & 5.46 & 5.24 & 9.76 & 25.60 & 7.49 & 6.55 & 1.61 & 0.23 & 1.44 & 2.94 & 1.39 \\
        20 & 0.1 & 0.1 & 92 & 1 & 1.18 & 15.00 & 10.45 & 854.89 & 3.62 & 5.31 & 2.77 & 7.45 & 11.60 & 5.32 & 5.95 & 1.63 & 0.26 & 1.46 & 2.51 & 1.54 \\
        20 & 0.1 & 0.3 & 264 & 1 & 3.41 & 18.73 & 13.36 & 934.99 & 3.67 & 5.56 & 3.21 & 10.37 & 14.50 & 7.72 & 4.92 & 1.47 & 0.16 & 1.33 & 2.29 & 1.70 \\
        20 & 0.1 & 0.3 & 273 & 0.5 & 3.41 & 16.40 & 14.47 & 16.90 & 4.62 & 5.87 & 9.18 & 15.61 & 53.60 & 12.55 & 4.73 & 1.85 & 0.60 & 1.63 & 0.53 & 2.17 \\
        20 & 0.1 & 0.4 & 360 & 1 & 4.42 & 12.62 & 14.87 & 0.83 & 5.28 & 5.92 & 4.75 & 12.62 & 30.20 & 10.04 & 6.59 & 1.58 & 0.22 & 1.41 & 3.52 & 1.35 \\
        20 & 0.1 & 0.4 & 360 & 0.5 & 4.42 & 11.66 & 15.09 & 0.45 & 5.34 & 5.64 & 6.45 & 11.66 & 32.30 & 8.81 & 7.46 & 1.60 & 0.26 & 1.44 & 2.08 & 1.21 \\
        20 & 0.1 & 0.5 & 443 & 1 & 5.29 & 11.91 & 15.28 & 0.54 & 5.29 & 5.58 & 4.65 & 11.91 & 27.60 & 9.25 & 5.42 & 1.58 & 0.27 & 1.41 & 3.61 & 1.64 \\
        20 & 0.1 & 0.5 & 443 & 0.5 & 5.30 & 9.79 & 15.47 & 0.75 & 5.18 & 5.41 & 4.19 & 9.79 & 20.70 & 7.32 & 5.20 & 1.49 & 0.11 & 1.34 & 4.68 & 1.63 \\
        20 & 0.1 & 0.6 & 521 & 1 & 6.01 & 11.60 & 15.58 & 0.59 & 5.27 & 5.56 & 4.62 & 11.60 & 29.00 & 9.18 & 5.98 & 1.65 & 0.30 & 1.48 & 2.74 & 1.56 \\
        20 & 0.1 & 0.6 & 521 & 0.5 & 6.01 & 9.42 & 15.76 & 0.73 & 5.18 & 5.40 & 3.79 & 9.42 & 19.40 & 7.11 & 4.76 & 1.49 & 0.11 & 1.34 & 2.60 & 1.78 \\
        20 & 0.1 & 0.7 & 596 & 0.5 & 6.55 & 9.33 & 16.00 & 0.77 & 5.17 & 5.39 & 4.20 & 9.33 & 20.40 & 7.05 & 5.35 & 1.47 & 0.09 & 1.33 & 2.66 & 1.56 \\
        20 & 0.1 & 0.8 & 664 & 1 & 6.89 & 11.76 & 15.96 & 0.51 & 5.32 & 5.65 & 5.41 & 11.76 & 33.50 & 9.52 & 6.65 & 1.76 & 0.44 & 1.56 & 4.44 & 1.48 \\
        20 & 0.1 & 0.8 & 665 & 0.5 & 6.88 & 9.09 & 16.14 & 0.50 & 5.27 & 5.44 & 4.28 & 9.09 & 20.40 & 6.86 & 6.32 & 1.68 & 0.30 & 1.49 & 2.50 & 1.49 \\
        20 & 0.5 & 0.4 & 305 & 1 & 4.72 & 9.60 & 14.50 & 1.01 & 5.11 & 5.41 & 2.77 & 9.60 & 16.10 & 7.25 & 4.14 & 1.53 & 0.11 & 1.37 & 0.82 & 2.09 \\
        20 & 0.5 & 0.6 & 440 & 1 & 6.47 & 9.02 & 15.28 & 0.87 & 5.15 & 5.41 & 2.68 & 9.02 & 15.30 & 6.69 & 4.97 & 1.46 & 0.10 & 1.32 & 2.64 & 1.68 \\
        21 & 0.01 & 0.6 & 568 & 1 & 6.40 & 6.56 & 14.90 & 0.54 & 5.22 & 5.28 & 1.35 & 6.56 & 7.33 & 4.89 & 4.21 & 1.54 & 0.21 & 1.39 & 1.29 & 2.07 \\
        21 & 0.01 & 0.6 & 568 & 5 & 6.39 & 8.08 & 14.84 & 0.55 & 5.21 & 5.29 & 2.32 & 8.08 & 23.20 & 8.08 & 5.64 & 1.49 & 0.12 & 1.35 & 2.33 & 1.51 \\
        21 & 0.01 & 0.6 & 568 & 0.5 & 6.40 & 10.31 & 14.82 & 0.76 & 5.18 & 5.45 & 5.73 & 10.31 & 26.30 & 7.68 & 5.77 & 1.44 & 0.10 & 1.31 & 3.69 & 1.43 \\
        21 & 0.1 & 0.6 & 525 & 1 & 6.58 & 11.82 & 14.67 & 0.59 & 5.27 & 5.57 & 4.68 & 11.82 & 28.70 & 9.30 & 6.63 & 1.55 & 0.23 & 1.39 & 3.54 & 1.32 \\
        21 & 0.1 & 0.6 & 525 & 0.5 & 6.58 & 9.70 & 14.81 & 0.71 & 5.19 & 5.42 & 3.83 & 9.70 & 19.90 & 7.29 & 5.15 & 1.53 & 0.13 & 1.38 & 3.02 & 1.68 \\
        21 & 0.5 & 0.6 & 443 & 1 & 7.11 & 9.02 & 14.37 & 0.99 & 5.15 & 5.53 & 2.69 & 9.02 & 15.50 & 6.77 & 5.30 & 1.60 & 0.16 & 1.43 & 1.66 & 1.70 \\
        22 & 0.01 & 0.6 & 572 & 1 & 6.97 & 6.93 & 14.05 & 0.59 & 5.20 & 5.29 & 1.49 & 6.93 & 7.83 & 5.15 & 4.73 & 1.56 & 0.17 & 1.40 & 1.37 & 1.87 \\
        22 & 0.01 & 0.6 & 572 & 5 & 6.98 & 7.97 & 13.98 & 0.46 & 5.27 & 5.35 & 1.95 & 7.97 & 19.00 & 7.92 & 4.88 & 1.51 & 0.21 & 1.36 & 1.65 & 1.75 \\
        22 & 0.01 & 0.6 & 572 & 0.5 & 6.97 & 10.43 & 13.99 & 0.66 & 5.22 & 5.48 & 5.31 & 10.43 & 25.50 & 7.72 & 7.62 & 1.66 & 0.19 & 1.48 & 3.02 & 1.22 \\
        22 & 0.1 & 0.6 & 530 & 0.5 & 7.18 & 9.88 & 13.99 & 0.67 & 5.21 & 5.43 & 3.84 & 9.88 & 19.60 & 7.41 & 5.96 & 1.51 & 0.16 & 1.36 & 2.60 & 1.44 \\
        22 & 0.5 & 0.6 & 448 & 1 & 7.75 & 9.08 & 13.57 & 1.08 & 5.09 & 5.36 & 2.74 & 9.08 & 16.20 & 6.88 & 5.20 & 1.42 & 0.08 & 1.29 & 1.64 & 1.56 \\
        23 & 0.01 & 0.6 & 577 & 1 & 7.56 & 6.98 & 13.31 & 0.52 & 5.23 & 5.32 & 1.48 & 6.98 & 8.08 & 5.21 & 4.69 & 1.57 & 0.21 & 1.41 & 2.76 & 1.89 \\
        23 & 0.01 & 0.6 & 576 & 5 & 7.58 & 6.35 & 13.32 & 0.48 & 5.23 & 5.23 & 1.04 & 6.35 & 6.06 & 4.80 & 3.50 & 1.50 & 0.16 & 1.35 & 2.26 & 2.43 \\
        23 & 0.01 & 0.6 & 577 & 0.5 & 7.56 & 11.02 & 13.24 & 0.57 & 5.26 & 5.53 & 5.92 & 11.02 & 28.30 & 8.19 & 8.15 & 1.82 & 0.50 & 1.60 & 2.28 & 1.24 \\
        23 & 0.1 & 0.6 & 533 & 1 & 7.80 & 12.45 & 13.13 & 0.54 & 5.30 & 5.61 & 5.27 & 12.45 & 30.40 & 9.57 & 6.74 & 1.60 & 0.27 & 1.43 & 4.88 & 1.34 \\
        23 & 0.1 & 0.6 & 533 & 0.5 & 7.80 & 10.03 & 13.25 & 0.63 & 5.22 & 5.44 & 4.12 & 10.03 & 21.50 & 7.59 & 6.08 & 1.57 & 0.20 & 1.41 & 2.23 & 1.46 \\
        23 & 0.5 & 0.6 & 451 & 1 & 8.44 & 9.16 & 12.86 & 0.74 & 5.18 & 5.43 & 2.83 & 9.16 & 16.60 & 6.95 & 4.11 & 1.56 & 0.19 & 1.41 & 3.37 & 2.15 \\
        24 & 0.01 & 0.6 & 580 & 5 & 8.18 & 6.45 & 12.65 & 0.44 & 5.25 & 5.25 & 1.03 & 6.45 & 6.00 & 4.89 & 3.83 & 1.59 & 0.20 & 1.42 & 1.86 & 2.34 \\
        24 & 0.01 & 0.6 & 581 & 0.5 & 8.16 & 11.41 & 12.57 & 0.50 & 5.30 & 5.56 & 6.18 & 11.41 & 29.40 & 8.48 & 7.58 & 1.70 & 0.36 & 1.51 & 2.86 & 1.26 \\
        24 & 0.1 & 0.6 & 537 & 1 & 8.44 & 12.59 & 12.48 & 0.75 & 5.23 & 5.61 & 5.78 & 12.59 & 31.20 & 9.49 & 4.87 & 1.50 & 0.14 & 1.36 & 1.66 & 1.75 \\
        24 & 0.1 & 0.6 & 537 & 0.5 & 8.44 & 10.92 & 12.57 & 0.59 & 5.24 & 5.48 & 4.82 & 10.92 & 24.80 & 8.23 & 7.28 & 1.65 & 0.29 & 1.47 & 6.77 & 1.27 \\
        24 & 0.5 & 0.6 & 453 & 1 & 9.17 & 9.17 & 12.24 & 1.00 & 5.11 & 5.39 & 2.93 & 9.17 & 16.70 & 6.96 & 5.36 & 1.52 & 0.09 & 1.37 & 2.28 & 1.61 \\
        25 & 0.01 & 0.1 & 103 & 0.5 & 1.70 & 24.93 & 8.65 & 110.83 & 4.11 & 5.48 & 3.71 & 9.73 & 17.40 & 7.08 & 7.03 & 1.92 & 0.66 & 1.69 & 2.26 & 1.51 \\
        25 & 0.01 & 0.2 & 205 & 1 & 3.38 & 24.72 & 9.92 & 109.61 & 4.15 & 5.64 & 7.79 & 13.45 & 43.00 & 10.49 & 10.00 & 1.80 & 0.43 & 1.60 & 5.70 & 1.00 \\
        25 & 0.01 & 0.3 & 306 & 1 & 4.96 & 9.38 & 11.32 & 0.64 & 5.21 & 5.39 & 2.46 & 9.38 & 13.70 & 7.06 & 5.33 & 1.51 & 0.12 & 1.36 & 2.27 & 1.60 \\
        25 & 0.01 & 0.4 & 404 & 1 & 6.43 & 7.63 & 11.62 & 0.51 & 5.25 & 5.35 & 1.66 & 7.63 & 9.37 & 5.72 & 4.09 & 1.57 & 0.21 & 1.41 & 1.21 & 2.17 \\
        25 & 0.01 & 0.4 & 404 & 0.5 & 6.43 & 11.74 & 11.56 & 0.48 & 5.32 & 5.58 & 5.81 & 11.74 & 29.00 & 8.80 & 6.72 & 1.73 & 0.38 & 1.54 & 2.01 & 1.44 \\
        25 & 0.01 & 0.5 & 497 & 1 & 7.72 & 7.67 & 11.85 & 0.53 & 5.24 & 5.36 & 1.73 & 7.67 & 9.59 & 5.75 & 4.39 & 1.52 & 0.20 & 1.37 & 2.12 & 1.97 \\
        25 & 0.01 & 0.5 & 497 & 0.5 & 7.72 & 11.83 & 11.79 & 0.50 & 5.31 & 5.61 & 6.28 & 11.83 & 29.80 & 8.75 & 7.02 & 1.68 & 0.33 & 1.50 & 2.09 & 1.34 \\
        25 & 0.01 & 0.6 & 585 & 1 & 8.79 & 7.46 & 12.07 & 0.49 & 5.26 & 5.39 & 1.65 & 7.46 & 9.09 & 5.58 & 4.80 & 1.55 & 0.23 & 1.40 & 1.27 & 1.83 \\
        25 & 0.01 & 0.6 & 584 & 5 & 8.82 & 6.56 & 12.07 & 0.42 & 5.27 & 5.27 & 1.15 & 6.56 & 6.64 & 4.96 & 4.40 & 1.63 & 0.23 & 1.45 & 1.21 & 2.08 \\
        25 & 0.01 & 0.6 & 585 & 0.5 & 8.79 & 11.45 & 11.99 & 0.51 & 5.30 & 5.57 & 5.52 & 11.45 & 27.10 & 8.53 & 7.09 & 1.76 & 0.45 & 1.56 & 2.41 & 1.39 \\
        25 & 0.01 & 0.7 & 669 & 1 & 9.60 & 7.42 & 12.24 & 0.50 & 5.25 & 5.35 & 1.64 & 7.42 & 9.19 & 5.57 & 5.24 & 1.58 & 0.23 & 1.42 & 2.19 & 1.70 \\
        25 & 0.01 & 0.7 & 669 & 0.5 & 9.60 & 11.86 & 12.17 & 0.49 & 5.31 & 5.59 & 6.91 & 11.86 & 33.50 & 8.86 & 7.37 & 1.66 & 0.28 & 1.48 & 3.03 & 1.27 \\
        25 & 0.01 & 0.8 & 747 & 1 & 10.10 & 7.52 & 12.34 & 0.51 & 5.24 & 5.34 & 1.70 & 7.52 & 9.40 & 5.63 & 4.50 & 1.58 & 0.21 & 1.41 & 1.45 & 1.98 \\
        25 & 0.01 & 0.8 & 747 & 0.5 & 10.10 & 11.36 & 12.29 & 0.60 & 5.26 & 5.54 & 5.70 & 11.36 & 27.00 & 8.37 & 7.27 & 1.79 & 0.54 & 1.58 & 2.23 & 1.37 \\
        25 & 0.1 & 0.1 & 95 & 1 & 1.76 & 13.86 & 8.96 & 1271.00 & 3.61 & 5.62 & 6.40 & 11.93 & 34.00 & 9.29 & 8.47 & 1.72 & 0.36 & 1.53 & 4.09 & 1.14 \\
        25 & 0.1 & 0.2 & 190 & 1 & 3.49 & 23.86 & 9.46 & 955.17 & 3.70 & 5.71 & 4.36 & 11.00 & 21.50 & 8.28 & 5.63 & 1.82 & 0.57 & 1.61 & 0.46 & 1.80 \\
        25 & 0.1 & 0.2 & 190 & 0.5 & 3.49 & 24.36 & 9.58 & 1085.14 & 3.65 & 5.62 & 5.08 & 12.42 & 26.90 & 9.68 & 6.87 & 1.79 & 0.45 & 1.58 & 2.48 & 1.45 \\
        25 & 0.1 & 0.3 & 283 & 1 & 5.13 & 19.98 & 10.77 & 1501.66 & 3.62 & 5.81 & 7.04 & 16.04 & 32.40 & 12.02 & 6.66 & 1.76 & 0.39 & 1.56 & 3.08 & 1.47 \\
        25 & 0.1 & 0.4 & 372 & 1 & 6.65 & 12.87 & 11.39 & 0.55 & 5.30 & 5.63 & 5.54 & 12.87 & 53.70 & 12.80 & 4.58 & 1.53 & 0.16 & 1.38 & 3.13 & 1.90 \\
        25 & 0.1 & 0.4 & 372 & 0.5 & 6.65 & 13.20 & 11.54 & 0.63 & 5.27 & 5.65 & 7.34 & 13.20 & 39.10 & 10.19 & 5.43 & 1.53 & 0.14 & 1.38 & 2.11 & 1.60 \\
        25 & 0.1 & 0.5 & 458 & 1 & 7.99 & 12.69 & 11.68 & 0.71 & 5.24 & 5.61 & 5.75 & 12.69 & 31.70 & 9.63 & 5.20 & 1.53 & 0.13 & 1.38 & 2.75 & 1.67 \\
        25 & 0.1 & 0.6 & 540 & 1 & 9.10 & 12.51 & 11.91 & 0.71 & 5.23 & 5.60 & 5.81 & 12.51 & 31.60 & 9.47 & 6.83 & 1.57 & 0.13 & 1.41 & 6.02 & 1.30 \\
        25 & 0.1 & 0.7 & 620 & 1 & 9.95 & 11.94 & 12.06 & 0.65 & 5.28 & 5.70 & 5.16 & 11.94 & 27.20 & 9.04 & 7.14 & 1.63 & 0.20 & 1.46 & 4.28 & 1.29 \\
        25 & 0.1 & 0.7 & 619 & 0.5 & 9.95 & 11.21 & 12.15 & 0.51 & 5.32 & 5.66 & 5.71 & 11.21 & 27.20 & 8.30 & 6.95 & 1.77 & 0.50 & 1.57 & 2.08 & 1.42 \\
        25 & 0.1 & 0.8 & 693 & 0.5 & 10.50 & 10.79 & 12.26 & 0.65 & 5.23 & 5.50 & 5.36 & 10.79 & 24.70 & 7.88 & 7.60 & 1.81 & 0.57 & 1.60 & 2.56 & 1.33 \\
        25 & 0.5 & 0.4 & 315 & 1 & 7.20 & 9.32 & 11.16 & 0.63 & 5.20 & 5.37 & 3.11 & 9.32 & 16.70 & 7.04 & 5.02 & 1.53 & 0.15 & 1.38 & 2.69 & 1.73 \\
        25 & 0.5 & 0.5 & 389 & 1 & 8.62 & 8.54 & 11.44 & 0.50 & 5.26 & 5.39 & 2.57 & 8.54 & 25.10 & 8.51 & 5.77 & 1.72 & 0.35 & 1.53 & 1.35 & 1.67 \\
        25 & 0.5 & 0.6 & 458 & 1 & 9.87 & 9.11 & 11.70 & 0.91 & 5.13 & 5.38 & 2.90 & 9.11 & 16.20 & 6.90 & 5.03 & 1.48 & 0.08 & 1.34 & 2.05 & 1.67 \\
        25 & 0.5 & 0.6 & 458 & 1 & 9.88 & 9.04 & 11.65 & 0.72 & 5.17 & 5.34 & 2.98 & 9.04 & 15.70 & 6.83 & 4.49 & 1.52 & 0.09 & 1.37 & 3.77 & 1.92 \\
        26 & 0.01 & 0.6 & 588 & 1 & 9.44 & 7.61 & 11.54 & 0.55 & 5.23 & 5.33 & 1.66 & 7.61 & 9.36 & 5.72 & 3.73 & 1.56 & 0.18 & 1.40 & 1.56 & 2.37 \\
        26 & 0.01 & 0.6 & 588 & 5 & 9.44 & 5.79 & 11.62 & 0.43 & 5.25 & 5.21 & 0.82 & 5.79 & 4.82 & 4.38 & 3.83 & 1.62 & 0.23 & 1.45 & 1.17 & 2.38 \\
        26 & 0.01 & 0.6 & 588 & 0.5 & 9.44 & 11.66 & 11.48 & 0.50 & 5.30 & 5.57 & 5.46 & 11.66 & 26.70 & 8.66 & 6.68 & 1.68 & 0.34 & 1.50 & 1.90 & 1.41 \\
        26 & 0.1 & 0.6 & 544 & 1 & 9.77 & 12.34 & 11.37 & 0.60 & 5.27 & 5.58 & 5.47 & 12.34 & 52.90 & 12.27 & 5.83 & 1.48 & 0.13 & 1.34 & 3.71 & 1.45 \\
        26 & 0.5 & 0.6 & 460 & 1 & 10.60 & 8.79 & 11.22 & 0.71 & 5.17 & 5.32 & 2.81 & 8.79 & 15.00 & 6.67 & 4.88 & 1.45 & 0.09 & 1.32 & 2.92 & 1.70 \\
        27 & 0.01 & 0.6 & 592 & 1 & 10.10 & 7.82 & 11.07 & 0.52 & 5.25 & 5.37 & 1.73 & 7.82 & 9.71 & 5.87 & 4.08 & 1.55 & 0.19 & 1.39 & 1.25 & 2.15 \\
        27 & 0.1 & 0.6 & 547 & 1 & 10.50 & 12.41 & 10.91 & 0.63 & 5.25 & 5.57 & 5.20 & 12.41 & 51.00 & 12.38 & 5.00 & 1.47 & 0.14 & 1.33 & 2.84 & 1.68 \\
        27 & 0.1 & 0.6 & 548 & 0.5 & 10.50 & 11.82 & 10.97 & 0.52 & 5.30 & 5.58 & 5.27 & 11.82 & 27.10 & 8.88 & 6.37 & 1.58 & 0.23 & 1.42 & 2.53 & 1.40 \\
        27 & 0.5 & 0.6 & 463 & 1 & 11.40 & 8.89 & 10.76 & 0.68 & 5.18 & 5.33 & 2.87 & 8.89 & 15.50 & 6.73 & 5.17 & 1.51 & 0.10 & 1.36 & 3.87 & 1.65 \\
        28 & 0.01 & 0.6 & 595 & 1 & 10.80 & 8.22 & 10.64 & 0.53 & 5.24 & 5.38 & 1.93 & 8.22 & 10.90 & 6.18 & 4.48 & 1.55 & 0.19 & 1.40 & 1.96 & 1.96 \\
        28 & 0.01 & 0.6 & 594 & 5 & 10.80 & 5.64 & 10.73 & 0.52 & 5.20 & 5.16 & 0.77 & 5.64 & 4.49 & 4.27 & 3.62 & 1.54 & 0.14 & 1.38 & 2.10 & 2.41 \\
        28 & 0.01 & 0.6 & 595 & 0.5 & 10.80 & 12.95 & 10.57 & 0.53 & 5.31 & 5.64 & 7.26 & 12.95 & 69.80 & 12.88 & 4.92 & 1.55 & 0.22 & 1.39 & 3.88 & 1.78 \\
        28 & 0.1 & 0.6 & 550 & 1 & 11.20 & 12.28 & 10.48 & 0.59 & 5.27 & 5.58 & 5.29 & 12.28 & 51.70 & 12.24 & 4.20 & 1.51 & 0.15 & 1.36 & 1.62 & 2.04 \\
        28 & 0.1 & 0.6 & 550 & 0.5 & 11.20 & 11.69 & 10.55 & 0.50 & 5.30 & 5.57 & 4.96 & 11.69 & 26.40 & 8.86 & 6.86 & 1.57 & 0.23 & 1.41 & 4.52 & 1.29 \\
        28 & 0.5 & 0.6 & 466 & 1 & 12.20 & 8.76 & 10.36 & 0.64 & 5.19 & 5.34 & 2.73 & 8.76 & 25.70 & 8.66 & 4.45 & 1.46 & 0.10 & 1.32 & 2.49 & 1.86 \\
        29 & 0.01 & 0.6 & 599 & 1 & 11.50 & 8.33 & 10.24 & 0.52 & 5.25 & 5.38 & 1.93 & 8.33 & 10.90 & 6.27 & 4.38 & 1.53 & 0.19 & 1.37 & 1.09 & 1.97 \\
        29 & 0.1 & 0.3 & 288 & 1 & 6.69 & 21.15 & 9.32 & 61.34 & 4.37 & 6.02 & 12.20 & 19.47 & 65.20 & 15.32 & 8.99 & 1.82 & 0.49 & 1.60 & 4.70 & 1.12 \\
        29 & 0.1 & 0.6 & 554 & 1 & 11.90 & 12.27 & 10.13 & 0.53 & 5.35 & 5.78 & 5.35 & 12.27 & 52.50 & 12.23 & 6.45 & 1.61 & 0.24 & 1.44 & 3.95 & 1.40 \\
        29 & 0.1 & 0.6 & 554 & 0.5 & 11.90 & 11.97 & 10.17 & 0.48 & 5.32 & 5.60 & 4.93 & 11.97 & 27.20 & 9.19 & 5.74 & 1.57 & 0.24 & 1.41 & 2.32 & 1.55 \\
        29 & 0.5 & 0.6 & 469 & 1 & 13.00 & 9.02 & 9.96 & 0.41 & 5.31 & 5.43 & 2.79 & 9.02 & 26.80 & 8.96 & 5.68 & 1.57 & 0.25 & 1.41 & 2.06 & 1.56 \\
        30 & 0.01 & 0.1 & 106 & 1 & 2.35 & 29.79 & 7.16 & 63.99 & 4.26 & 5.62 & 3.74 & 11.07 & 19.40 & 8.28 & 6.05 & 1.80 & 0.54 & 1.59 & 1.24 & 1.66 \\
        30 & 0.01 & 0.2 & 211 & 0.5 & 4.66 & 29.70 & 8.61 & 117.05 & 4.23 & 6.03 & 11.70 & 18.32 & 59.80 & 14.10 & 9.06 & 1.82 & 0.56 & 1.61 & 3.79 & 1.12 \\
        30 & 0.01 & 0.3 & 314 & 1 & 6.86 & 13.59 & 9.27 & 0.67 & 5.26 & 5.64 & 5.15 & 13.59 & 50.20 & 13.53 & 5.54 & 1.51 & 0.14 & 1.36 & 3.93 & 1.55 \\
        30 & 0.01 & 0.3 & 314 & 0.5 & 6.86 & 14.20 & 9.33 & 0.62 & 5.29 & 5.69 & 7.23 & 14.20 & 38.40 & 10.86 & 5.94 & 1.51 & 0.14 & 1.36 & 4.28 & 1.44 \\
        30 & 0.01 & 0.4 & 415 & 1 & 8.90 & 8.94 & 9.53 & 0.48 & 5.28 & 5.44 & 2.15 & 8.94 & 12.20 & 6.77 & 4.85 & 1.64 & 0.24 & 1.46 & 1.43 & 1.90 \\
        30 & 0.01 & 0.4 & 415 & 5 & 8.90 & 6.12 & 9.57 & 0.42 & 5.26 & 5.24 & 0.92 & 6.12 & 5.35 & 4.63 & 3.94 & 1.60 & 0.22 & 1.43 & 1.52 & 2.29 \\
        30 & 0.01 & 0.4 & 415 & 0.5 & 8.90 & 13.92 & 9.48 & 0.70 & 5.25 & 5.66 & 7.46 & 13.92 & 71.80 & 13.85 & 5.56 & 1.59 & 0.14 & 1.43 & 2.12 & 1.62 \\
        30 & 0.01 & 0.5 & 511 & 1 & 10.70 & 8.57 & 9.72 & 0.63 & 5.20 & 5.36 & 1.92 & 8.57 & 10.70 & 6.42 & 4.11 & 1.48 & 0.14 & 1.33 & 1.56 & 2.04 \\
        30 & 0.01 & 0.5 & 511 & 0.5 & 10.70 & 13.76 & 9.65 & 0.64 & 5.28 & 5.67 & 7.72 & 13.76 & 38.50 & 10.40 & 6.12 & 1.56 & 0.15 & 1.40 & 3.97 & 1.44 \\
        30 & 0.01 & 0.6 & 602 & 1 & 12.20 & 8.50 & 9.88 & 0.60 & 5.21 & 5.36 & 2.02 & 8.50 & 11.40 & 6.38 & 5.17 & 1.53 & 0.17 & 1.38 & 1.31 & 1.68 \\
        30 & 0.01 & 0.6 & 602 & 5 & 12.20 & 5.06 & 10.01 & 0.49 & 5.20 & 5.14 & 0.63 & 5.06 & 3.67 & 3.81 & 3.60 & 1.48 & 0.16 & 1.34 & 1.35 & 2.34 \\
        30 & 0.01 & 0.6 & 602 & 0.5 & 12.20 & 13.72 & 9.82 & 0.69 & 5.25 & 5.65 & 7.76 & 13.72 & 74.40 & 13.64 & 6.28 & 1.53 & 0.14 & 1.38 & 5.70 & 1.38 \\
        30 & 0.01 & 0.7 & 690 & 1 & 13.40 & 8.59 & 10.02 & 0.63 & 5.20 & 5.36 & 2.07 & 8.59 & 11.80 & 6.48 & 5.19 & 1.55 & 0.15 & 1.39 & 1.59 & 1.69 \\
        30 & 0.01 & 0.7 & 690 & 0.5 & 13.40 & 13.55 & 9.96 & 0.70 & 5.25 & 5.64 & 7.52 & 13.55 & 71.70 & 13.47 & 4.88 & 1.49 & 0.12 & 1.35 & 5.03 & 1.74 \\
        30 & 0.01 & 0.8 & 773 & 1 & 14.10 & 8.37 & 10.11 & 0.61 & 5.21 & 5.35 & 1.97 & 8.37 & 11.00 & 6.29 & 3.99 & 1.50 & 0.15 & 1.35 & 6.30 & 2.13 \\
        30 & 0.01 & 0.8 & 773 & 0.5 & 14.10 & 13.55 & 10.05 & 0.70 & 5.25 & 5.64 & 7.50 & 13.55 & 72.30 & 13.48 & 5.49 & 1.49 & 0.13 & 1.35 & 2.56 & 1.55 \\
        30 & 0.1 & 0.1 & 98 & 1 & 2.44 & 27.05 & 7.46 & 1310.05 & 3.63 & 5.72 & 8.38 & 14.26 & 46.60 & 11.34 & 7.26 & 1.77 & 0.52 & 1.57 & 3.91 & 1.36 \\
        30 & 0.1 & 0.2 & 194 & 0.5 & 4.82 & 29.15 & 7.99 & 1028.65 & 3.67 & 5.67 & 5.93 & 13.64 & 32.70 & 10.56 & 5.85 & 1.82 & 0.57 & 1.61 & 0.59 & 1.73 \\
        30 & 0.1 & 0.5 & 471 & 1 & 11.10 & 12.67 & 9.60 & 0.62 & 5.26 & 5.58 & 5.22 & 12.67 & 51.70 & 12.65 & 5.45 & 1.52 & 0.22 & 1.37 & 4.03 & 1.58 \\
        30 & 0.1 & 0.5 & 471 & 0.5 & 11.10 & 12.83 & 9.65 & 0.62 & 5.27 & 5.61 & 5.44 & 12.83 & 30.80 & 9.93 & 5.34 & 1.53 & 0.14 & 1.38 & 3.92 & 1.63 \\
        30 & 0.1 & 0.6 & 556 & 1 & 12.70 & 12.91 & 9.76 & 0.63 & 5.26 & 5.60 & 5.74 & 12.91 & 56.40 & 12.88 & 6.40 & 1.61 & 0.15 & 1.44 & 2.92 & 1.42 \\
        30 & 0.1 & 0.6 & 557 & 0.5 & 12.70 & 12.35 & 9.81 & 0.51 & 5.31 & 5.61 & 5.32 & 12.35 & 28.30 & 9.36 & 6.43 & 1.62 & 0.25 & 1.45 & 3.01 & 1.42 \\
        30 & 0.1 & 0.7 & 639 & 0.5 & 13.90 & 12.39 & 9.95 & 0.54 & 5.30 & 5.62 & 6.22 & 12.39 & 30.30 & 9.16 & 5.27 & 1.56 & 0.22 & 1.40 & 1.82 & 1.67 \\
        30 & 0.5 & 0.3 & 244 & 1 & 7.77 & 14.54 & 8.65 & 0.58 & 5.34 & 5.82 & 7.33 & 14.54 & 69.30 & 14.39 & 6.89 & 1.71 & 0.32 & 1.52 & 3.65 & 1.39 \\
        30 & 0.5 & 0.4 & 322 & 1 & 10.10 & 9.53 & 9.23 & 0.61 & 5.27 & 5.63 & 3.00 & 9.53 & 28.90 & 9.46 & 4.52 & 1.54 & 0.16 & 1.38 & 2.84 & 1.92 \\
        30 & 0.5 & 0.5 & 398 & 1 & 12.10 & 9.07 & 9.42 & 0.49 & 5.27 & 5.42 & 2.67 & 9.07 & 14.40 & 6.86 & 3.88 & 1.52 & 0.20 & 1.37 & 1.43 & 2.22 \\
        30 & 0.5 & 0.6 & 471 & 1 & 13.90 & 9.14 & 9.63 & 0.63 & 5.20 & 5.36 & 2.89 & 9.14 & 27.70 & 9.07 & 5.46 & 1.48 & 0.11 & 1.34 & 2.42 & 1.54 \\
        30 & 0.5 & 0.7 & 543 & 1 & 15.30 & 9.53 & 9.76 & 0.69 & 5.19 & 5.40 & 3.21 & 9.53 & 17.40 & 7.20 & 5.45 & 1.49 & 0.13 & 1.34 & 1.52 & 1.55 \\
        30 & 0.5 & 0.8 & 610 & 1 & 16.20 & 9.42 & 9.50 & 0.62 & 5.21 & 5.39 & 3.04 & 9.42 & 29.20 & 9.35 & 4.71 & 1.53 & 0.14 & 1.37 & 2.02 & 1.84 \\
\enddata
\end{deluxetable*}
\end{longrotatetable}
Notes. The columns are the initial mass $M$ (Column 1), the initial metallicity $Z$ (Column 2), the initial rotation rate $\Omega$ (Column 3) and corresponding velocity at the equator $V_{\rm equ}$ (Column 4), the ``Dutch'' wind scale factor $\eta_{\rm wind}$ (Column 5), the initial total angular momentum $J_{\rm ini}$ (Column 6), final mass $M_{\rm f}$ (Column 7), final star ages (Column 8), final radius $R_{\rm f}$ (Column 9), final effective temperature $\log T_{\rm eff}$ (Column 10) and luminosity $\log L$ (Column 11) in logarithmic form, final angular momentum in He core $J_{\rm He~core}$ (Column 12), He core mass at the end of calculation $M_{\rm He~core}$ (Column 13), final angular momentum in CO core $J_{\rm CO~core}$ (Column 14), CO core mass at the end of calculation $M_{\rm CO~core}$ (Column 15), final angular momentum in Fe core $J_{\rm Fe~core}$ (Column 16), Fe core mass at the end calculation $M_{\rm Fe~core}$ (Column 17), the compactness parameter $\xi_{2.5}$ (Column 18), the NS mass $M_{\rm NS}$ (Column 19), the average surface magnetic field strength $B_{\rm NS}$ (Column 20), and NS rotation period $P_{\rm NS}$ (Column 21). He, CO and Fe core masses have been measured using
$10\%$ mass fraction point X(H) $\leq 0.1$ and X(He) $\geq 0.1$ for $M_{\rm He~core}$, X(He) $\leq 0.1$ and X(C+O) $\geq 0.1$ for $M_{\rm CO~core}$, X(Si)$\leq 0.1$ and X(iron) $\geq 0.1$ for $M_{\rm Fe~core}$. For all cases, “iron” includes any species with A > 46.
\label{TABpreSN}
\end{document}